\documentclass[aps,prd,superscriptaddress,10pt,showpacs,notitlepage]{revtex4}
	\usepackage{graphicx}
	\usepackage{amsmath,amssymb,mathrsfs}
\usepackage{epsf}
\usepackage{epsfig}

\usepackage{color}

\usepackage{amsmath}

\usepackage{amssymb}

\numberwithin{equation}{section}

\usepackage{subfigure}

\newcommand{\be}{\begin{equation}}
\newcommand{\ee}{\end{equation}}
\newcommand{\bea}{\begin{eqnarray}}
\newcommand{\eea}{\end{eqnarray}}

\newcommand{\bb}{\bibitem}
 
\newcommand{\eqn}{\begin{eqnarray}}
\newcommand{\eqnx}{\end{eqnarray}}

\begin{document}
\title{Radial vibrations of BPS skyrmions}

\author{C. Adam}
\affiliation{Departamento de F\'isica de Part\'iculas, Universidad de Santiago de Compostela and Instituto Galego de F\'isica de Altas Enerxias (IGFAE) E-15782 Santiago de Compostela, Spain}
\author{M. Haberichter}
\affiliation{Department of Physics, and Research and Education Center for Natural
Sciences, Keio University, Hiyoshi 4-1-1, Yokohama, Kanagawa 223-8521,
Japan}
\author{T. Romanczukiewicz}
\affiliation{Institute of Physics,  Jagiellonian University,
Lojasiewicza 11, Krak\'{o}w, Poland}
\author{A. Wereszczynski}
\affiliation{Institute of Physics,  Jagiellonian University,
Lojasiewicza 11, Krak\'{o}w, Poland}

\begin{abstract}
We study radial vibrations of spherically symmetric skyrmions in the BPS Skyrme model. Concretely, we numerically solve the linearised field equations for small fluctuations in a skyrmion background, both for linearly stable oscillations and for (unstable) resonances. This is complemented by numerical solutions of the full nonlinear system, which confirm all the results of the linear analysis. In all cases, the resulting fundamental excitation provides a rather accurate value for the Roper resonance, supporting the hypothesis that the BPS Skyrme model already gives a reasonable approximate description of this resonance.  Further, for many potentials additional higher resonances appear, again in agreement with known experimental results.
\end{abstract}
\maketitle 

\section{Introduction}
The Skyrme model \cite{skyrme} is a nonlinear field theory of scalar (pion) fields which is considered a good candidate for the low-energy effective field theory (EFT) of QCD \cite{thooft} and, as such, should provide a realistic description of the physical properties of nucleons, atomic nuclei and nuclear matter. In a first approximation, nucleons and nuclei are described by topological soliton solutions which the model supports ("skyrmions")
\cite{skyrme}, \cite{manton}. This description may be further improved by the quantization of certain symmetry modes (most importantly, spin and isospin \cite{adkins} - \cite{wood}, which are good quantum numbers of nuclei), or by the inclusion of Coulomb energy contributions. The Skyrme model is very successful on a qualitative level, because it incorporates in a  natural way many nontrivial properties of nuclei (e.g., baryon number conservation, nuclear binding, rotational bands (excitation spectra), etc.), as should be expected for a unified EFT description. 

Quantitatively, the original version of the model as introduced by Skyrme typically leads to a precision of about 30\% when applied to nucleons and light nuclei (although there are situations where a much better precision may be achieved, see e.g. \cite{lau}). When a description of both light and heavy nuclei is attempted, the original Skyrme model faces the problem that the predicted nuclear binding energies are too high. This last problem, however, may be easily remedied by considering a more general class of models which contain the original proposal as a submodel. Under some mild assumptions (Poincare invariance and the existence of a standard hamiltonian), the most general Skyrme model is (here $U$ is the SU(2)-valued Skyrme field, and the $\lambda_i$ are dimensionful coupling constants)
\begin{equation} \label{gen-lag}
\mathcal{L}=
 \mathcal{L}_0+
\mathcal{L}_2+ \mathcal{L}_4 +\mathcal{L}_6
\end{equation}  
where
\begin{equation}
\mathcal{L}_2=-\lambda_2 \mbox{Tr}\; (L_\mu L^\mu), \;\;\; \mathcal{L}_4=\lambda_4 \mbox{Tr} \; ([L_\mu , L_\nu]^2), \;\;\;  L_\mu \equiv U^\dagger \partial_\mu U
\end{equation}
are the kinetic (sigma model) and Skyrme terms, constituting the original Skyrme model. Further, $\mathcal{L}_0 = -\lambda_0 \mathcal{U} (U)$ is a potential (non-derivative) term, where in this paper we shall exclusively consider potentials which only depend on the trace of the Skyrme field ${\rm tr}\, U$, thereby breaking the full chiral symmetry ${\rm SU}(2)_L \times {\rm SU}(2)_R$ down to the isospin symmetry ${\rm SU}(2)_{\rm diag}$. 
 Finally, the sextic term \cite{omega-B} - \cite{sextic-skyrme}
\begin{equation} \label{BPSmodel}
 \mathcal{L}_6= - \lambda_6 \mathcal{B}_\mu \mathcal{B}^\mu 
\end{equation}
is just the square of the baryon current (here, $B$ is the baryon number)
\begin{equation} \label{top-curr}
\mathcal{B}^\mu = \frac{1}{24\pi^2} \epsilon^{\mu \nu \rho \sigma} \mbox{Tr} \; L_\nu L_\rho L_\sigma, \quad B\equiv 
\int d^3 x \mathcal{B}^0.
\end{equation}
 Both the original Skyrme model $\mathcal{L}_{\rm Sk} = \mathcal{L}_2 + \mathcal{L}_4$ and the massive Skyrme model $\mathcal{L}_{\rm mSk} = \mathcal{L}_{\rm Sk} - \lambda_0 \mathcal{U}_\pi$ (where the pion mass potential $\mathcal{U}_\pi = (1/2)({\bf 1} -{\rm tr}\, U)$ is added) share the problem of the too high binding energies. One simple way to ameliorate this problem is to add further terms to the potential like, e.g., higher powers of the pion mass potential \cite{BPSM}, \cite{bjarke1}, $\mathcal{L}_0 = \sum_a \lambda_{0,a}\, \mathcal{U}_\pi^a$. These higher powers introduce a short-range repulsion between individual nucleons (i.e., $B=1$ skyrmions), thereby reducing the overall binding energies.

The second way to cure the binding energy problem consists in including both a potential and the sextic term $\mathcal{L}_6$ into the model and in choosing values for their coupling constants such that they provide significant contributions to the total skyrmion (nuclear) masses. This possibility is related to the fact that the submodel
\begin{equation} \label{BPSmodel}
\mathcal{L}_{BPS} \equiv \mathcal{L}_6+\mathcal{L}_0 \equiv - \lambda_6 \mathcal{B}_\mu \mathcal{B}^\mu- \lambda_0 \mathcal{U}
\end{equation}
(the so-called BPS Skyrme model \cite{BPS}, \cite{review}) has the BPS property, that is, an energy bound linear in the baryon number, $E\ge \mathcal{C}_\mathcal{U}|B|$ (here $\mathcal{C}_\mathcal{U}$ is a constant which depends on the potential), and skyrmion solutions saturating the bound for arbitrary $B$. Classical binding energies are, therefore, zero, and small nonzero binding energies may be generated, either already for the BPS submodel, by  including further energy contributions (spin and isospin excitations, Coulomb energy \cite{bind}), or by adding more terms to the lagrangian \cite{Marl1} - \cite{Sp2}, leading to the general lagrangian (\ref{gen-lag}).  
There exist some further proposals to solve the binding energy problem, which require, however, to go beyond the field content of the Skyrme model (e.g., by including vector mesons \cite{SutBPS}, \cite{rho1}) with the ensuing complications this implies.

In addition to the spin and isospin excitations, there exist further collective degrees of freedom like, e.g., vibrational modes (which are, in general, not related to symmetries, i.e., they are not moduli), whose excitations will be relevant for the description of the spectra of nuclei and other physical properties of  nuclear matter.  The identification of the correct vibrational modes of a given skyrmion solution is, in general, a difficult problem (for a recent calculation see, e.g., \cite{halcrow}), but for skyrmions with spherically symmetric baryon and energy densities there exists the much simpler possibility of purely radial vibrations (monopole excitations). In the $B=1$ sector, radial vibrations will be related to the so-called Roper resonance \cite{hajduk} - \cite{BPS-rot-vib}, whereas for higher $B$ radial vibrations may be relevant for the description of the giant monopole resonance (a spherically symmetric collective vibration which is found for many heavier nuclei) or for the determination of the compression modulus of nuclear matter.

Within the Skyrme model context, a skyrmion may be expanded into purely radially symmetric vibration modes only for skyrmions with a spherically symmetric energy density. For the standard Skyrme model (and for generalizations (\ref{gen-lag}) with all four terms present) this means that such a mode expansion is possible only for the $B=1$ skyrmion, because only this skyrmion is radially symmetric. For non-spherical higher $B$ skyrmions, radial modes may either be described in an approximation (e.g., by collective modes, like the uniform rescaling (Derrick mode), or by approximating a crystal-type large $B$ skyrmion by a lattice of point nucleons or other substructures and by identifying the vibration modes of this lattice), or by a description which goes beyond the excitation of a single skyrmion (e.g., by averaging over differently oriented non-spherical skyrmions corresponding to the same nucleus). The BPS submodel (\ref{BPSmodel}), on the other hand, allows for an axially symmetric ansatz leading to skyrmions with a radially symmetric energy density for arbitrary baryon number $B$, so radial excitation modes may be studied for any $B$ in this case. 

In this paper, we shall study  the radial vibrational excitations in the BPS Skyrme model for arbitrary $B$ (a detailed investigation of the radial vibrations of $B=1$ skyrmions in generalized Skyrme models and its application to the Roper resonance will be published in a forthcoming paper).
Radial excitations may be studied using various methods with different levels of precision and rigor \cite{Campbell} - \cite{bizon}. The simplest description assumes that the excitation may be described by a uniform rescaling (Derrick scaling) of the field variables, $U(t,\vec x) = U_0 (\vec x/\rho (t))$, where $U_0$ is the static skyrmion solution. Inserting this rescaled field into the action and expanding up to second order in the small fluctuation $q \equiv \rho -1$, leads to the harmonic oscillator action for the variable $q$, and to the corresponding harmonic oscillator frequency. The variable $q$ may then be promoted to a quantum mechanical variable, leading to the quantum harmonic oscillator. This corresponds to a truncated quantization of the field $U$, where the infinitely many degrees of freedom of the field are approximated by the finitely many d.o.f. of certain collective modes. In the concrete case just described, there is only one mode (the Derrick mode $\rho$), but some further collective modes like, e.g., collective (rigid) rotations of the skyrmion or non-radial vibrational modes  may, in principle, be included in this truncated quantization. 
The uniform rescaling is, in general, not a true vibrational eigenmode of the skyrmion, but at most an approximation, whose accuracy should be checked by comparison with more precise results. It is, however, implicitly or explicitly used in many descriptions of nuclear matter, in particular in macroscopic descriptions where nuclear matter is modeled as a (perfect or viscous) fluid.

The true linear eigenmodes (eigenmodes for small fluctuations) of a skyrmion are calculated by inserting the small perturbation $U = U_0 (\vec x)+ \delta U(t, \vec x)$ into the action and by expanding up to second order in $\delta U$ (the simplified expressions for $U_0$ and for $\delta U$
implied by spherical symmetry will be given in Section III). This procedure leads to a linear, Schr\"odinger-type equation for $\delta U$ in an effective potential $Q(r)$, whose eigenstates and eigenvalues provide the true linear excitations and their frequencies. Depending on the value of a certain mass threshold in the theory (which is related to the value of the effective potential $Q(r)$ at infinity), there may exist zero, finitely many or infinitely many eigenmodes. Here, the condition for the existence of an eigenmode is always that its eigenfrequency must lie below the mass threshold. These eigenmodels are of physical relevance both in a classical and in a quantum context. Classically, they describe the vibrational eigenmodes of the skyrmion (nucleon or nucleus). Within semi-classical quantization, each eigenmode is quantized as an independent harmonic oscillator. Each mode $\delta U_i$ , therefore, contributes a term $(1/2) \hbar \omega_i$ to the mass of the quantized skyrmion (here, the scattering states also contribute, so the mass of the quantum skyrmion must be renormalized in all cases). Further, the excitation energies of excited states (nucleus-meson bound states) are given by $\hbar n_i \omega_i$ (where neither $n_i$ nor $\omega_i$ should be too large for the semi-classical approximation to be valid). Concretely, the lowest excitation energy is $\hbar \omega_1$.

Eigenmodes below the mass threshold are linearly stable. Nonlinearities in the theory, on the other hand, effectively couple several eigenmodes leading to a total frequency $\sum_i n_i\omega_i$ above the mass threshold, which allows energy to be radiated away. This requires, however, the presence of sufficient energy to begin with, or a subsequent injection of additional energy into the system (e.g., via energy non-conserving boundary conditions). In the quantum theory, the linearly stable eigenmodes correspond to stable bound states.  Again, these stable bound states may be broken up into their constituents if enough energy is pumped into the system.

Certain frequencies above the mass threshold, sometimes, provide strongly enhanced contributions to the time evolution of a weakly perturbed  skyrmion. In the linearized analysis, these frequencies correspond to resonances or quasinormal modes, that is, to solutions for the linear fluctuation equations for complex frequencies. The corresponding ``eigenmodes", therefore, have an exponentially decaying instead of a time-independent norm. Physically, they correspond to exponentially damped vibrations in a classical context or to unstable bound states (resonances) in a quantum description. [We remark that resonance modes within the BPS Skyrme model for a certain class of potentials have been studied numerically in \cite{arpad}.]

Finally, the full nonlinear time evolution may be investigated for the classical field theory. 
When taking a weakly perturbed Skyrmion as initial
condition, it is expected that the frequencies of the normal and/or quasinormal modes dominate the time evolution at least for not too large evolution times. In the large time limit, on the other hand, the leading behaviour of a solution may be completely different and will be determined by the asymptotic properties of the full nonlinear system. A corresponding quantum description of the full nonlinear field theory (beyond the semi-classical quantization approximation) is much more difficult to find (or even define), as the underlying model is perturbatively non-renormalizable.

\vspace*{0.2cm}
Our paper is organized as follows. In Section II, we consider some scalar field theories in 1+1 dimensions as simple toy models where all the features mentioned in the preceding paragraphs may be investigated in a simpler setting. Concretely, we consider both the uniform scaling approximation and the full linear fluctuation analysis (normal and quasinormal modes) and study the accuracy and adequacy of the former.
Further, we calculate the full nonlinear time evolution numerically.
In Section III we do the same analysis for the BPS Skyrme model. First, we consider a particular limiting potential where both the scaling solution and the linear fluctuation analysis may be done analytically. Then we introduce a more general class of potentials where some steps have to be done numerically. We also numerically calculate the full nonlinear time evolution and compare with the results of the linearized analysis.
Section IV contains our conclusions, where also physical implications of our results are discussed. Some more technical discussions are relegated to three appendices.

  
\section{(1+1) dimensional examples}
\subsection{The collective mode approximation}
As a first step, we want to compare the accuracy of the Derrick (uniform rescaling) collective mode with the known true oscillational modes of some simple theories, concretely the $\phi^4$ and sine-Gordon models. The following analysis can, in principle, be extended to other solitons, but it is difficult to obtain it in a general form. The Lagrangian in 1+1 d has the form
\begin{equation}
 L=\int dx\; \left\{\frac12\phi_t^2-\frac12\phi_x^2-U(\phi)\right\}
\end{equation} 
where $\phi$ is a real scalar field and $U$ is a potential.
For the collective mode approximation, we insert the ansatz $\phi(x,t)=\Phi\left(x/\rho(t)\right)$ into the lagrangian, where $\Phi(x)$ is a static solution to the evolution equation (i.e., a soliton).
After changing the variables $y=x/\rho$ we obtain:
\begin{equation}
 L = \int \rho dy\;\left\{\frac{y^2\Phi_y^2}{2\rho^2}\dot{\rho}^2-\frac{\Phi_y^2}{2\rho^2}-U(\Phi)\right\},
\end{equation} 
and after integration
\begin{equation}
 L=\frac{\dot\rho^2}{2\rho}A-\frac{M}{2}\left(\frac{1}{\rho}+\rho\right)
\end{equation} 
where 
\begin{equation}
 A = \int dy\;y^2\Phi_y^2
\end{equation} 
and 
\begin{equation}
 M=\int dy\;\Phi_y^2=2\int dy\; U(\Phi(y))
\end{equation} 
is usually referred to as the mass of the soliton.\\
Assuming a small perturbation around the static solution $\rho=1+q$ with $|q|\ll1$ we obtain the Lagrangian for the harmonic oscillator:
\begin{equation}
 L = -M + \frac{A\dot{q}^2}{2}-\frac{Mq^2}{2}
\end{equation} 
with the frequency:
\begin{equation}
 \omega^2 = M/A.
\end{equation} 

\subsubsection{$\phi^4$ example}
\begin{figure}
\centering
\includegraphics[width=0.4\linewidth,angle=0]{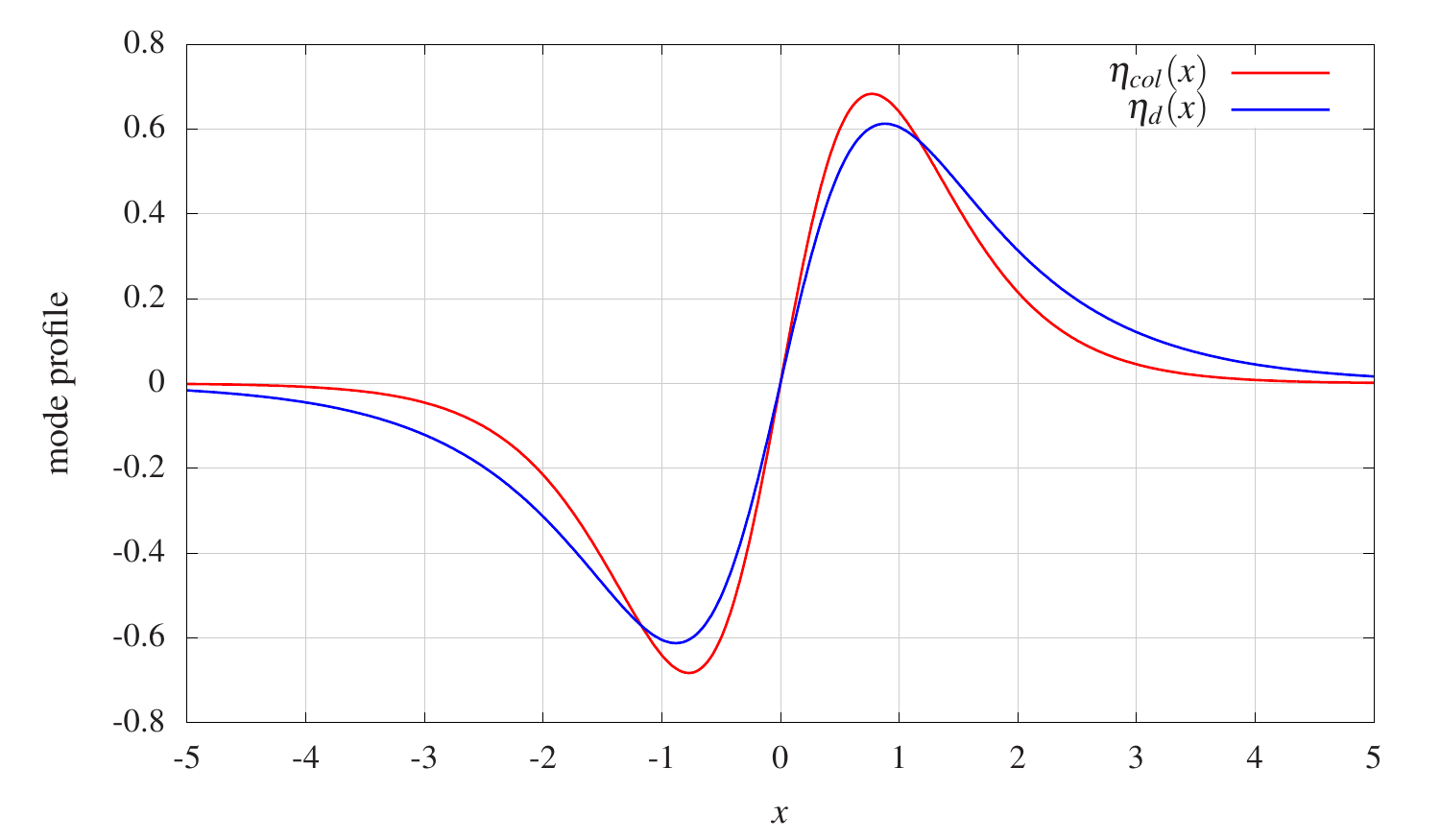}
\caption{\small Normalized modes (true oscillational and collective) for the $\phi^4$ model.}\label{Figuremodes1}
\end{figure}

In the $\phi^4$ case
\begin{equation}
 U(\phi)=\frac{1}{2}(\phi^2-1)^2,\;\;\Phi(x)=\tanh x
\end{equation} 
there exists a single oscillational mode 
\begin{equation}
 \eta_d(x)=\sqrt{\frac32}\frac{\tanh x}{\cosh x},\;\;\omega_d^2=3.
\end{equation} 
The mass threshold in this model is $m^2=4$. The integrals give $A=\frac{\pi^2-6}{9}$ and $M=\frac{4}{3}$. The collective excitation frequency is $\omega_c^2 = 
\frac{12}{\pi^2-6}\approx3.101$, which is pretty close to the accurate value of 3. Moreover, the profile of the assumed mode,
\begin{equation}
 \tilde\eta_{col}(x) = -\left.\frac{d}{dq}\tanh\left(\frac{x}{1+q}\right)\right|_{q=0}=\frac{x}{\cosh^2x}
\end{equation} 
or, after normalization, 
\begin{equation}
 \eta_{col}(x)=\frac{3}{\sqrt{\pi^2-6}}\frac{x}{\cosh^2x},
\end{equation} 
is also not very different from the true oscillational mode, see Figure \ref{Figuremodes1}.
The collective mode can be decomposed as 
\begin{equation}
 \eta_{col}(x) = (\eta_{col}, \eta_d)\eta_d(x) + \int_{-\infty}^{\infty}\!dk\;(\eta_{col},\eta_{k})\eta_k(x)
\end{equation} 
where 
\begin{equation}
 (\eta_{col}, \eta_d)=\pi\sqrt{\frac{3}{8(\pi^2-6)}}=0.977985
\end{equation} 
is the projection of the collective mode on the discrete oscillational mode.
Note that the collective mode consists of the oscillational mode to 98\%.
\begin{equation}
 \eta_{k}(x) = \frac{3\tanh^2x-3ik\tanh x-1-k^2}{\sqrt{(k^2+1)(k^2+4)}}e^{ikx}
\end{equation} 
are scattering modes for wave number $k=\pm\sqrt{\omega^2-4}$, and the projection of the collective mode onto the scattering modes can be expressed as
\begin{equation}
(\eta_{col},\eta_{k}) = \frac{3ik^2}{4\sinh\left(\frac{k\pi}{2}\right)\sqrt{(\pi^2-6)(k^2+1)(k^2+4)}}.
\end{equation} 
The linearized evolution of the excited collective mode $\phi(x,0)=\Phi(x)+A\eta_{col}(x),\phi_t(x,0)=0$ can be obtained from
\begin{equation}
 \phi(x,t) =\Phi(x)+A(\eta_{col}, \eta_d)\eta_d(x)\cos(\omega_d t) + 
A\int_{-\infty}^{\infty}\!dk\;(\eta_{col},\eta_{k})\eta_k(x)\cos(\omega_k t) +\mathcal{O}(A^2).
\end{equation} 
The scattering modes describe waves moving away from the soliton. After a long time only the oscillational mode will remain. 
This mode however will also decay via nonlinear coupling to scattering modes. 

\subsubsection{Sine-Gordon example}
In the sine-Gordon (sG) case
\begin{equation}
 U(\phi)=1-\cos^2\phi,\;\;\Phi(x)=4\tan^{-1}\left(\exp(x)\right)
\end{equation} 
there is no oscillational mode.
The mass threshold in this model is $m^2=1$. The integrals give $A=\frac{2\pi^2}{3}$ and $M=8$. The collective excitation frequency is $\omega_c^2 
= \frac{12}{\pi^2}\approx1.21585$, which is above the mass threshold. The collective excitation mode 
\begin{equation}
 \eta_{col}(x) = \frac{\sqrt6}{\pi}\frac{x}{\cosh(x)}
\end{equation} 
can be decomposed in the eigenmodes of the sG spectrum:
\begin{equation}
 \eta_k(x) = \frac{ik-\tanh x}{\sqrt{1+k^2}}e^{ikx},\;\;k^2=\omega^2-1
\end{equation} 
as
\begin{equation}
 \eta_{col}(x) = -\frac{1}{2\pi}\int_{-\infty}^{\infty}\!\!dk\;\frac{\sqrt{6}\;\eta_k(x)}{\cosh\left(\frac{k\pi}{2}\right)\sqrt{1+k^2}}.
\end{equation} 

Note that in the sine-Gordon model there is no natural candidate which would take over the evolution of the excited collective mode. 
Therefore, we can conclude that such a collective mode does not exist. The evolution is determined by scattering modes radiating away the energy. 
The long time evolution will be dominated by modes with $k\to0$ (or $\omega\to 1$) which propagate very slowly and take a long time to 
be radiated out. 
We will discuss this problem in more detail in the following section.
\vspace*{0.3cm}

From the above analysis it is clear that the collective excitation approach can be misleading. Sometimes it works quite well as in the $\phi^4$ model, where the 
profile of the collective mode is very similar to the true oscillational mode. However, it can also fail completely, as in the sine-Gordon model, where the collective approach 
gave some profile and frequency which does not reproduce any feature of the full evolution of the system.

\subsection{The double sine-Gordon model}
The two models of the last subsection are rather special and, thus, do not allow to study the full range of phenomena entailed by nonlinear field theories, in general. 
Let us, therefore, consider the double sine-Gordon model
\begin{equation}
  \mathcal{L} = \frac{1}{2}\phi_t-\frac{1}{2}\phi_x^2-1+\cos(\phi)+\epsilon\cos(2\phi).
\end{equation} 
with Euler-Lagrange equation
\begin{equation}
 \phi_{tt}-\phi_{xx} + \sin(\phi)+2\epsilon\sin(2\phi)=0
\end{equation} 
as an example of a more general model.
The static solution can be written as
\begin{equation}
 \phi_0(x) = \pi+2\tan^{-1}\left[\frac{\sinh\left(\sqrt{1+4\epsilon}\;x\right)}{\sqrt{1+4\epsilon}}\right].
\end{equation} 
Small perturbations around the static solution have mass $m^2=U''(\phi_{vac})=1+4\epsilon$. For $\epsilon=0$, the model is just the sine-Gordon model.
For $\epsilon=-1/4$ the model becomes massless. One internal, oscillational  mode exists for $\epsilon>0$.

One of the techniques to study the evolution of small perturbations around static solution is the linear approximation. We express the field as $\phi = \phi_0(x)+\xi(x,t)$. Next, we expand the Euler-Lagrange equation assuming that $\xi$ is small, and keep only terms proportional to the first power of $\xi$. The resulting field 
equation 
usually has the form of a Schr\"odinger equation, and the solution can be decomposed in terms of the modes $\xi(x,t)=e^{i\omega t}\eta_k(x)$, where $k=k(\omega)$ is a 
wave number. The linear Schr\"odinger equation can have discrete or continuous spectrum or both. The existence of discrete modes plays a very important role 
for the late time evolution. \\

\begin{figure}
\centering
\includegraphics[width=1\linewidth,angle=0]{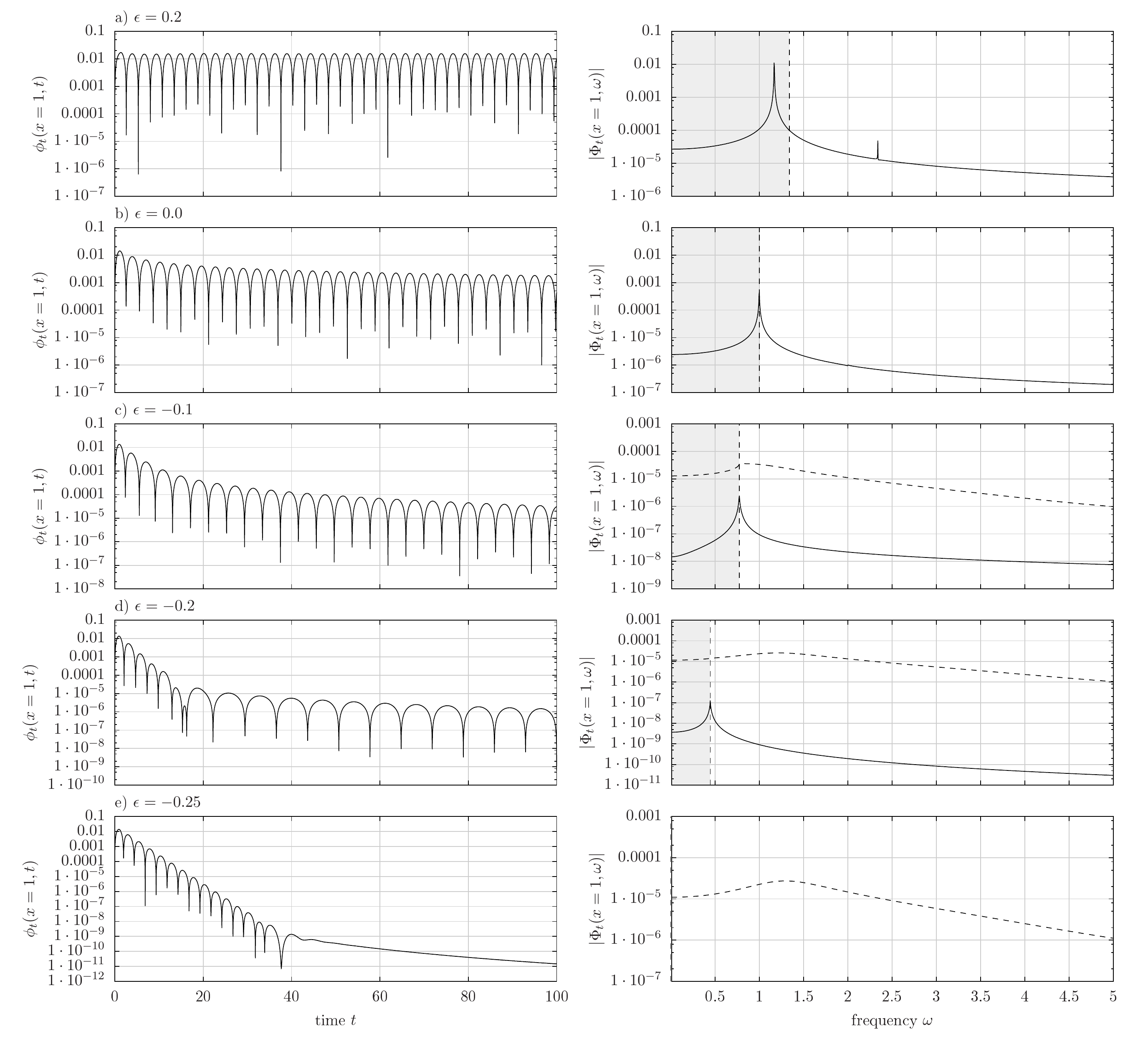}
\caption{\small Relaxation of the field for squeezed soliton initial conditions and their spectra for different values of the parameter $\epsilon$. 
The solid lines represent FFT (Fast Fourier Transform) for $t\in[100,1800]$, the dashed include earlier times $t\in[0,1800]$. The gray areas 
represent the non-propagating region below the mass threshold.
a) $\epsilon=0.2$ oscillational mode excited, almost constant amplitude, small radiation via second harmonics; b) $\epsilon=0$ pure sG 
case, slow modes on the threshold $\omega=1$ dominate long time evolution; c) $\epsilon=-0.1$, fast decay of the resonance mode and 
remaining tail at the mass threshold; d) similar to c) but longer decay of the resonance and clearly seen difference in frequency e) 
$\epsilon=-0.25$ massless case with decaying resonance and polynomial tail described by Bizo\'n \cite{bizon}.}\label{Figurespectra1}
\end{figure}

\textit{Oscillational mode.} Usually, the perturbation in the continuous spectrum is radiated out and the discrete mode remains. After squeezing the soliton, the 
vibrational or oscillational mode is being excited. This mode remains for quite a long time. In fact, in the linearized approximation the discrete mode remains 
excited forever and oscillates with constant amplitude and frequency. However, including higher nonlinear terms shows that this mode usually couples to 
scattering modes and loses its energy via radiation through higher harmonics. If the oscillational mode is not very much below the mass threshold, i.e.,  
$\omega_d>m/2$, already the $2\omega_d$ mode carries away energy. As Manton et al \cite{MantonMerabet}  showed, this 
leads to the decay of the mode:
\begin{equation}
 \frac{dE}{dt}\sim -A^4\Rightarrow \frac{dA}{dt} \sim -A^{3}\Rightarrow A\sim t^{-1/2}.
\end{equation}
An example of this process is depicted in Figure \ref{Figurespectra1} a). For $\epsilon=0.2$ there exists a single oscillational mode with the frequency 
$\omega=1.2681$. In the spectrum, one can see the second harmonics which carries away the energy from the perturbation. Because of the small amplitude, this 
process is very slow, and only a very small  change of the amplitude can be noticed during the simulation.\\

\textit{Mass threshold.} If there are no oscillational modes in the spectrum, another possibility arises. The field can be expressed in term of the 
eigenmodes:
\begin{equation}
 \xi(x,t)=\int_{-\infty}^{\infty} dk\; c(k)e^{i\omega(k)t} \eta_k(x)+c.c., 
\end{equation} 
where the $c(k)$ represent the mode excitations. If the potential tends to a constant $m^2$ at infinity, the wave number obeys the standard 
dispersion relation $k^2=\omega^2-m^2$. The waves with high frequency move faster, and the waves close to the threshold move slowly. After a long time we can 
expect that only the waves near the mass threshold remain. Expanding the frequency around $k=0$, the solution can be expressed in the form:
\begin{equation}
 \xi(x,t)\approx e^{imt}\int_{-\infty}^{\infty} dk\; c(k)e^{\frac{ik^2t}{2m}}\eta_k(x)+c.c..
\end{equation} 
The exponent, in the limit of large times, tends to $\sqrt{2m\pi i/t}\,\delta(k)$, and the long time behavior of the solution can be found as
\begin{equation}
 \xi(x,t)\sim\frac{\cos\left(mt+\pi/4\right)}{\sqrt{t}}.
\end{equation}
Note that the leading term has the same behavior as in the previous case $t^{-1/2}$, however its origin is completely different. The previous decay was nonlinear, 
and here the whole analysis used the linear approximation. In the previous case, the evolution was described by a single mode. Here, we actually have no single mode 
but rather a combination of many modes with no well defined shape. This process dominates Figures \ref{Figurespectra1} b), c), d).

The above analysis is not valid when $m\to 0$. In this case, the explicit form of $\eta_k(x)$ as $k\to 
0$ is needed.\\

\textit{Resonance mode.} The other conjecture that was used was that $c(k)$ had no poles in the complex plane. This assumption does not hold when there are 
resonance modes in the spectrum. In this case, after integration, other frequencies remain in the long time evolution of the system. The poles are in the 
complex plane, so the imaginary part is responsible for the damping. The quasi-normal modes behave like:
\begin{equation}
 \xi(x,t)\sim e^{-\Gamma t}\cos(\omega t).
\end{equation} 
Usually, the resonance mode with the smallest damping $\Gamma$ remains for the longest time. The presence of this mode can be seen in Figures \ref{Figurespectra1} 
c), d), e). In the first two cases, the mode decays faster than the threshold modes, and the previously described mechanism dominates the long time evolution. In the 
last case, the field is massless. The Schr\"odinger potential has an especially simple form:
\begin{equation}
 V_{\epsilon=-1/4}(x)=\frac{6x^2-2}{(1+x^2)^2}.
\end{equation} 
This potential has a resonance mode with the frequency $\omega=1.2664+0.4441\,i$ which well fits the behavior of the full nonlinear case.\\
In this model, we can see only a single quasinormal mode. However, it is possible to have an infinite number of such modes. 
In \cite{Forgacs}, 
Forgacs et al show for
BPS monopoles how a small perturbation evolves when there is an infinite
number of resonance modes on a finite frequency range. 

Note: there is no resonance mode in the case of the pure sine-Gordon or $\phi^4$ model. \\

\textit{Nonlinear behavior.} For massive fields, another possibility can occur. Sometimes, even a low amplitude perturbation can have nonlinear properties. For 
example, the $\phi^4$ model exhibits oscillons which can last for very long time. They are nonintegrable counterparts of breathers in the sine-Gordon equation. Their 
basic frequency is below the mass threshold, implying that there can be no propagation at this frequency.  Usually, the oscillons radiate via higher harmonics, but the 
radiation is extremely small. When they radiate they lose energy, which leads to a decrease of the amplitude and an increase of the frequency. In higher dimensions, 
there exist some critical frequency above which the oscillon disappears very quickly. In 1+1d there is no such frequency, and the oscillon slowly evolves with its
frequency tending to the mass threshold from below. 

\begin{figure}
\centering
\includegraphics[width=1\linewidth,angle=0]{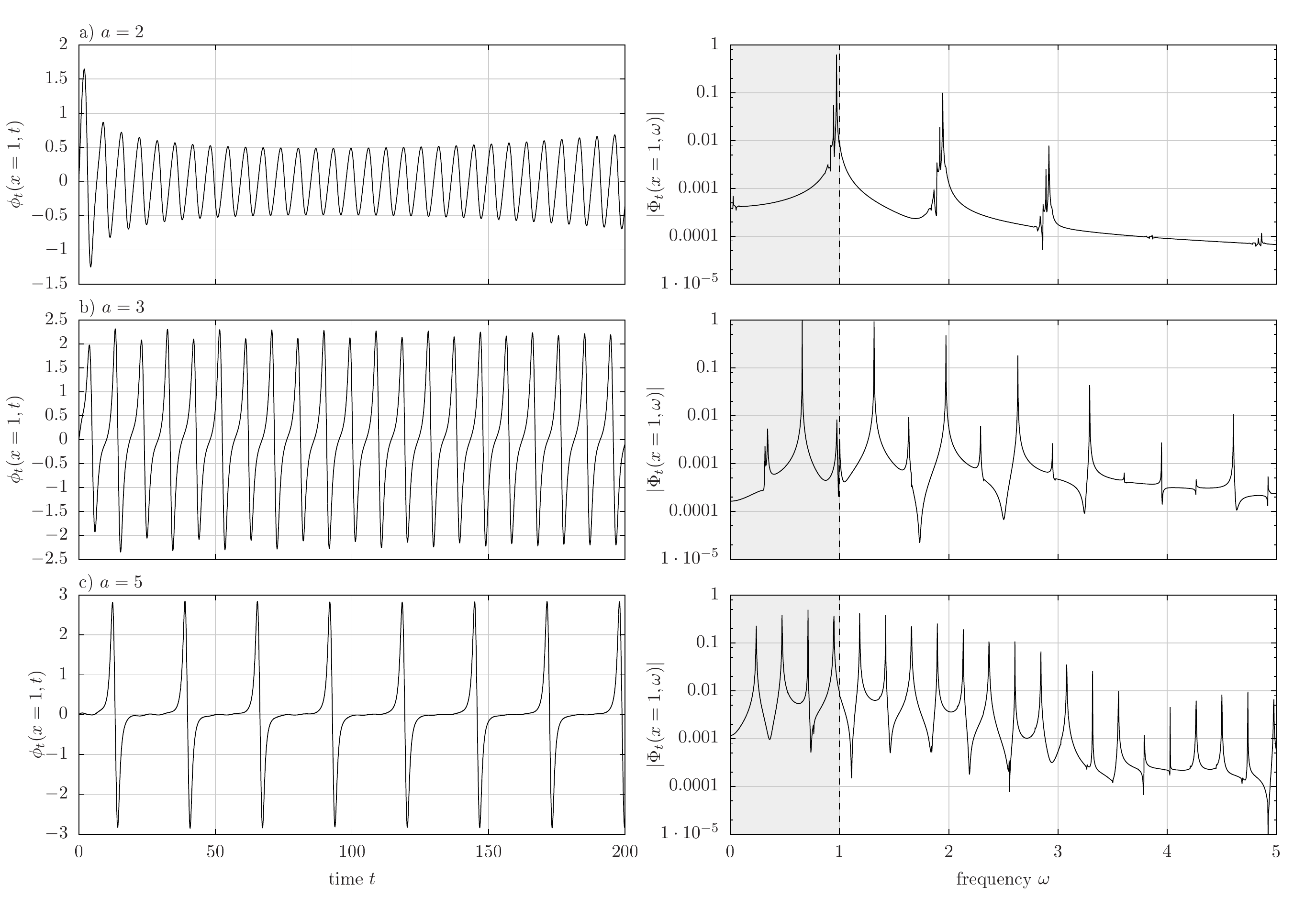}
\caption{\small Relaxation of the field for wobbler initial data for sine-Gordon (\ref{eq:wobbler}) for three different initial separations. 
The nonlinear nature of these objects, with basic frequency below the mass threshold and many higher harmonics, is clearly seen.}\label{FigurewobblersG}
\end{figure}

In the case of the pure sine-Gordon model, it is possible to find an exact analytic form of a breather with a soliton in the center. Such three-soliton solutions are 
well known. However, in the case when $\epsilon\neq0$ such periodic solutions do not exist.  They may appear as weakly unstable bound states of three solitons. We 
can prepare such a solution 
\begin{equation}
 \phi(x,t=0) = \phi_0(x)+\phi_0(x-a)+\phi_0(x+a)-2\pi\label{eq:wobbler}
\end{equation} 
and see how this configuration evolves. For $\epsilon=0$, the situation is more or less clear. The configuration is not a pure three soliton state, so some additional energy can be radiated out, but a nearly periodic solution eventually develops from these initial data. One of the manifestations of the nonlinear 
origin of this configuration is a quite complicated spectrum with the basic frequency below the mass threshold and visible higher harmonics, see Figure 
\ref{FigurewobblersG}. Note that in the sine-Gordon model
there are no oscillational modes, so any odd perturbation can be expressed as a linear combination of scattering modes with $\omega>1$. The persistence of such a
configuration is guaranteed by the complete integrability of the sine-Gordon model. However, even a small  $|\epsilon|=0.01$ spoils the integrability, and the evolution is 
governed by one of the previously described scenarios for the generic case. This may appear strange, because even for models which are clearly non-integrable ($\phi^4$ 
or $\phi^6$) oscillons exist. However, oscillons are two-soliton states and the configuration described by (\ref{eq:wobbler}) is a three-soliton state. 

For massless fields ($m=0$), another possibility is a nonlinear polynomial tails, as was pointed out by Bizo\'n \cite{bizon}. 

\vspace*{0.2cm}
For generic initial conditions, the long-time behavior of the field is determined by the existence of oscillational modes, a mass threshold 
and 
resonances (quasi-normal modes). The most persistent states from the linear approximation are the oscillational modes. Resonance modes decay exponentially fast, so 
their presence can be seen when either the oscillational modes or the slow modes on the threshold are not excited, or when there exist no oscillational modes, or for 
massless fields. Sometimes, the initial conditions can give rise to some long-living non-linear structures, like oscillons or polynomial tails. 
The above analysis is quite general, and squeezing the soliton is just one of the ways to perturb it. The numerical values of the excitations of the 
long-living modes depend on the way how the soliton was excited, but the leading behavior after long time is rather universal. 

However, we cannot see any reason why the squeezing of the soliton would generate some well defined frequency. We think that the model defines some time scale 
(such as mass of the scalar field or oscillational modes) which more or less agrees with the time scale defined by Derrick collective mode.

\begin{figure}
\centering
\includegraphics[width=0.5\linewidth,angle=0]{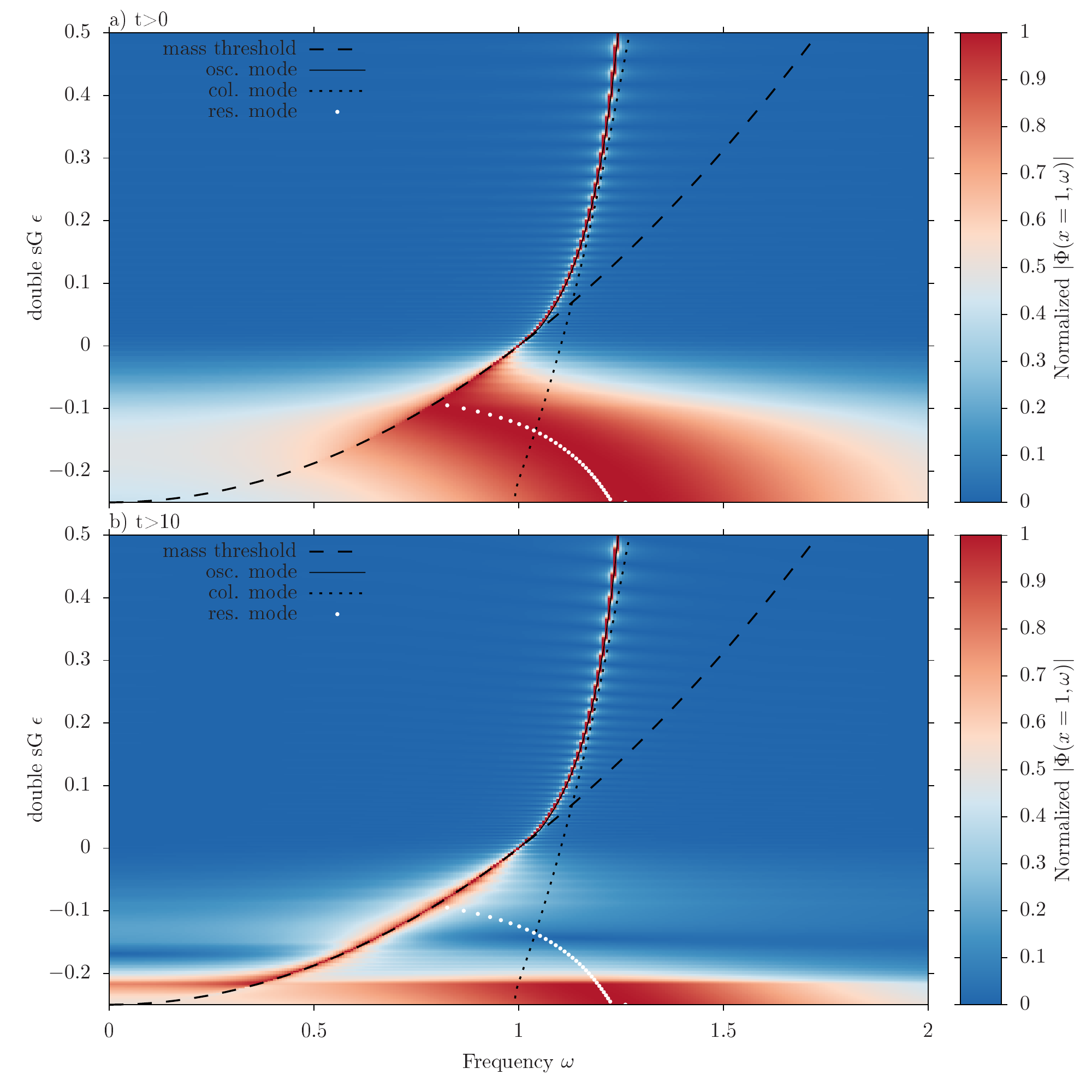}
\caption{\small Fast Fourier Transform (FFT) of the field at $x=1$ obtained for the time evolution of squeezed soliton initial data. The upper plot is obtained for all times and the bottom plot for $t>10$ to exclude the decayed resonance mode. The spectra were normalized in such a way that the maximum of the power spectrum for each 
$\epsilon$ is 1. For $\epsilon>0$, the evolution is clearly dominated by the oscillational mode. For $\epsilon<0$, both the resonance 
mode and mass threshold evolution dominate the behavior of the field. The resonance mode decays faster and, in the bottom plot it continues to be visible only for $\epsilon$ 
close to $-1/4$. The collective Derrick mode provides a reasonable approximation only for $\epsilon>0$, where it seems to be quite close to the
oscillational mode. However, for $\epsilon<0$, it misses both the mass threshold and the resonance mode.}\label{Figurecollective}
\end{figure}

Finally, also for the double sine-Gordon model we may find the frequency obtained from the collective mode approximation and compare it with the frequencies dominating the long 
time evolution of the perturbation around the soliton, see Figure \ref{Figurecollective}. For $\epsilon>0$, there exists an oscillational mode which dominates the long 
time behavior. For $-1/4<\epsilon < 0$ the most dominating frequency is the frequency of the mass threshold. For the massless case $\epsilon=-1/4$, the resonance 
mode dominates. From the figure we can see that the collective mode provides a reasonable approximation only in the vicinity of $\epsilon=0.3$. However, this is probably just a coincidence. 

\section{The BPS Skyrme model}
The Lagrangian of the BPS Skyrme model in standard form is \cite{BPS}
\begin{equation}
\mathcal{L}_{BPS} =-\lambda^2\pi^4 \mathcal{B}_\mu \mathcal{B}^\mu- \mu^2 \mathcal{U},
\end{equation}
but here we prefer to introduce
coupling constants with units of mass $\boldsymbol{m}$ and length $\boldsymbol{l}$,
\begin{equation}
\lambda^2 = \boldsymbol{m} \boldsymbol{l}^{3}\, ,\quad \mu^2 = \boldsymbol{m}\boldsymbol{l}^{-3}.
\end{equation}
The Skyrme field parametrization is
\begin{equation}
U= \exp (i\xi \vec\tau \cdot \vec n) = \cos \xi \, {\bf 1} + i \sin \xi \, \vec \tau \cdot \vec n
\end{equation}
where $\xi$ is a real field, $\vec\tau $ are the Pauli matrices, and
\begin{equation}
\vec n = (\sin \Theta \cos \Phi ,\sin \Theta \sin \Phi ,\cos \Theta )
\end{equation}
is a three-component unit vector. Our metric convention is $(+,-,-,-)$. Further, the potential $\mathcal{U}(\xi)$ is a function of the profile function $\xi$ only, where in the following we shall consider only potentials which have their unique vacuum at $\xi =0$ (i.e., at $U={\bf 1}$).

We emphasize that the BPS submodel (\ref{BPSmodel}) is just a certain limit of the generalized model (\ref{gen-lag}), where only the latter    
should be considered a genuine low-energy EFT of strong interaction physics (e.g., the term $\mathcal{L}_2$ is indispensable for pion propagation). The coupling constants in the full model (\ref{gen-lag}), however, should probably be such that $\mathcal{L}_6$ and $\mathcal{L}_0$ give the main contributions to the static skyrmion energies (the nuclear masses) - a ``near-BPS Skyrme model". This explains why certain (in particular, static) properties of the full near-BPS Skyrme model may be studied in the BPS submodel (which frequently allows for an analytical treatment), at least approximately. Mathematically, the fact that the BPS submodel should be considered only as a limit is reflected in its lack of a well-defined Cauchy problem. The sextic term $\mathcal{L}_6$ is the square of the bayon current, so, loosely speaking, the time evolution in a region on a given time slice might face problems whenever certain components of the baryon current are zero in that region. The baryon current is an antisymmetric product of gradients of the three Skyrme fields, so it is zero whenever these gradients are linearly dependent. This problem is immaterial for static configurations, and static BPS skyrmions may be calculated easily by solving the BPS equation. We shall find that, within the assumptions we use, the time evolution (for sufficiently small time intervals) of small radially symmetric perturbations about axially symmetric BPS skyrmions is well-defined, as well, and not affected by that problem.    

Concretely, for static configurations we  use the axially symmetric ansatz for general baryon number $B$,
\begin{equation} \label{ax-sym}
\xi = \xi (r) \, , \quad \Theta = \theta \, , \quad \Phi = B\phi ,
\end{equation}
which leads to a spherically symmetric energy density and baryon density. The three Skyrme field gradients are obviously linearly independent, so the baryon density is zero only if the Skyrme field takes its vacuum value $\xi =0$. 
\subsection{Small perturbation approximation - oscillating modes}
We want to consider radial (monopole) excitations, therefore we now use the ansatz $\xi = \xi (r,t)$$, \Theta = \theta$, $\Phi = B\phi $ for the time-dependent Skyrme field. This ansatz is compatible with the full four-dimensional Euler-Lagrange equations and leads to the reduced action \cite{arpad}
\begin{equation}
S= 4\pi \boldsymbol{m} \int dt dr \left( \frac{B^2 \boldsymbol{l}^3}{4r^2} \sin^4 \xi \left(\dot\xi^2 -\xi'^2 \right) -r^2\boldsymbol{l}^{-3} \mathcal{U} \right).
\end{equation}
The Euler-Lagrange (EL) equation for $\xi$ is
\begin{equation}
\frac{B^2 \boldsymbol{l}^6}{2r^2} \sin^4 \xi (\ddot \xi - \xi'') +\frac{B^2 \boldsymbol{l}^6}{r^2} \sin^3 \xi \cos \xi (\dot\xi^2 -\xi'^2)
+\frac{B^2\boldsymbol{l}^6}{r^3} \sin^4 \xi \, \xi' + r^2 \mathcal{U}_{,\xi} =0.
\end{equation}
Now we insert $\xi = \xi_0 (r) + \bar\eta (t,r)$, where $\xi_0 (r)$ is the static BPS solution and $\bar\eta (t,r)$ is the fluctuation field, and expand up to first order in $\bar\eta$ to get the following linear equation for $\bar\eta$
\begin{eqnarray}
&& \frac{1}{2}\sin^4 \xi_0 (\ddot {\bar\eta} - \bar\eta'') + (-2\sin^3 \xi_0 \cos \xi_0 \xi'_0 + \frac{1}{r}\sin^4 \xi_0 )\bar\eta' + \nonumber \\
&& \left(-2\sin^3 \xi_0 \cos \xi_0 \xi_0'' + (\sin^4 \xi_0 - 3 \sin^2 \xi_0 \cos^2 \xi_0 )\xi'{}_0^2 +\frac{4}{r} \sin^3 \xi_0 \cos \xi_0 \xi'_0\right)\bar\eta + \frac{r^4}{B^2 \boldsymbol{l}^6}\mathcal{U}_{,\xi\xi}\vert \bar\eta =0 . \label{lin-eq}
\end{eqnarray}
Here and below, the vertical bar notation means evaluation at the skyrmion solution, i.e., $\mathcal{U}_{,\xi\xi}\vert \equiv \mathcal{U}_{,\xi\xi}\vert_{\xi = \xi_0}$ etc.
$\bar\eta$ must be zero at $r=0$ in order to maintain the skyrmion topology.
Further, for potentials $\mathcal{U}(\xi)$ with a less than sextic approach to the vacuum (i.e., such that $\lim_{\xi \to 0} \mathcal{U} \sim \xi^b $ for $b<6$), the skyrmion solutions $\xi_0 (r)$ are compactons, i.e., take their vacuum value $\xi_0(r) =0$ for $r\ge R$, where $R$ is the compacton radius. For compactons we impose that $\bar \eta (r)=0$ for $r\ge R$, as well. If $\mathcal{U}_{,\xi\xi} (\xi =0) \not= 0$, then the condition $\bar\eta (R)=0$ follows immediately from the linearized equation (\ref{lin-eq}). Whether or under which conditions $\bar \eta (r)=0$ for $r\ge R$ is maintained by the full non-linear time evolution, is a different question, which we shall investigate numerically. 

The Euler-Lagrange equation gets much simpler if we use the new field variable
\begin{equation}
d\chi = \sin^2 \xi d\xi \quad \Rightarrow \quad \chi = \frac{1}{2} (\xi - \frac{1}{2} \sin 2\xi)
\end{equation}
leading to the action
\begin{equation}
S= 4\pi \boldsymbol{m} \int dt dr \left( \frac{B^2 \boldsymbol{l}^3}{4r^2} \left(\dot\chi^2 -\chi'^2 \right) -r^2\boldsymbol{l}^{-3} \mathcal{U} \right)
\end{equation}
and EL equation
\begin{equation}
\ddot \chi - \chi'' + \frac{2}{r} \chi' + \frac{2r^4}{B^2 \boldsymbol{l}^6}\mathcal{U}_{,\chi} =0.
\end{equation}  
With $\chi = \chi_0 (r) + \eta (t,r)$ the fluctuation equation now becomes
\begin{equation} \label{eta-fluc-eq}
\ddot \eta - \eta'' + \frac{2}{r} \eta' + \frac{2r^4}{B^2 \boldsymbol{l}^6}\mathcal{U}_{,\chi\chi}\vert \eta =0.
\end{equation}
The eigenmodes and angular eigenfrequencies for small fluctuations may be calculated by inserting the ansatz $\eta (t,r) = \cos \omega t \, \eta (r)$ into the fluctuation equation, leading to the eigenmode equation
\begin{equation} \label{fluct-eq}
\eta'' - \frac{2}{r} \eta' = \left( \frac{2r^4}{B^2 \boldsymbol{l}^6}\mathcal{U}_{,\chi\chi}\vert - \omega^2 \right) \eta .
\end{equation}
Using the variable $\chi$, the symmetry-reduced BPS equation for the axially symmetric ansatz reads
\begin{equation} \label{BPSeq}
\chi' = \pm \frac{2r^2}{B\boldsymbol{l}^3} \sqrt{\mathcal{U}} ,
\end{equation}
where in the following we shall choose the minus sign, corresponding to the skyrmion boundary conditions $\chi (0)=(\pi /2)$ and $\chi (\infty )=0$ (or $\chi (R)=0$ for compactons).

For simplicity, we shall mainly consider potentials of the type $\mathcal{U} = \vert \chi \vert^{2a}$ in this paper. The corresponding BPS skyrmion solutions are compactons for $0<a<1$, whereas they decay like $e^{-c r^3}$ for $a=1$ and with some inverse powers of $r$ for $a>1$. We remark that, for general $\chi$, these potentials do not correspond to well-defined, single-valued functions on target space (on SU(2), which as a manifold is just $\mathbb{S}^3$), as they should. They are, however, well-defined target space functions if $\chi$ is restricted to $-(\pi/2) \le \chi \le (\pi /2)$ (i.e., $\xi$ to $-\pi \le \xi \le \pi$), which will be the case in all cases we consider. 

\subsubsection{Step function potential}
In general, the eigenmode equation is too complicated to find exact analytical solutions, but in the case of the step function potential $\mathcal{U} = \Theta (\chi)$ this is possible. This potential produces skyrmions with an exactly constant baryon density inside the skyrmion (nucleon or nucleus), suddenly jumping to zero at the compacton boundary, which is not realistic from a physical perspective, but still may produce useful approximations under some circumstances.  
For the step function potential, the solution to the BPS equation (\ref{BPSeq}) is
\begin{equation}
\chi = \frac{2}{3B\boldsymbol{l}^3}(R^3 - r^3)
\end{equation}
where $R$ is the compacton radius. The condition $\chi (0)=(\pi /2)$ leads to
\begin{equation}
R^3 = \frac{3\pi B}{4}\boldsymbol{l}^3 \quad \Rightarrow \quad V\equiv \frac{4\pi}{3}R^3 = \pi^2 B\boldsymbol{l}^3
\end{equation}
(here $V$ is the skyrmion volume) and to the BPS skyrmion energy
\begin{equation}
E = 4\pi \boldsymbol{m} \int_0^R dr (2r^2 \boldsymbol{l}^{-3} \mathcal{U}) = 2\pi^2 B \boldsymbol{m}.
\end{equation}
For the step function potential, the eigenmode equation does not depend directly on the BPS skyrmion, because $\mathcal{U}_{,\chi\chi}=0$ (it depends indirectly on the solution via the compacton boundary condition),
\begin{equation}
\eta'' - \frac{2}{r} \eta' + \omega^2 \eta =0.
\end{equation}
This equation is very similar to the ODE for the spherical Bessel functions. Indeed, with the transformations $s=\omega r$, $\eta = s^2 \psi$, this equation transforms into
\begin{equation}
\psi'' + \frac{2}{s} \psi' + \left( \omega^2 - \frac{2}{s^2}\right) \psi =0
\end{equation}
which is the ODE for spherical Bessel functions for angular momentum $\ell=1$. The general solution is a linear combination of the spherical Bessel functions of the first and second kind, $j_1(s)$ and $y_1(s)$, but the boundary condition $\eta (0)=0$ eliminates $y_1$, so our solution is $\psi(s)= j_1(s)$ or
\begin{equation} \label{eta-sol}
\eta (r) = (\omega r)^2 j_1 (\omega r) = \sin (\omega r) - (\omega r) \cos (\omega r).
\end{equation}
The second boundary condition $\eta (R)=0$ leads to a quantization of $\omega$. Indeed, 
\begin{equation}
\omega_n = \frac{z_n}{R}
\end{equation}
where $z_n$ is the $n$'th zero of the spherical Bessel function $j_1(z)$. From (\ref{eta-sol}) we may find a more explicit expression for the $z_n$,
\begin{equation}
z_n = \tan z_n.
\end{equation}
This seems to have an additional solution $z_0 =0$, but this leads to $\eta \equiv 0$ and is, therefore, not acceptable. If we restrict the $\tan$  function to the principal branch (p.b.), $-(\pi/2) < z< (\pi/2)$ then the zeros are defined via
\begin{equation}
\left( \tan (z_n - n\pi) \right)_{\rm p.b.} = z_n  \, , \quad n=1,2, \ldots
\end{equation}
Numerically, $z_1 = 4.4934$, $z_2 = 7.7252$, whereas for large $n$, $z_n \sim (n+\frac{1}{2})\pi$. The corresponding excitation energies are
\begin{equation}
E_n = \hbar \omega_n = \frac{\hbar}{R} z_n .
\end{equation}
If lengths are measured in fm and energies (masses) are measured in MeV, then $\hbar = 197.33 \, {\rm MeV}\, {\rm fm}$, and
\begin{equation}
E_n = \frac{197.33}{R}z_n \, {\rm MeV}
\end{equation}
where $R$ is given in units of fm. 
In particular, the lowest excitation energy is (here $R=B^\frac{1}{3} R_1$, $R_1 \equiv R(B=1)$)
\begin{equation}
E_1 = \frac{197.33}{B^\frac{1}{3}R_1}z_1 \, {\rm MeV} =886.7 B^{-\frac{1}{3}}R_1^{-1} {\rm MeV}.
\end{equation}
It is a curious observation that, for large $n$, the $E_n$ approach the harmonic oscillator energies $E_n \sim \frac{\pi \hbar}{R}(n+\frac{1}{2})$. This does {\em not} mean that these energies form the excitations of one (approximately oscillator) d.o.f. Instead, each $\eta_i$ corresponds to an independent d.o.f. which produces its independent harmonic oscillator spectrum in the semi-classical approximation.
 
\begin{figure} \label{step-pot-fig}
\includegraphics[height=6cm]{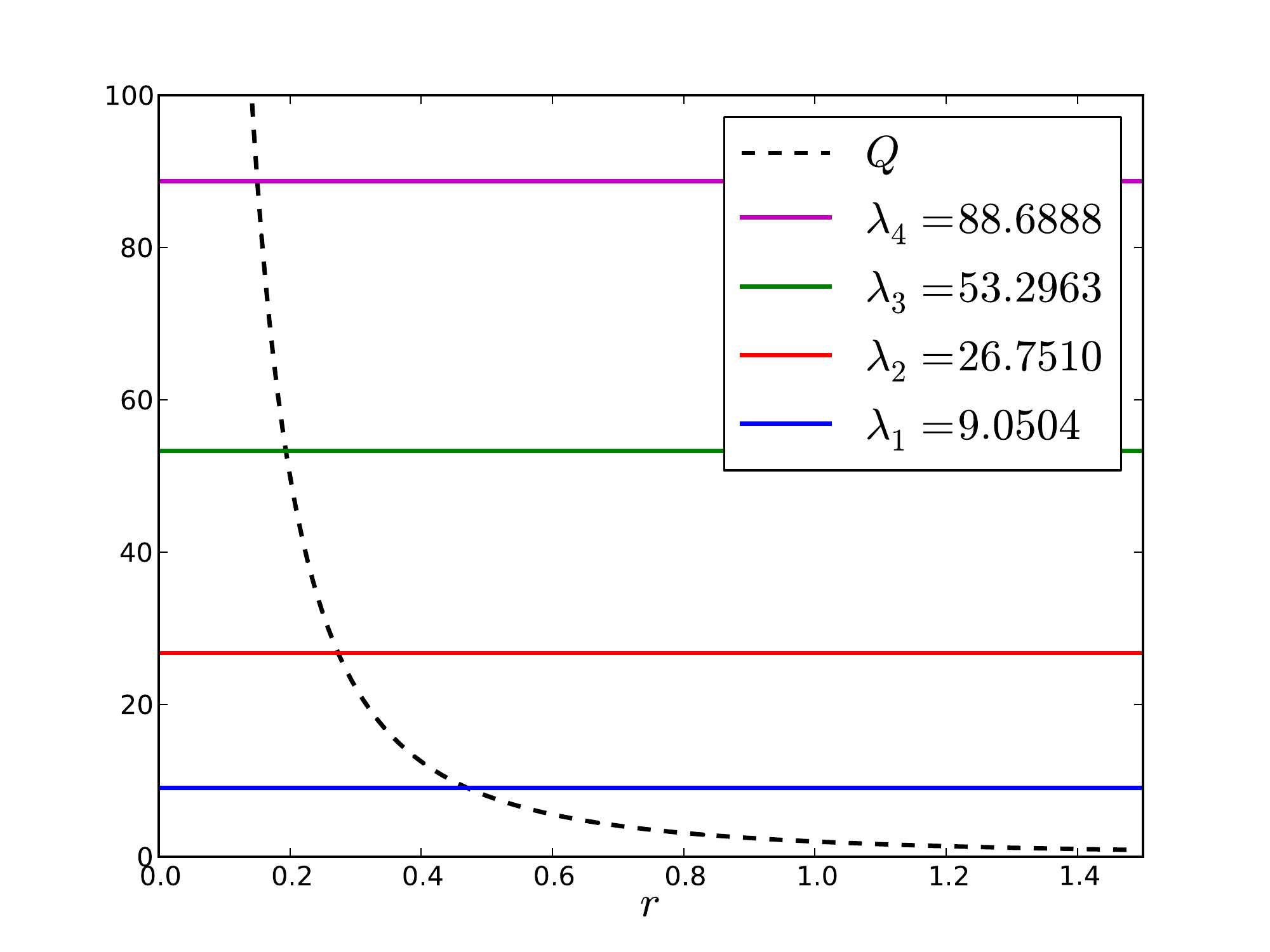}
\caption{Effective potential $Q$ with the numerically calculated energy levels $\lambda_i = \omega_i^2$ for the step-function Skyrme potential, for the parameter value $\boldsymbol{l}^6 = 2$, corresponding to $R_1 = 1.494$.}
\end{figure}

For a numerical determination of the eigenvalues,
 it is convenient to recast the problem as a Sturm-Liouville problem in the normal form (see Appendix A). 
The corresponding effective potential for the step-function potential is
\be
Q=\frac{2}{x^2}
\ee
and the eigenvalues (which are discrete as a consequence of the compacton boundary condition) agree with the exact ones within the numerical precision, see Figure 5.

Finally, we want to point out that the case of the step function potential has a certain pathology which is related to the rather singular character of this potential. That is to say, as follows easily from Eq. (\ref{col-freq}), the frequency of the collective mode, $\omega_c = \sqrt{15}/R$, is below the lowest oscillation frequency $\omega_1 =  z_1/R$. The simple reason which makes 
this possible is that the wave function $\eta_{\rm col}$ for the collective mode 
does not obey the boundary condition $\eta_{\rm col} (r=R) = 0$. Indeed, 
\begin{equation}
\eta_{\rm col}  = \frac{d}{dq} \left. \frac{\pi}{2 R^3}\left( R^3 - \frac{r^3}{(1+q)^3}\right) \right|_{q=0} = \frac{3\pi}{2R^3}r^3 .
\end{equation}
We emphasize that this {\em only} happens for the step function potential. For all other potentials which give rise to compactons, the collective mode obeys $\eta_{\rm col} (R)=0$ which implies $\omega_c >\omega_1$. 

\subsubsection{Compacton case - $\mathcal{U}=\chi^{2a}, \, a<1$}
The step function potential allowed for an exact calculation of its linear oscillation spectrum, but is not expected to give a realistic description of nuclear matter. We should, therefore, analyse other potentials. Qualitatively, depending on their approach $\mathcal{U}\sim \xi^b$ to the vacuum (at $\xi=0$) they lead to three different cases: 1) compactons for $b\in[0,6)$; 2) exponentially localised solutions $b=6$; 3) power-like localised skyrmions for $b>6$.  Here, for simplicity, we consider potentials of the type $\mathcal{U}=\chi^{2a}$ for $0<a<1$ (observe that $b=6a$, and that the step function potential may be formally recovered in the limit $a\to 0$). Then, the BPS equation is
\be
\frac{B \boldsymbol{l}^3}{2r^2} \chi_r=   \pm \chi^{a},
\ee
leading to the solution 
\be \label{BPS-sol-comp}
\chi = \frac{\pi}{2}\left( 1 - \frac{r^3}{R^3}\right)^\frac{1}{1-a} ,
\ee
where the compacton radius $R$ is related to the dimensionful coupling constant $\boldsymbol{l}$ by
\be
R^3 = \frac{3B}{2(1-a)}\left( \frac{\pi}{2}\right)^{1-a} \boldsymbol{l}^3 .
\ee
In our numerical calculations, we shall choose the fixed value $\boldsymbol{l}^6 = 2$ for simplicity. For physical applications, on the other hand, the compacton radius $R$ (or some other radial observable like, e.g., root-mean-square (RMS) radii w.r.t. the energy or baryon charge densities) should be fitted to a certain value, which will lead to different values for $\boldsymbol{l}$ for different potentials. 
Concretely, we shall need the RMS energy density radii for later comparison, which, for the potentials considered, are related to the compacton radii like
\be \label{E-RMS}
R_E^2 \equiv \left< r^2\right> = E^{-1} \int d^3 x r^2 \rho_ E (x)= R^2\,   \Gamma(5/3) \frac{(c+1)\Gamma (c+1)}{\Gamma(c+8/3) }
\, , \quad  c \equiv \frac{2a}{1-a}   
\ee
where 
\be
\rho_E =  \boldsymbol{m}  \left. \left( \frac{B^2 \boldsymbol{l}^3}{4r^4} \chi'^2  + \boldsymbol{l}^{-3} \mathcal{U} \right) \right|
\ee
is the energy density (per volume unit) of the BPS skyrmion.

Our numerical calculation of the linear spectrum now consists in the following steps. First of all, the BPS solution (\ref{BPS-sol-comp}) is inserted into the linear fluctuation equation (\ref{fluct-eq}), leading to a Sturm-Liouville type ODE. This Sturm-Liouville equation is then transformed into its normal form (see appendix A for details), leading to the effective potential
\be
Q=\frac{2}{r^2} + \frac{4a(2a-1)}{B^2 \boldsymbol{l}^6} \frac{r^4}{\left[ \left( \frac{\pi}{2} \right)^{1-a} -\frac{r^3}{\frac{3 B {\bf l}^3}{2(1-a)} }\right]^2}
\ee
For the numerical calculation, we then assume $B=1$ and $\boldsymbol{l}^6=2$, leading to
\be
Q=\frac{2}{r^2} + 2a(2a-1) \frac{r^4}{\left[ \left( \frac{\pi}{2} \right)^{1-a} -\frac{\sqrt{2}(1-a)r^3}{3} \right]^2}
\ee
(oscillation frequencies $\bar \omega_n$ for general values of $B$ and $\boldsymbol{l}$ may easily be calculated from the oscillation frequencies $\omega_n$ for the fixed values $B=1$ and $\boldsymbol{l}^6 =2$ via $2^{(1/6)} \omega_n = B^{(1/3)}\boldsymbol{l} \bar \omega_n$). 
The Sturm-Liouville equation in normal form is then solved numerically
with the SLEIGN2 package \cite{sleign2}, and we also
double-checked our calculations with
the MATSLISE package \cite{matslise}.

\begin{figure} 
\includegraphics[height=6cm]{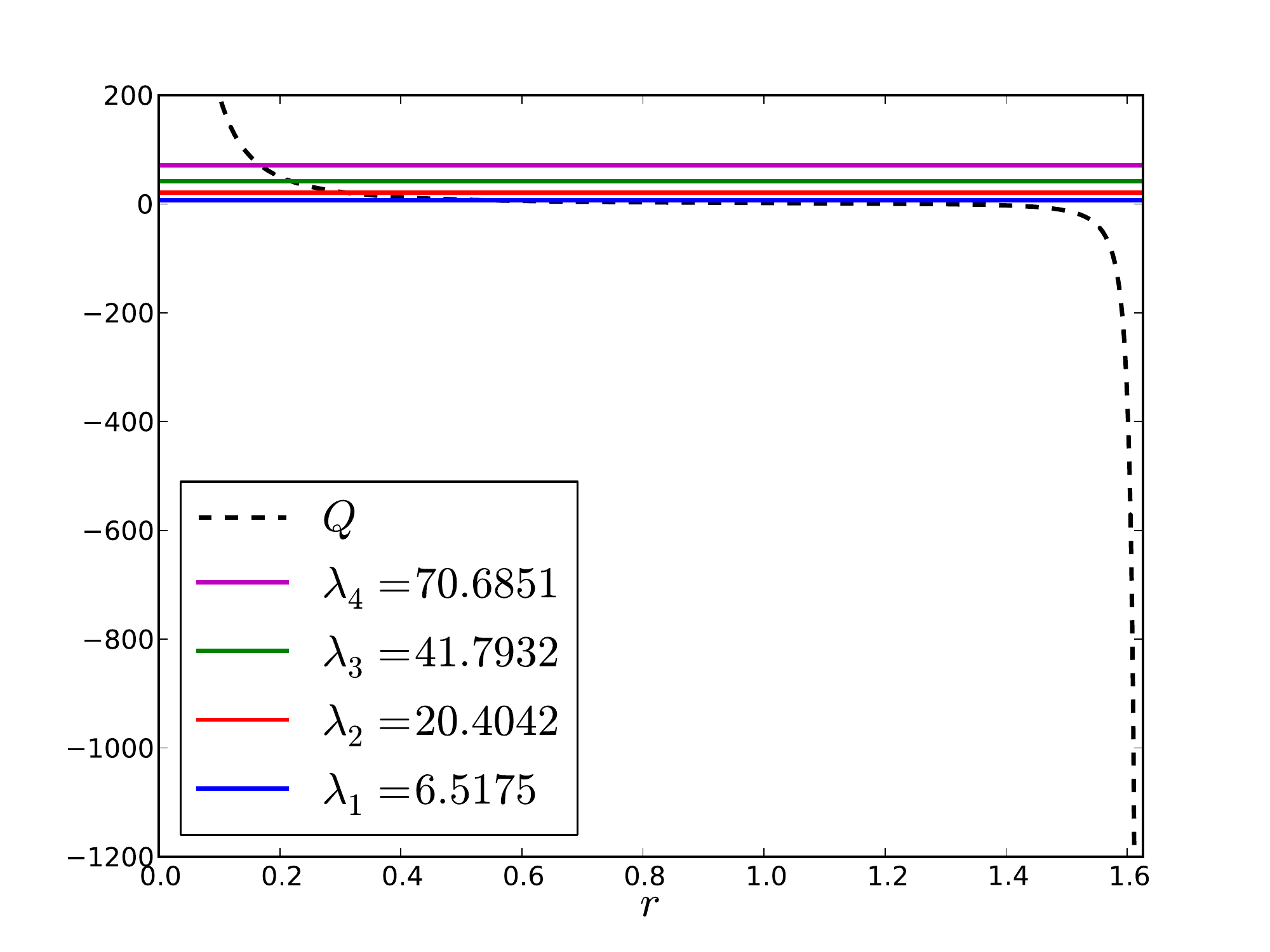}
\includegraphics[height=6cm]{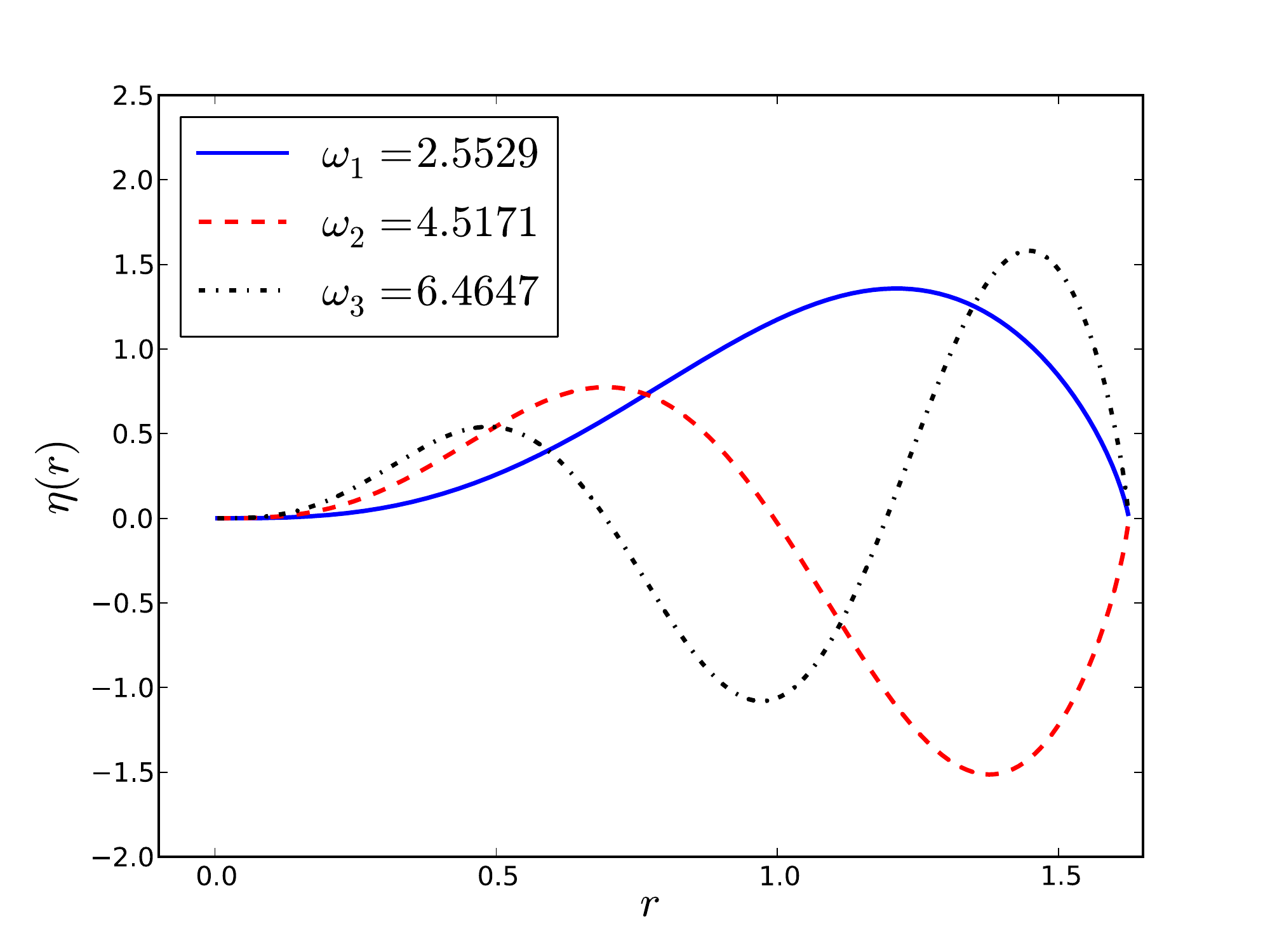}
\includegraphics[height=6cm]{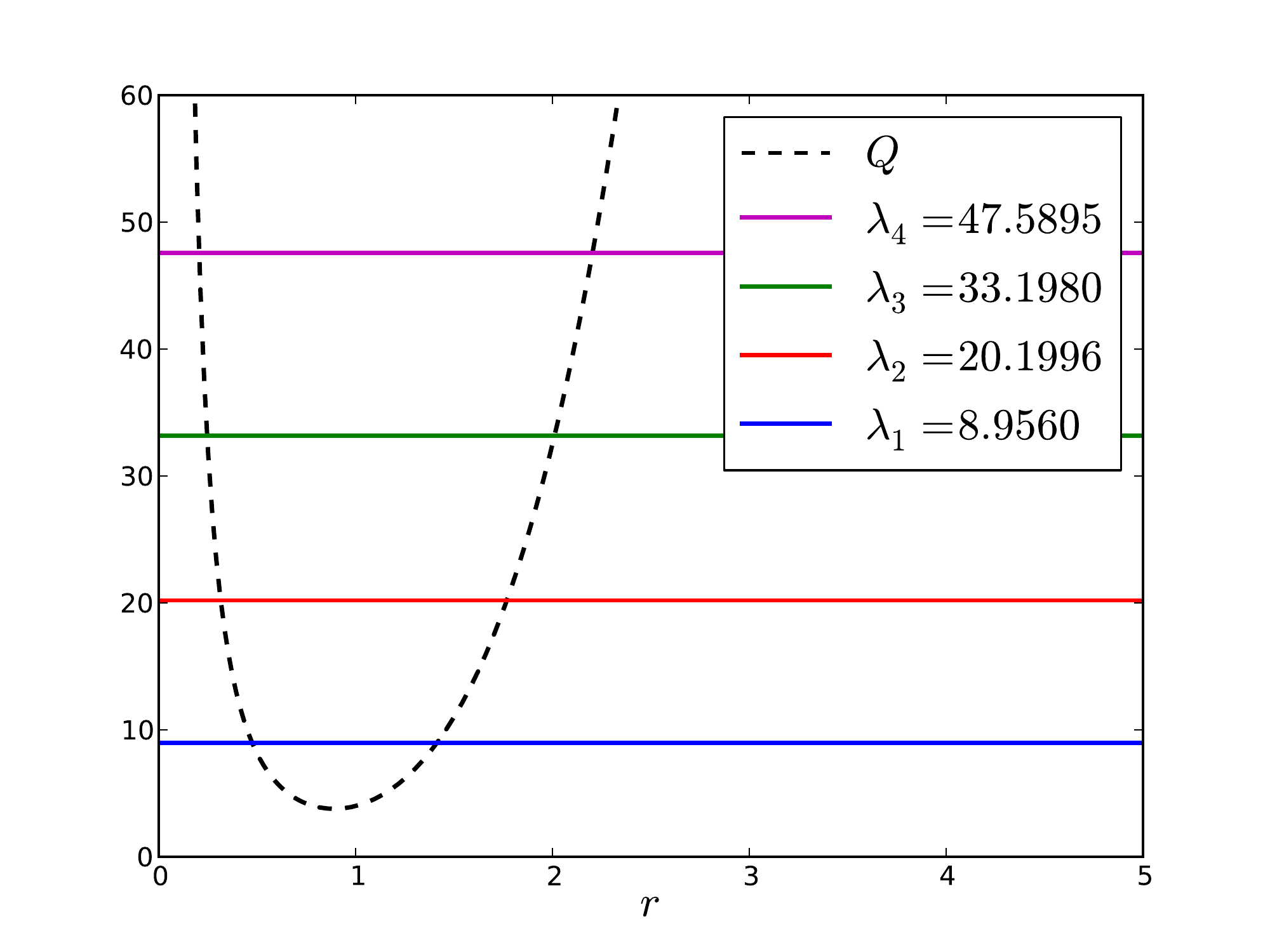}
\includegraphics[height=6cm]{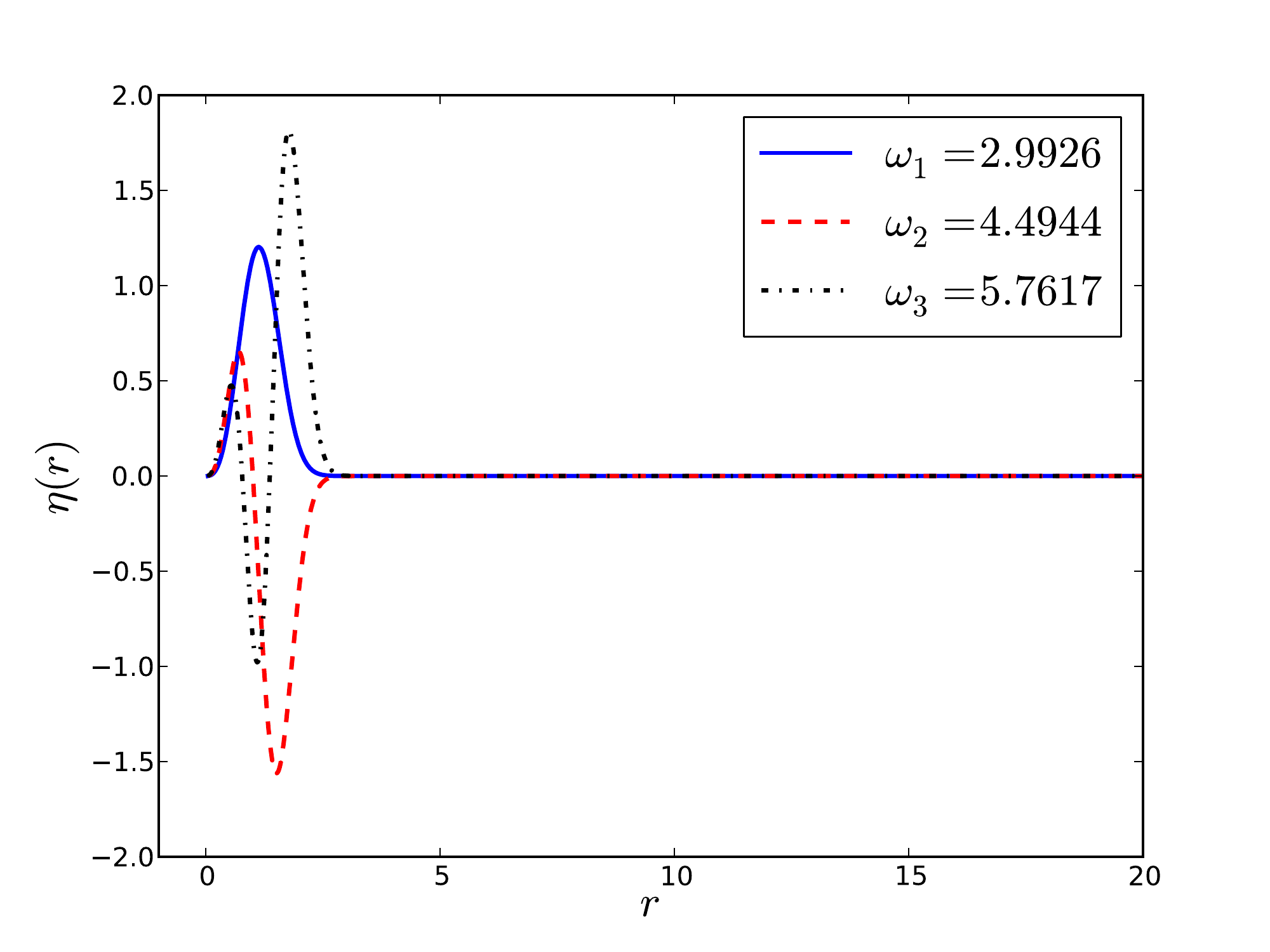}
\caption{1-2) Effective potential $Q$ with the first few energy levels for the potential $\mathcal{U}=\chi^{2/3}$. 3-4) Effective potential $Q$ with the first few energy levels for the potential $\mathcal{U}=\chi^2$. In both cases, the scale constant is $\boldsymbol{l}^6 =2$.} \label{Fig-6}
\end{figure}

Concretely, we show the numerical results for $\mathcal{U}= \chi^\frac{2}{3}$, i.e., $a=\frac{1}{3}$, corresponding to a quadratic, pion-mass type approach to the vacuum. The resulting compacton, compacton radius and effective potential are
\be
\chi = \left( \frac{2}{3}\right)^3 \left( \frac{1}{B \boldsymbol{l}^3} (R^3-r^3)\right)^{3/2},
\ee
\be
 R=\left( \frac{9}{4} \left( \frac{\pi}{2}\right)^{2/3} \right)^{1/3} B^{1/3} \boldsymbol{l}
\ee
and
\be
Q=\frac{2}{r^2} -  \frac{9}{4} \frac{r^4}{(R^3-r^3)^2} .
\ee
The results of the numerical calculation are shown in Figure \ref{Fig-6}. 
In general, for all potentials which lead to compact BPS skyrmions, one has an infinite number of discrete excitation levels. This is a direct consequence of the compact nature of these skyrmions with a fixed boundary. 
\subsubsection{Exponentially localised skyrmions - $\mathcal{U}=\chi^2$}
Now, we investigate a potential which leads to more than exponentially localised BPS skyrmions i.e., $\mathcal{U}=\chi^2$. This potential is, in fact, quite special as the corresponding BPS model action is {\it quadratic} in the scalar field $\chi$,
\be
S=4\pi \boldsymbol{m} \int dt dr \left( \frac{B^2\boldsymbol{l}^3}{4r^2} (\dot{\chi}^2-\chi'^2)  -r^2 \boldsymbol{l}^{-3} \chi^2 \right).
\ee
As a consequence, the field equation for the field $\chi$  is linear
\be
\ddot{\chi}-\chi''+\frac{2}{r} \chi' +\frac{4r^4}{B^2 \boldsymbol{l}^6} \chi=0
\ee
and a linear superposition of solutions is a valid way of constructing new solutions. Obviously, in order to stay in a fixed topological sector, such a superposition cannot change the boundary conditions. 

This potential leads to a {\it non-compacton} solution. Indeed, the BPS equation is 
\be
\frac{B \boldsymbol{l}^3}{2r^2} \chi_r=  \pm \sqrt{\mathcal{U}} = \pm \chi
\ee
and ($\chi(0)=(\pi/2), \; \chi(\infty)=0$)
\be
\chi = \frac{\pi}{2} e^{-\frac{2r^3}{3B \boldsymbol{l}^3} }.
\ee
Qualitatively, this more than exponential localisation can be understood as the existence of an effective
mass $m_{\rm eff}$ which grows with distance 
\be
\chi = \frac{\pi}{2} e^{-\frac{m_{\rm eff} r}{3}}, \;\;\; m_{\rm eff}\equiv \frac{2r^2}{B \boldsymbol{l}^3}.
\ee
In other words, there is a position-dependent mass threshold. It diverges with $r\to \infty$, providing an infinite mass threshold. 

The eigenmode equation reads
\be \label{a1-eigenmode}
\eta''-\frac{2}{r} \eta'=\left( \frac{4r^4}{B^2 \boldsymbol{l}^6} -\omega^2\right)\eta
\ee
and the effective potential in normal form is
\be
Q=\frac{2}{r^2} + \frac{4r^4}{B^2 \boldsymbol{l}^6}.
\ee
In agreement with the observed infinite mass threshold, this potential grows without bound for $r\to \infty$, giving rise to an infinite number of discrete eigenvalues, see Figure \ref{Fig-6}. 

As the BPS skyrmions are no longer compactons, there is no compacton radius available to characterize the size of the soliton. We want to use, again, 
the energy RMS radius, 
\be
R_E^2 = \left< r^2\right> \equiv \frac{1}{E} 4\pi \int_0^\infty dr r^2 \cdot \rho_E \cdot r^2
\ee
where $\rho_E$ is the energy density, leading to the energy
$
E=4\pi \int_0^\infty dr r^2 \cdot \rho.
$
This gives
\be
E=2\pi^3 \boldsymbol{m} B \frac{1}{4}
\ee
and
\be
R_E^2= \boldsymbol{l}^2 B^{2/3} 4 \int dx x^4 e^{-4x^3/3} = \boldsymbol{l}^2 B^{2/3} 4 \cdot 0.1863 = \boldsymbol{l}^2 B^{2/3} 0.7452 .
\ee
Hence,
\be
R_E = 0.8632 \; \boldsymbol{l} \; B^{1/3}.
\ee
It is interesting to note that this case is qualitatively similar to the compacton case. The spectrum is discrete and infinite. 

Note that the perturbation equation, as usually happens in a linear theory, is completely independent of the form of the background solution. Thus, the eigenmodes equation is valid in the nontrivial topological background (static soliton) as well as in the topologically trivial case, describing waves propagating around the vacuum $\chi=0$. As a consequence, we observe a "confinement" of radial waves with a given frequency $\omega$ in a segment $[0,R_{\rm crit}]$ where $R_{\rm crit}$ is the larger root of
\be
Q(R_{\rm crit}) -\omega^2 \equiv \frac{2}{R_{\rm crit}^2} + \frac{4R_{\rm crit}^4}{B^2 \boldsymbol{l}^6} - \omega^2 =0
\ee  
which, for large frequency, may be approximated by
\be
R_{\rm crit}^2 = \frac{B\boldsymbol{l}^3}{2}\omega - \frac{1}{\omega^2} + \ldots
\ee
Indeed, for $r>R_{\rm crit}$, the "potential" term $Q-\omega^2$ in the Sturm-Liouville equation in normal form (the Schr\"odinger type equation) changes sign, and the waves are exponentially suppressed. This means that, from an initial perturbation, no wave can propagate to infinity. Instead, they are completely "confined" within a finite distance corresponding to the highest frequency (higher frequency perturbations can propagate to larger distances). 
\subsubsection{Compacton as a limit}
\begin{figure}
\includegraphics[height=6cm]{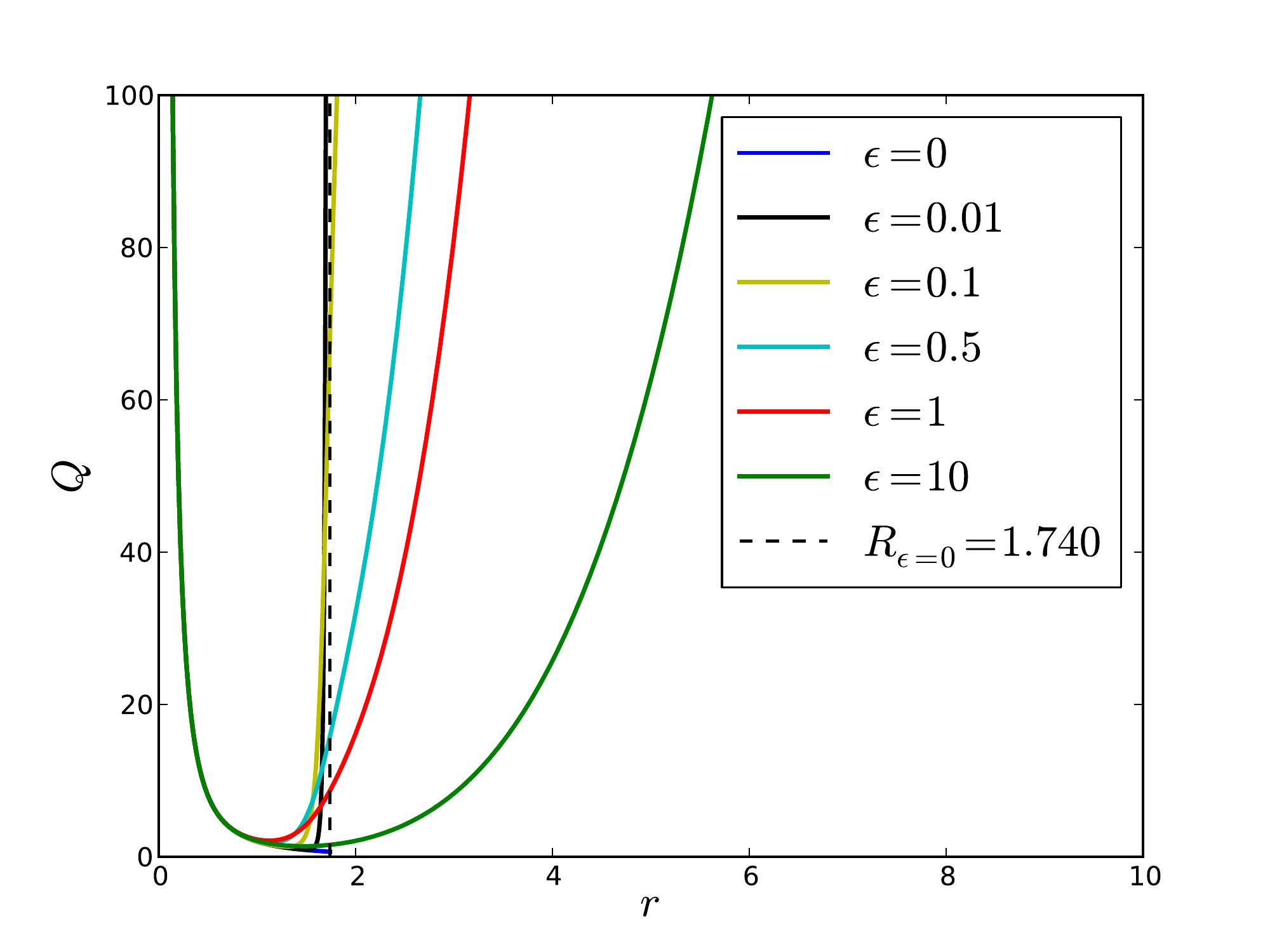}
\caption{Effective potentials $Q$ for $\mathcal{U}_\epsilon = \sqrt{\epsilon^2+\chi^2} - \epsilon$, for the scale constant $\boldsymbol{l}^6 =2$.}
\label{Fig-7}
\end{figure}

Compactons are less regular objects than solitons with infinite tails. It is, therefore, an interesting question and a nontrivial consistency check for our results on compactons whether they can be recovered from a certain limit of the results for solitons with tails. Here we shall see in a specific example that this is, indeed, the case. Concretely, we choose the following one-parameter family of potentials,

\be
\mathcal{U}_\epsilon = \sqrt{\epsilon^2+\chi^2} - \epsilon .
\ee
For $\epsilon > 0$, this potential behaves like $\chi^2$ close to the vacuum, and the BPS skyrmions have an exponential tail, like the ones in Section III.A.3. In the limit $\epsilon \to 0$, on the other hand, the potential is $|\chi |$ and leads to compact BPS skyrmions. 
The soliton solution for non-zero $\epsilon$ in implicit form reads
\be
2 \sqrt{\sqrt{\epsilon^2+\chi^2} + \epsilon} -\sqrt{2\epsilon} \; \mbox{arc tanh} \left( \frac{1}{\sqrt{2\epsilon}} \sqrt{\sqrt{\epsilon^2+\chi^2} - \epsilon}\right) = - \frac{2}{3B \boldsymbol{l}^3} (\beta +r^3)
\ee
where $\beta$ is an integration constant which is given by
\be
2 \sqrt{\sqrt{\epsilon^2 + \left(\frac{\pi}{2} \right)^2} + \epsilon} -\sqrt{2\epsilon} \; \mbox{arc tanh} \left( \frac{1}{\sqrt{2\epsilon}} \sqrt{\sqrt{\epsilon^2+\left(\frac{\pi}{2} \right)^2} - \epsilon}\right) = - \frac{2}{3B \boldsymbol{l}^3} \beta .
\ee
Or, putting everything together
\bea
2 \sqrt{\sqrt{\epsilon^2+\chi^2} + \epsilon} -\sqrt{2\epsilon} \; \mbox{arc tanh} \left( \frac{1}{\sqrt{2\epsilon}} \sqrt{\sqrt{\epsilon^2+\chi^2} - \epsilon}\right) &=& \\
2 \sqrt{\sqrt{\epsilon^2 + \left(\frac{\pi}{2} \right)^2} + \epsilon} -\sqrt{2\epsilon} \; \mbox{arc tanh} \left( \frac{1}{\sqrt{2\epsilon}} \sqrt{\sqrt{\epsilon^2+\left(\frac{\pi}{2} \right)^2} - \epsilon}\right) - \frac{2}{3B \boldsymbol{l}^3} r^3
\eea
Then, the effective potential is
\be
Q=\frac{2}{r^2} + \frac{2r^4}{B^2 \boldsymbol{l}^6} \frac{\epsilon^2}{(\epsilon^2+\chi^2)^{3/2}} .
\ee
As can be seen in Figure \ref{Fig-7}, the effective potentials for nonzero $\epsilon$ converge to the effective potential for the compacton potential $|\chi |$,  maintaining the infinite wall at a finite $r=R$ (the compacton radius for the case $\epsilon =0$). The limit, therefore, reproduces the wave functions $\eta_i$ with the boundary conditions $\eta_i (R)=0$ and the corresponding frequencies.
\subsubsection{Power-like localised skyrmions - $\mathcal{U}=\chi^{2a}, \;\; a>1$}
\begin{figure}
\includegraphics[height=6cm]{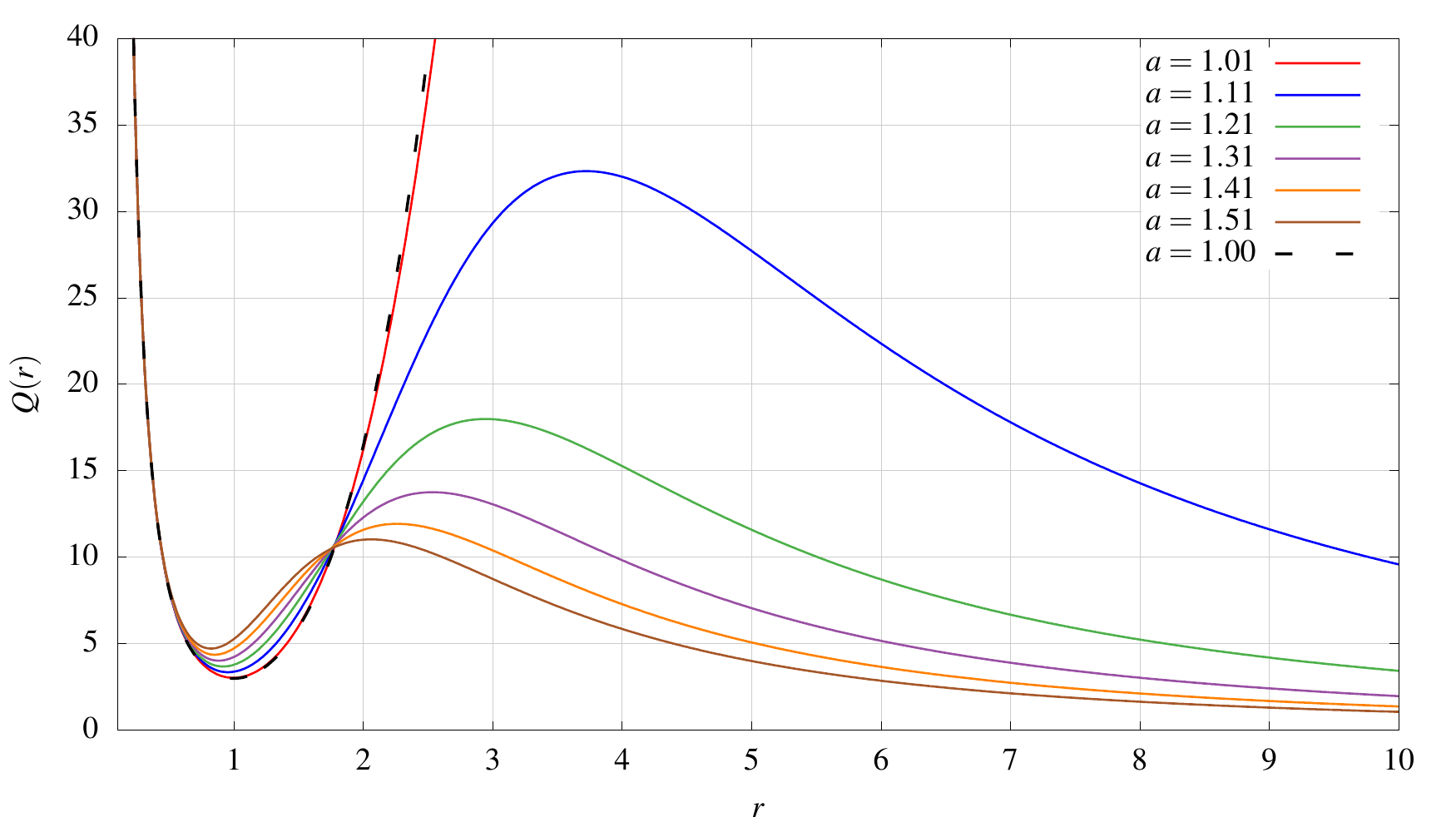}
\caption{Effective potential $Q$ for $\mathcal{U}=\chi^{2a}$, $a>1$, for the scale constant $\boldsymbol{l}^6 =2$.}
\label{Fig-8}
\end{figure}

Skyrmions with power-like hair are given, e.g.,  by the following class of potentials $\mathcal{U}=\chi^{2a}, \; a>1$. The BPS equation is
\be
\chi'=\pm \frac{2r^2}{B \boldsymbol{l}^3} \chi^a
\ee
leading to the solution 
\be
\chi=\left( \frac{3}{2} \frac{B \boldsymbol{l}^3}{a-1} \cdot \frac{1}{r^3+\frac{3}{2} \frac{B \boldsymbol{l}^3}{a-1} \left( \frac{\pi}{2} \right)^{1-a}}  \right)^{\frac{1}{a-1}} .
\ee
The effective potential reads
\be
Q=\frac{2}{r^2} +2\cdot 2a(2a-1)\frac{r^4}{B^2 \boldsymbol{l}^6} \left( \frac{3}{2} \frac{B \boldsymbol{l}^3}{a-1} \cdot \frac{1}{r^3+\frac{3}{2} \frac{B \boldsymbol{l}^3}{a-1} \left( \frac{\pi}{2} \right)^{1-a}}  \right)^2
\ee
So, for $r\rightarrow \infty$
\be
Q \sim \frac{1}{r^2}
\ee
for any value of the parameter $a$. As the effective potential tends to 0, the Sturm-Liouville problem cannot have any discrete states with positive energy i.e., no oscillating modes, see Figure \ref{Fig-8}.

Again, we want to define the energy RMS radii for later use,
\be
R_E^2 = \left<r^2 \right> =R^2 \, \Gamma(5/3) \frac{\Gamma (c-5/3)(c-1)}{\Gamma(c)}
\, , \quad c \equiv \frac{2a}{a-1}
\ee
where $R$ is related to the parameter $\boldsymbol{l}$ like the compacton radius is for $a<1$, i.e., 
\be \label{pseudo-comp}
R^3  =\frac{3 B}{2(a-1)} \left( \frac{\pi}{2}\right)^{1-a} \boldsymbol{l}^3 .
\ee

\vspace*{0.2cm}

\noindent Generically, for $a>1$ the effective potential has a repulsive core $\sim r^{-2}$ and a decaying tail $\sim r^{-2}$. This excludes any discrete spectrum. Potentials with power-like localisation are, therefore, qualitatively different from the two previously discussed cases. What can happen, instead, (and does happen for the cases considered here, see Figure 8) is the emergence of a small potential barrier in the region of finite $r$. This may lead, in turn, to the appearance of resonance modes instead of vibrational modes, see next section. Furthermore, one cannot exclude the possibility that for some potentials with this power-like approach to the vacuum, a small negative well emerges. If this well is sufficiently deep, it may support a bound state with negative energy, that is, a purely imaginary frequency.  The corresponding mode is exponentially growing and can be interpreted as a sign of an instability in the model. In all the examples considered here (for power-like localized skyrmions), however, no such instability occurs. 

\subsection{Resonance modes}
Resonance modes or quasi-normal modes correspond to modes with a complex frequency, so for their calculation it is useful to consider the complexified version of the linear fluctuation equation (\ref{eta-fluc-eq}). The details of its solution are given in appendix B, here we just want to make some comments.

First of all, one should be aware that our choice of potentials is rather arbitrary. In contrast to the perturbative $\mathcal{L}_{024}$ Skyrme model, where the quadratic part of the potential (i.e., its form close to the vacuum) is uniquely dictated by the mass of the pionic excitations, there is no such phenomenological condition for the BPS potential. Of course, it should provide reasonable values for observables. Quite interestingly, some properties of nuclei like, e.g., the binding energies of atomic nuclei, only weakly depend on the potential. Also for thermodynamical properties of dense (high pressure) nuclear matter, the potential only gives subleading corrections, while the sextic term always dominates. This gives some freedom in the choice of the potential. Obviously, different potentials will lead to different quantitative predictions also for oscillations and resonances where, however, some general, qualitative features of vibrations in the BPS Skyrme model are potential independent. Let us discuss them in some detail.

\begin{enumerate}
\item {\it The role of the vacuum approach:} $\mathcal{U} \sim \chi^{2a}$ as $\chi \rightarrow 0$.
\\
For power-like localized BPS Skyrmions, $a>1$, the effective potential of the linearized problem tends to 0 at spatial infinity, and no mass threshold is present. This excludes the existence of oscillating modes, while it opens the possibility for the formation of quasi-normal modes. 
\\
For exponentially localized BPS Skyrmions, $a=1$, the effective potential diverges at spatial infinity. This results in the appearance of infinitely many oscillating modes and the absence of resonances.
\\
Finally, for $a<1$ (compact BPS Skyrmions), the effective potential also diverges, but at a finite distance (compacton radius) and, again, infinitely many oscillating modes are observed. This is confirmed by the full (not linearized) numerical computation, as the linearization is not completely trustworthy for compactons close to their boundary. 
\\
A very peculiar feature of the BPS model is the fact that the effective potential never approaches a finite, non-zero value at spatial infinity. Therefore, it is not possible to have a finite number of oscillating modes in the BPS Skyrme theory.  

\item {\it The existence of a barrier for the effective potential - narrow and wide resonances.}
\\
For any potential $\mathcal{U}$ which generates power-like localized BPS Skyrmions, a potential barrier may form for the effective potential $Q$. Its height and width depend on the details of the potential $\mathcal{U}$ (and probably can be related to some specific features of $\mathcal{U}$). As we found, the bigger the barrier, the more {\it narrow} resonance modes exist, as their decay is strongly suppressed by tunneling through a powerful barrier. However, beside these narrow resonances, there are also {\it wide} quasi-normal states with $\Omega^2 > Q_{\rm max}$. As they can go through the barrier relatively freely, they are short-lived excitations.  
\\
The number of narrow and wide resonance modes can, therefore, be controlled by a suitable choice of $\mathcal{U}$ and is not related to a particular vacuum approach (besides the fact that it must lead to power-like localized BPS Skyrmions) and, therefore, is independent of the corresponding (mean-field) equation of state,  which is affected by the close to vacuum  behavior. 
\\
The existence of the barrier is a qualitatively different feature, as compared to the perturbative massless Skyrme model $\mathcal{L}_{24}$. There, $Q_{\rm eff}$ also tends to 0 at  spatial infinity, but it develops a little well at an intermediate distance, which allows for a resonance mode (but there is no negative energy bound state). In principle, one cannot exclude a situation where a BPS potential would provide a well rather than a  barrier. However, the existence of such a potential is an open question.
\end{enumerate}     

It is worth to notice that, for our choice of the potentials, one can have an arbitrary number of narrow resonances by assuming $a$ sufficiently close to 1. The reason for this is that any of the narrow resonance modes originates in a particular oscillating mode of the $a=1$ model. Quasi-normal modes with $\Omega^2 > Q_{\rm max}$ develop into wide resonances which quite quickly (as $a$ increases) interact with the spectrum, leading to very complicated FFT structures. Once $a$ further grows, more narrow resonances transform into wide ones.  In the full dynamics, these wide quasi-normal modes decay very quickly, and the vibration is soon dominated by the few first still narrow resonances. Finally, only wide quasi-normal states remain. 

\subsubsection{Resonances for the potentials $\mathcal{U}=\chi^{2a}, \;\; a>1$}
\begin{figure}
\includegraphics[height=8cm]{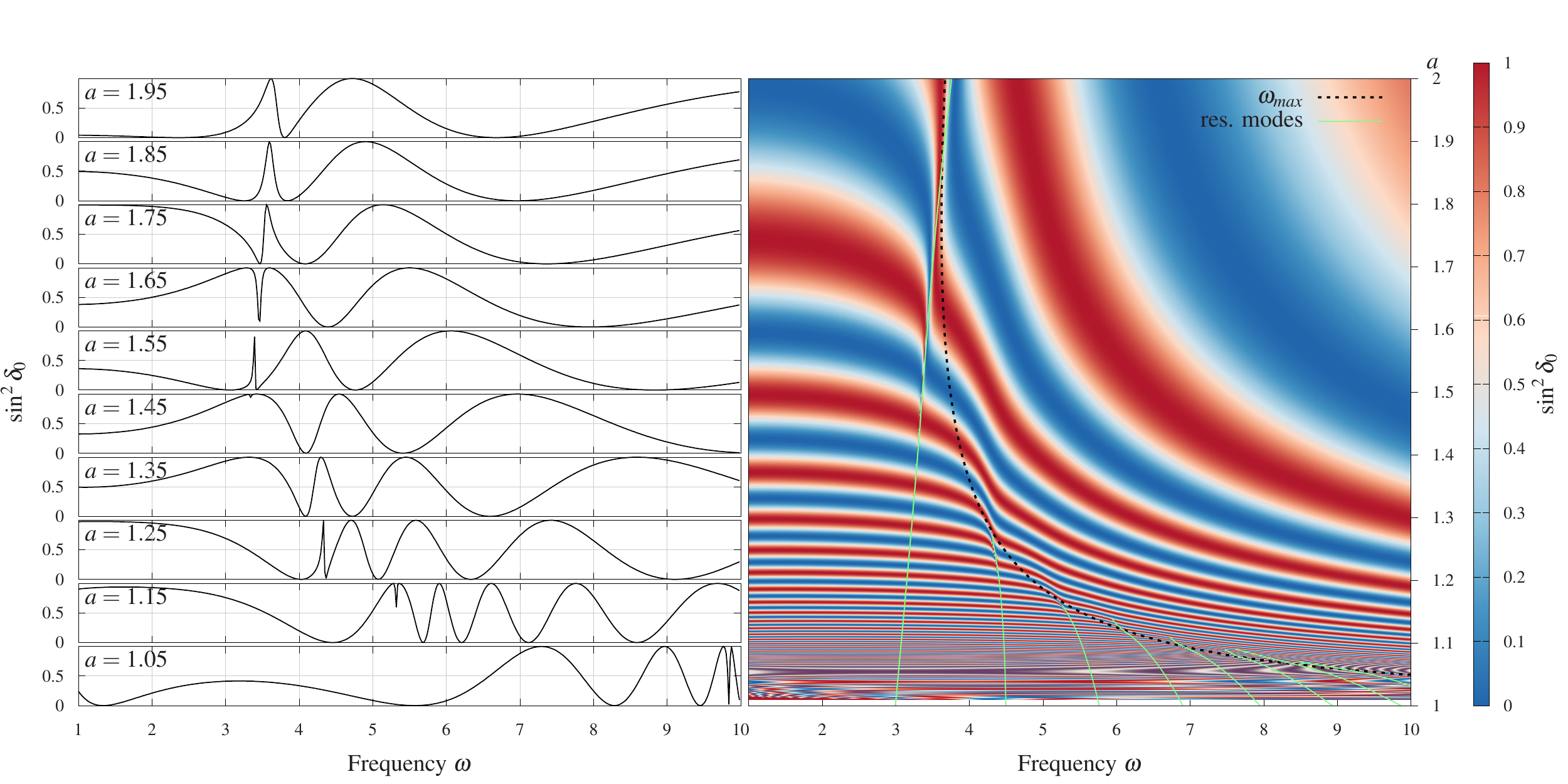}
\caption{Cross section multiplied by $\omega^2/4\pi$: $\sin^2\delta_0$. Maxima indicate approximately real values of the resonant frequency. 
We can see the drift of the resonance as the barrier is crossed. When the frequency is larger than $\omega_{\rm max} =\sqrt{Q_{\rm max}}$, the resonance becomes 
wide. Green lines show the position of narrow resonances (too narrow to be seen on the scan). We remark that in the left plot the lowest (fundamental) resonance for small values of $a$ is too narrow to be visible.}\label{FigureXSection}
\end{figure}

In the case of $a=1$, the linearized effective potential $Q(x)$ tends to infinity as $x\to\infty$. However, for $a>1$, the potential vanishes at spatial infinity. 
For $a>1$ but very close to 1, the potential $Q(x)$ initially differs insignificantly from the $a=1$ case. The bound state solution for $a=1$ 
is a good approximation of a solution for small deviations from $a=1$. However, at some point the potential $Q$ reaches its maximum and vanishes at 
$x\to\infty$. It also forms a barrier with maximum $Q_{\rm max}$. This finite barrier allows tunneling. Bound modes from $a=1$ become 
resonances or quasi-normal modes as $a>1$. The larger  $a$, the smaller is the height and the width of the barrier.
Resonances with frequencies below $\sqrt{Q_{\rm max}}$ need to tunnel through that barrier. This results in small $\Gamma$ and long half-life of 
the resonances. These resonances are sometimes too narrow to be seen on the general scan of the cross section (see Figure~\ref{FigureXSection}). 
Resonances with frequencies above  $\sqrt{Q_{\rm max}}$ are not trapped inside the potential well and have much larger values of $\Gamma$. They become wide and dissolve as $a$ grows.

\begin{figure}
\includegraphics[height=6cm]{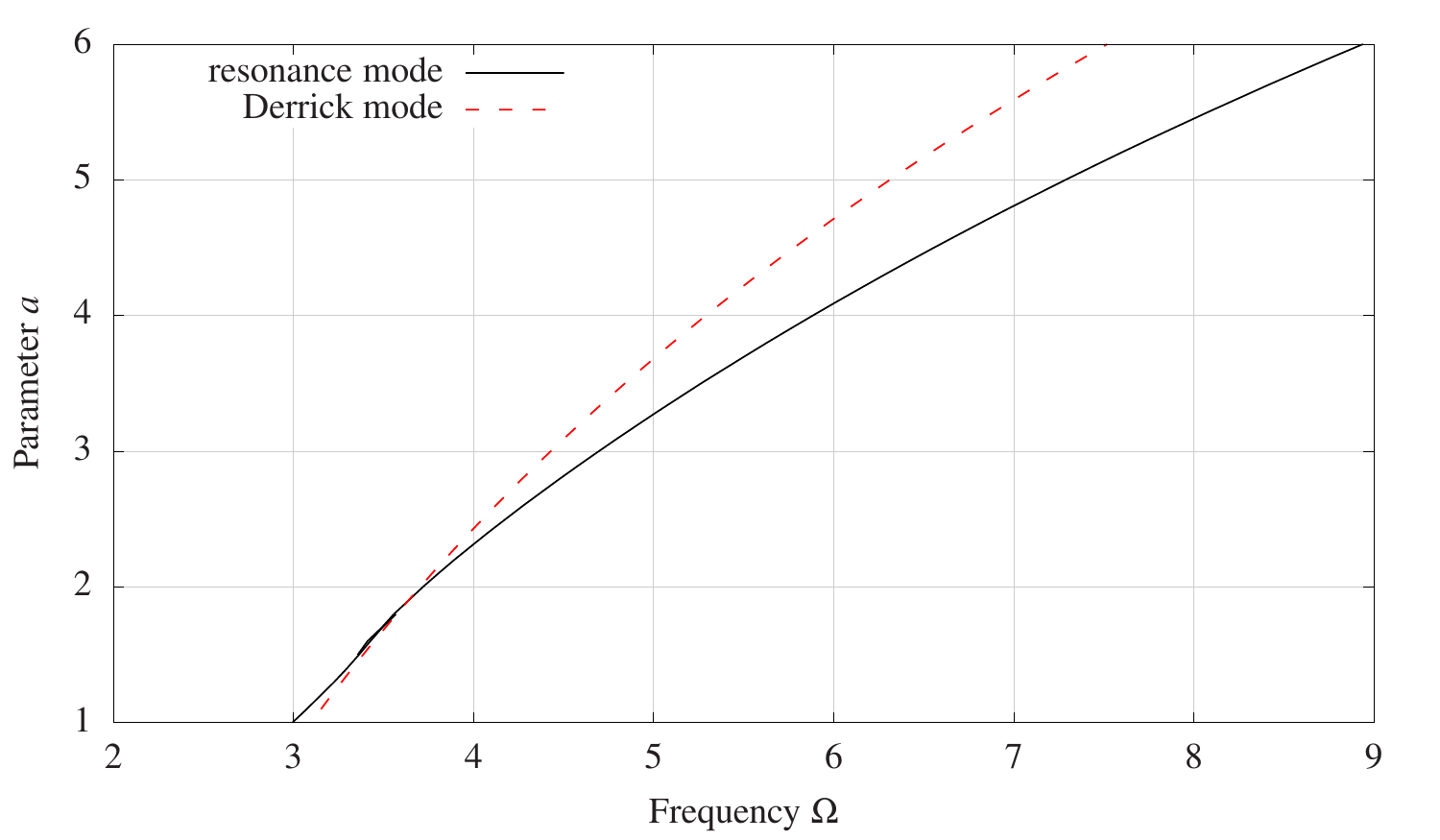}
\includegraphics[height=6cm]{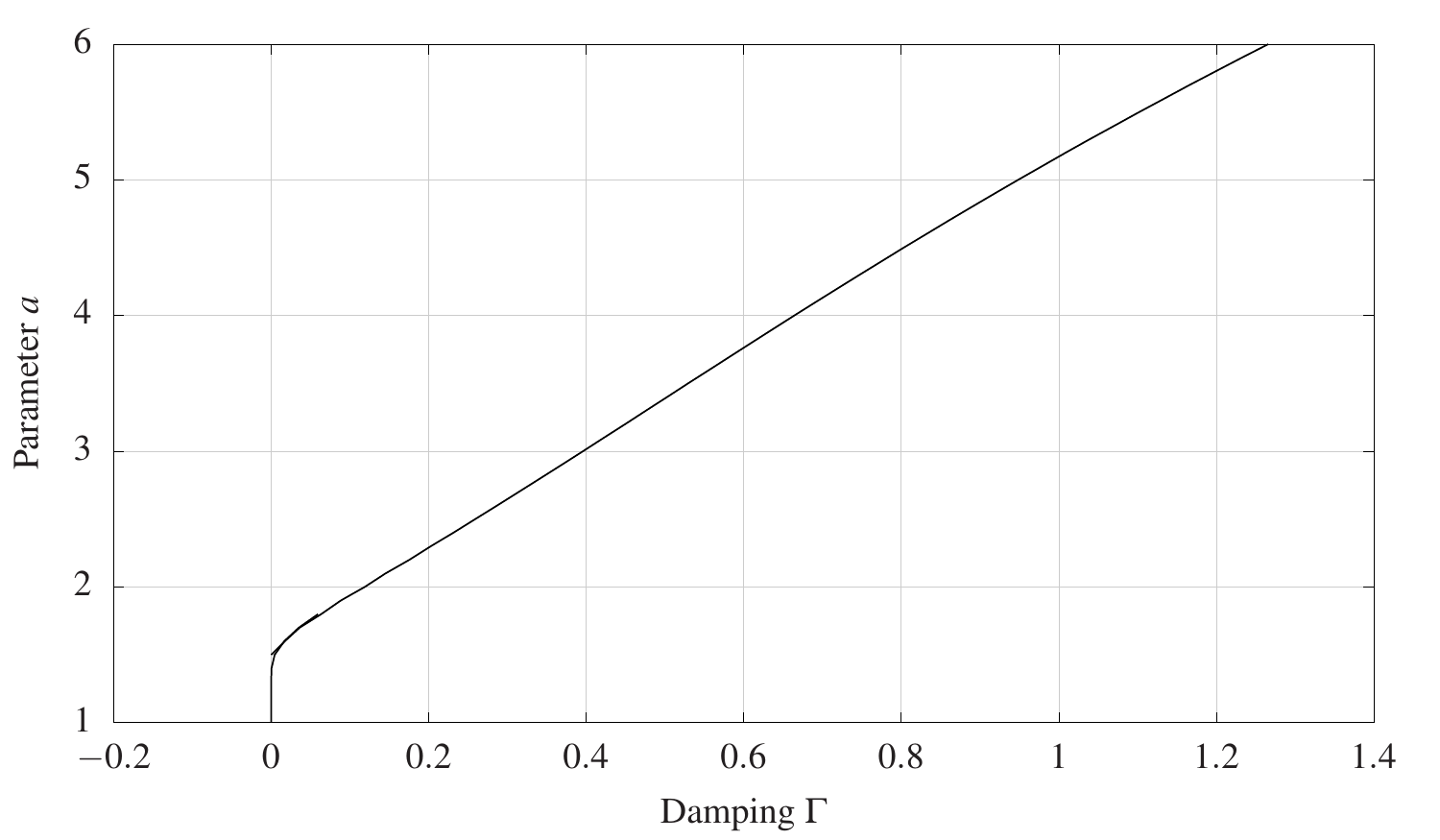}
\caption{Resonance frequencies and decay widths for the first (fundamental) resonance for different values of $a$. In the first plot we also compare with the frequency of the collective (Derrick) mode.}\label{FigureOmegaGamma}
\end{figure}

For later use, we also give the numerical values of the first (lowest) resonance mode $\omega_1 = \Omega_1 + i \Gamma_1$ in Table 1, for different 
values of $a$ and plot both of them in Figure \ref{FigureOmegaGamma}. 
\begin{table}
\begin{tabular}{lll}
\,  $a$   &     \; \; $\Omega_1 $ &     \; \; $\Gamma_1 $    \\
\hline
1.8  \; \; &   3.568028  \;\;    & $ 5.95148 \cdot 10^{-2}$ \\
1.7  &   3.491566  &      $3.51183 \cdot 10^{-2}$ \\
1.6  &   3.425363  &     $1.66694 \cdot 10^{-2}$ \\
1.5  &   3.361944  &     $4.46469 \cdot 10^{-3}$ \\
1.4  &   3.295956  &     $3.78620 \cdot 10^{-4}$ \\
1.3  &   3.224497  &      $ 2.45495\cdot 10^{-6}$ \\
1.2   &  3.149303    &    $1.26204 \cdot 10^{-11}$ \\
1.1  &   3.071850   &     $<\, 10^{-11}$ \\
1.0   &  2.992676   &     $0.00000$
\end{tabular}
\caption{Frequencies and decay widths of the first resonance, for different values of $a$.}
\end{table}

\subsubsection{Potential $\mathcal{U}=(1-\cos \xi)^6$ }
As mentioned already, resonance modes for the class of potentials $\mathcal{U}=(1-\cos \xi)^\alpha$, $\alpha >3$, have been studied numerically in \cite{arpad}. So,
for completeness, let us also analyze the potential $\mathcal{U}=(1-\cos \xi)^6$. Close to the vacuum, it behaves like $\xi^{12}$, i.e., like $\chi^4$, leading to power-like localized skyrmions. We, therefore, expect no oscillations but possible resonances. 
The BPS equation is (in the original $\xi$ profile function)
\be
\frac{B \boldsymbol{l}^3}{2r^2} \sin^2 \xi \xi'=\pm (1-\cos \xi)^3
\ee
leading to the solution 
\be
\xi = 2\mbox{ arccot } \left( \frac{2^{1/3} r}{B^{1/3} \boldsymbol{l}}\right) .
\ee
Now, we want to move back to the perturbation equation with the $\chi$ variable. To do this, one has to express $\mathcal{U}_{\chi \chi} $ in terms of $\xi$, 
\be
\mathcal{U}_{\chi \chi} = \mathcal{U}_{\xi \xi} \left( \frac{d\xi}{d\chi} \right)^2  + \mathcal{U}_{\xi} \frac{d^2 \xi}{d\chi ^2}
\ee
and, of course,
\be
\mathcal{U}_{\chi \chi} |_{\chi} \left. = \mathcal{U}_{\xi \xi} \left( \frac{d\xi}{d\chi} \right)^2  + \mathcal{U}_{\xi} \frac{d^2 \xi}{d\chi ^2} \right|_{\xi}
\ee
So, we do not need to express $\xi$ as a function of $\chi$. Now,
\be
 \frac{d\xi}{d\chi} =  \frac{1}{\sin^2 \xi}, \;\;\;\;\;  \frac{d^\xi}{d\chi^2} = -2 \frac{\cos \xi}{\sin^5 \xi}
\ee
and
\be
\mathcal{U}_{ \xi}= 6(1-\cos \xi)^5 \sin \xi ,\;\;\;\; \mathcal{U}_{\xi \xi} = 30 (1-\cos \xi)^4\sin^2 \xi +6(1-\cos \xi)^5 \cos \xi
\ee
Then,
\be
\mathcal{U}_{\chi \chi} = 6\frac{(1-\cos \xi)^4}{\sin^4 \xi} [ 5\sin^2 \xi - \cos \xi (1-\cos \xi)]
\ee
For the solution we find
\be
\cos \xi = \frac{\cot^2 \frac{\xi}{2} -1}{\cot^2 \frac{\xi}{2} +1}=\frac{y^2-1}{y^2+1}
\ee
\be
1-\cos \xi = \frac{2}{y^2+1}
\ee
\be
\sin \xi = \frac{2\cot \frac{\xi}{2} }{\cot^2 \frac{\xi}{2} +1}=\frac{2y}{y^2+1}
\ee
where $y \equiv \frac{2^\frac{1}{3} r}{B^\frac{1}{3} \boldsymbol{l}}$. So, the effective potential is
\be
Q=\frac{2}{r^2} + \frac{2r^4}{B^2 \boldsymbol{l}^6 }
\frac{12}{\left( \frac{2^\frac{1}{3} r}{B^\frac{1}{3} \boldsymbol{l} } \right)^4} 
\frac{\left( 1+9\left( \frac{2^\frac{1}{3} r}{B^\frac{1}{3} \boldsymbol{l} }
\right)^2\right)}{\left( 1+\left( \frac{2^\frac{1}{3} r}{B^\frac{1}{3} \boldsymbol{l} }
\right)^2\right)^2}
\ee
This effective potential approaches zero for $r\to \infty$,  so it cannot have linearly stable states for discrete eigenvalues, as expected. 
On the other hand, it develops a small barrier, so one or several resonance states may occur. Both a linear simulation and a full non-linear numerical study reveal that this is, indeed, the case. Concretely, we find for the first resonance $\omega_1 = \Omega_1 + i \Gamma_1$ in the linearized problem  $\Omega_1 = 5.118$, $\Gamma_1 = 0.8643$, and half-life $T_1 = 0.8020$. From the full non-linear simulation we get
the results shown in Table 2. 
\begin{table}
\begin{tabular}{lll}
  $r$   &     \; \; $\Omega_1 $ &     \; \; $\Gamma_1 $    \\
\hline
1  \; \; &   5.1322  \;\;    &  0.8051 \\
2  &   5.1281  &      0.8181 \\
3  &   5.1296  &     0.8200 \\
\end{tabular}
\caption{Frequencies and decay widths of the first resonance, for the potential $(1-\cos \xi)^6$. The values for $r$ are the radii for which the fit has been performed.}
\end{table}
This should be compared to the values reported in \cite{arpad}, $\Omega_1^{\rm IL} =4.71$ and $T_1^{\rm IL} = 0.83$. We find that the half-lives (decay widths) essentially agree, whereas there is a difference of about 10\% for the frequencies.

\subsection{The collective mode approximation}
Let us now investigate the collective scaling (Derrick mode) approximation for the BPS model. This means that we use the variational ansatz
\be
\chi (r,t)= \chi_0 \left( \frac{r}{\rho(t)} \right)
\ee
and plug it into the BPS action. Then we get
\be
L=\frac{\dot{\rho}^2}{\rho^3} A - \frac{1}{\rho^3} B - \rho^3 C
\ee
where
\be
A= 4\pi \boldsymbol{m} \int \frac{B^2 \boldsymbol{l}^3}{4}  \chi_{0 ,r}^2 dr 
\ee
\be
B= 4\pi \boldsymbol{m} \int \frac{B^2 \boldsymbol{l}^3}{4r^2}  \chi_{0,r}^2 dr
\ee
\be
C= 4\pi \boldsymbol{m} \boldsymbol{l}^{-3}\int \mathcal{U}(\chi_0(r)) r^2dr
\ee
Obviously, $B=C$ and their sum gives the soliton energy, $B+C=E_0$. Now, we expand it around 1 i.e., $\rho(t)=1+q(t)$, where $q$ is a small parameter. It gives
\be
L= -E_0+A \dot{q}^2 -\frac{9}{2} E_0q^2
\ee
and the oscillation frequency is 
\be
\omega_c = \sqrt{\frac{9E_0}{2A}}
\ee
Using the BPS equation we can write 
\bea
A=4\pi \boldsymbol{m} \boldsymbol{l}^{-3}\int \mathcal{U}(\chi_0(r)) r^4dr =\frac{1}{2} E_0 R_E^2\\
E_0= 2 \cdot 4\pi \boldsymbol{m} \boldsymbol{l}^{-3}\int \mathcal{U}(\chi_0(r)) r^2dr
\eea
and
\be \label{col-freq}
\omega_c = 3 \sqrt{\frac{\int \mathcal{U}(\chi_0(r)) r^2dr}{\int \mathcal{U}(\chi_0(r)) r^4dr}} =\frac{3}{R_E}
\ee
where $R_E$ is the energy RMS radius defined in (\ref{E-RMS}). 
Then, e.g., for the power-like localised solutions we find
\be
\omega_c= \frac{3}{R} \sqrt{\frac{a-1}{a+1} \frac{\Gamma \left( \frac{2a}{a-1} \right) }{\Gamma \left( \frac{5}{3} \right) \Gamma \left( \frac{a+5}{3(a-1)} \right) }}
\ee
where $R$ is defined in (\ref{pseudo-comp}). 
In the limit $a=1$, e.g., we get (for $B=1$, $\boldsymbol{l}^6 = 2$)
\be
\omega_c(a=1)=3^{2/3}\sqrt{\frac{2}{\Gamma \left( \frac{5}{3}\right) }} \approx 3.096
\ee
which is only $3\%$ above the true lowest frequency $\omega_1 =2.993$.

\subsection{Full numerical time evolution}
To test our predictions, we have solved the equations of motion numerically. We have applied a method of lines with 4th order Runge-Kutta method for 
time-stepping, and a five-point stencil method to discretize spacial derivatives. For the boundary condition at $r=0,$ we assumed a vanishing perturbation and 
its derivative to match the singularity expansion.\\
We have tested several initial conditions, but for the scans we have used the following initial conditions:
\begin{equation}
 \chi(x,0) = \chi_0(x)\left(1+\frac{Ar^4}{1+r^4}e^{-(r-1)^2}\right),\qquad\chi_t(x,0)=0,
\end{equation} 
where $\chi_0(x)$ is a static skyrmionic solution. We have decided not to use the squeezed skyrmion initial condition, since the numerical evolution 
became unstable, probably due to the violation of the singular expansion for $r=0$. Unless stated otherwise, $A=10^{-4}$.

\begin{figure}
\includegraphics[height=12cm]{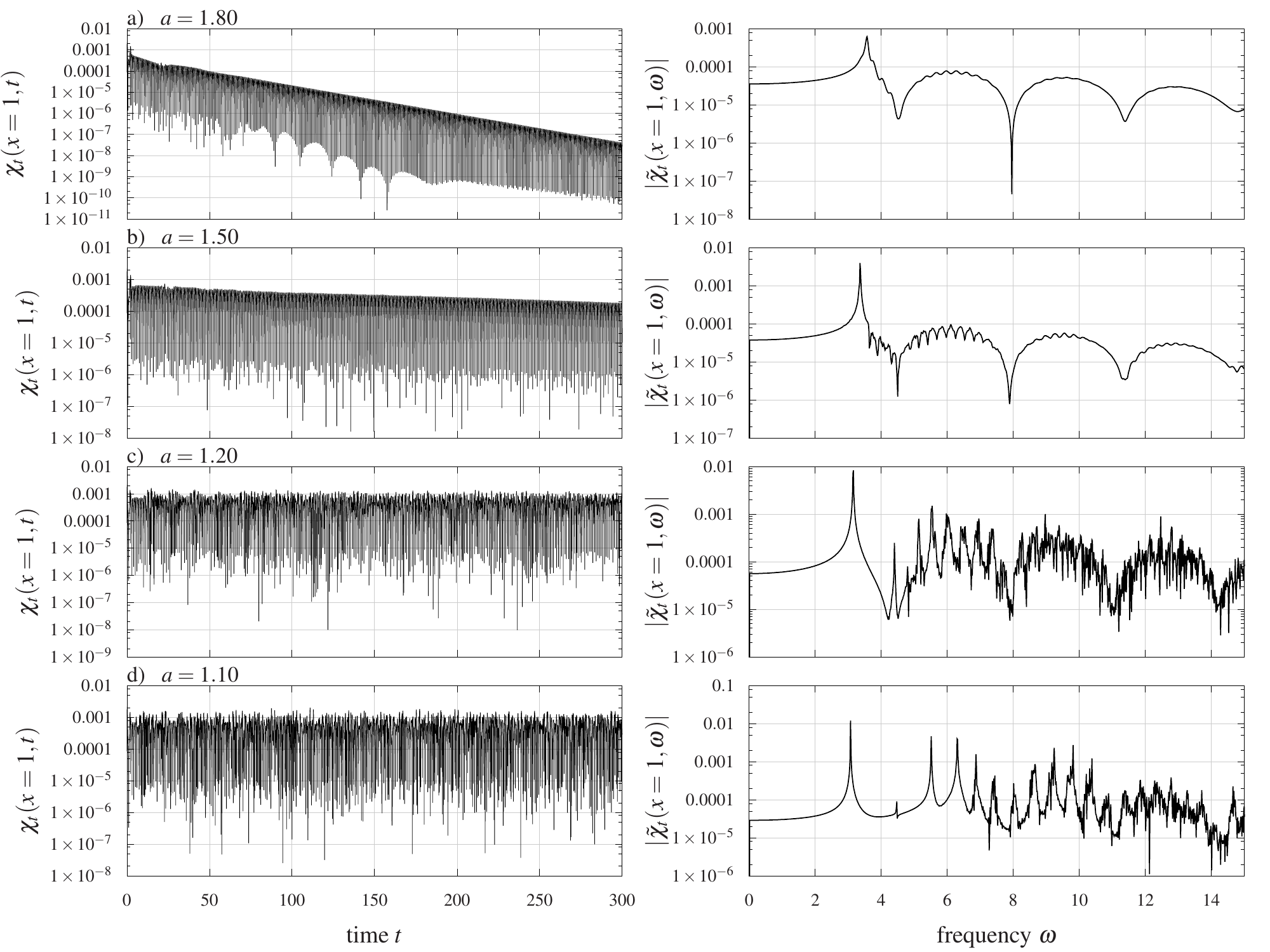}
\caption{Evolution of the field and power spectra for potentials $\mathcal{U}=\chi^{2a}$ for $a>1$. For $a=1.10$ there is a number of narrow excited
resonances, resulting in oscillations with almost constant amplitude, whereas for $a=1.80$ there is a single and much wider resonance which 
manifests itself with decaying oscillations.}\label{Figurescanalargea}
\end{figure}

Our aim was to find the most strongly excited frequencies. Therefore, we applied Fast Fourier Transform (FFT) to the field $\chi_t(x)$ gathered at $x=1$.
For $a>1$, we could see many resonances as discussed in the previous section. For $a$ values close to 1 we observed many narrow 
resonances (Figure~\ref{Figurescanalargea}). As $a$ increases, the potential barrier decreases and  the narrow resonances become wider.
For $a=1.80$ we could observe only the lowest narrow resonance corresponding to the lowest bound state for $a=1$.

\begin{figure}
\includegraphics[height=12cm]{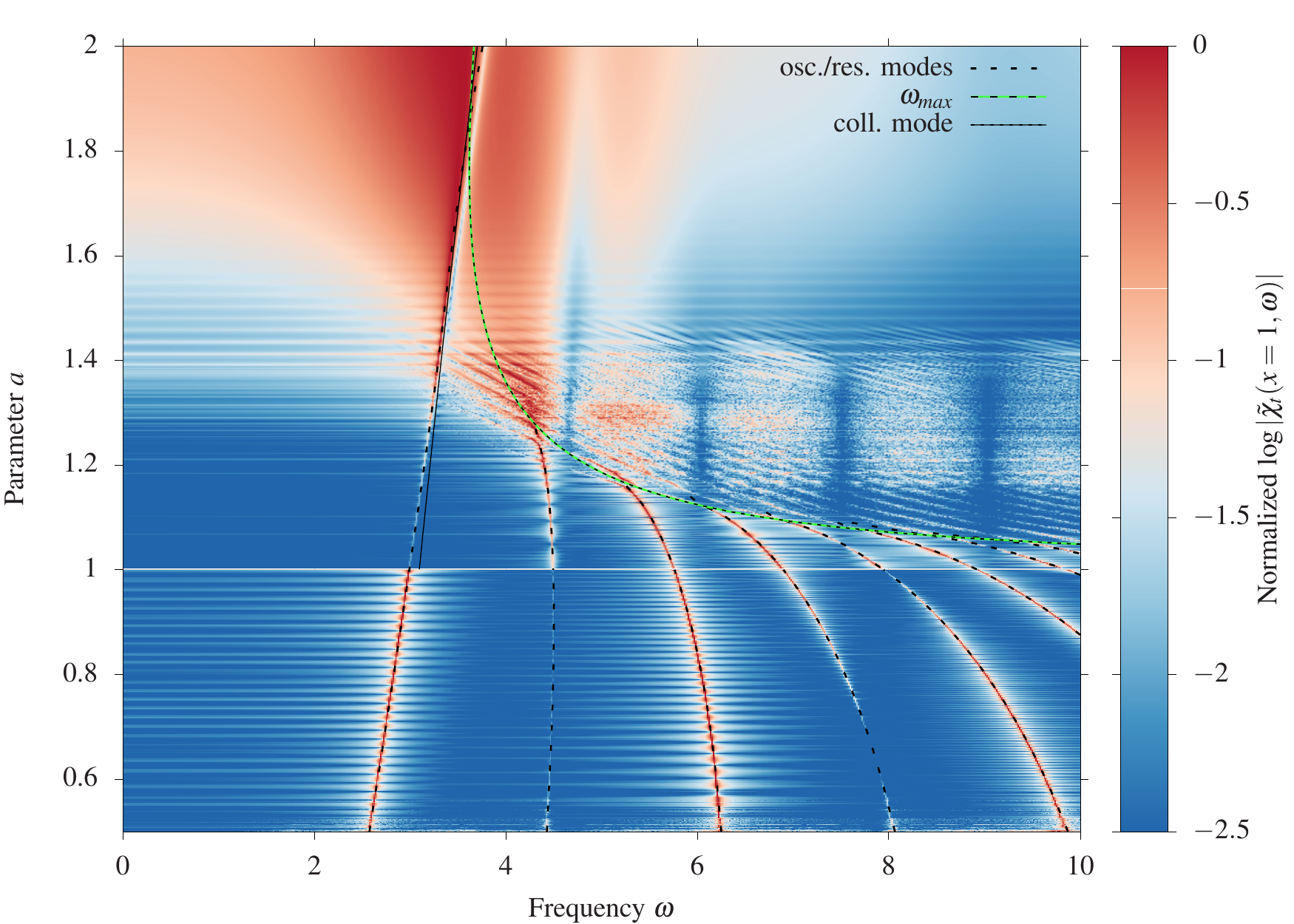}
\caption{The power spectra - scan for different $a$. It is clearly seen that the peaks from the full numerical time evolution precisely coincide with the frequencies of the bound modes for $a<1$ (for compactons) and with the resonance frequencies for $a>1$. It is also visible that for $\omega > \omega_{\rm max}$ the resonances quickly broaden and disappear. For comparison, we also plot the Derrick mode which is close to the first resonance. }\label{FiguremasterPlot}
\end{figure}

We gather all the results in a single scan showing  spectra for different values of $a$: $0.5<a<2.0$ (Figure~\ref{FiguremasterPlot}). We have normalized 
the spectra so that the 
largest maximum would be equal to 1 to have a better comparison between narrow and wide peaks.
We have also added the numerically obtained (from ODEs) narrow resonance positions and the Derrick mode. 
The resonant structure resembles the structure shown for the linearized cross section Figure~\ref{FigureXSection}. However, the narrow peaks are now clearly 
visible even for $\omega^2<Q_{\rm max}$. Discrete Fourier transform in a sense averages over the frequencies, and even if the dominant frequency does 
not match the exact value of the probing frequency, it influences its neighbors and is easily seen on the scan. 
Moreover, the narrow resonances are the ones which persist the longest and have the largest contribution in the spectrum.
Cross sections, as seen in Figure~\ref{FigureXSection}, are calculated also for discrete frequencies, but if a narrow resonance is located between 
those frequencies it may not be seen. 
Another discrepancy between the cross section and the power spectrum is that wide resonances are clearly seen on the cross section, but they do not contribute much in the power spectrum, because their life time is too short.

\begin{figure}
\includegraphics[height=12cm]{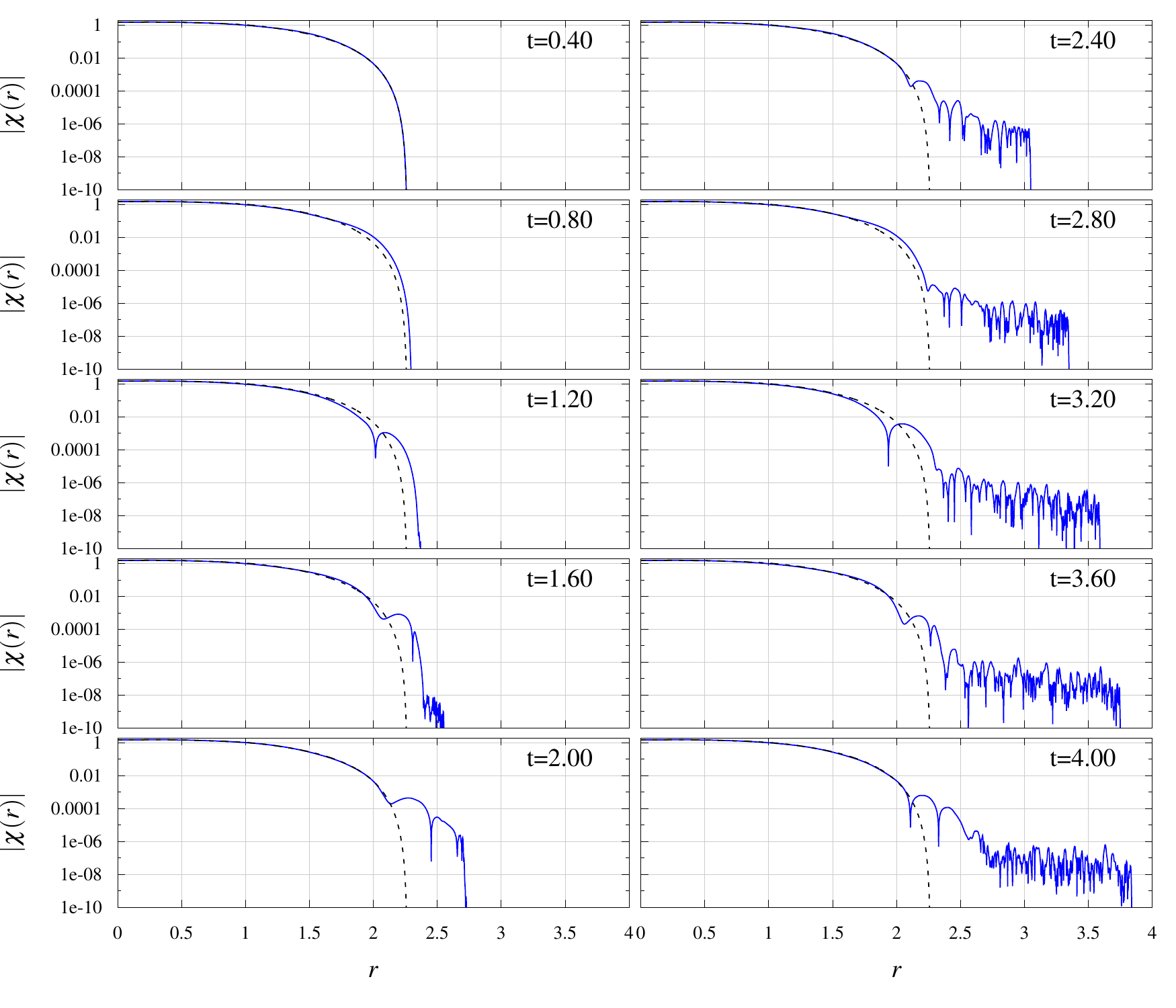}
\caption{Evolution of the perturbed compacton with $A=0.1$ and $a=0.8$. It can be seen that some radiation escapes the compacton although, at least for the chosen parameter values, the amplitude of the radiation is very small (observe the logarithmic scale; in the log plot, we actually plot $|\chi |$). Also, the frequency of the radiation outside the compacton seems to increase, and the radiation seems to "freeze" at some finite distance.}\label{Figurecompacton}
\end{figure}

\begin{figure}
\includegraphics[height=12cm]{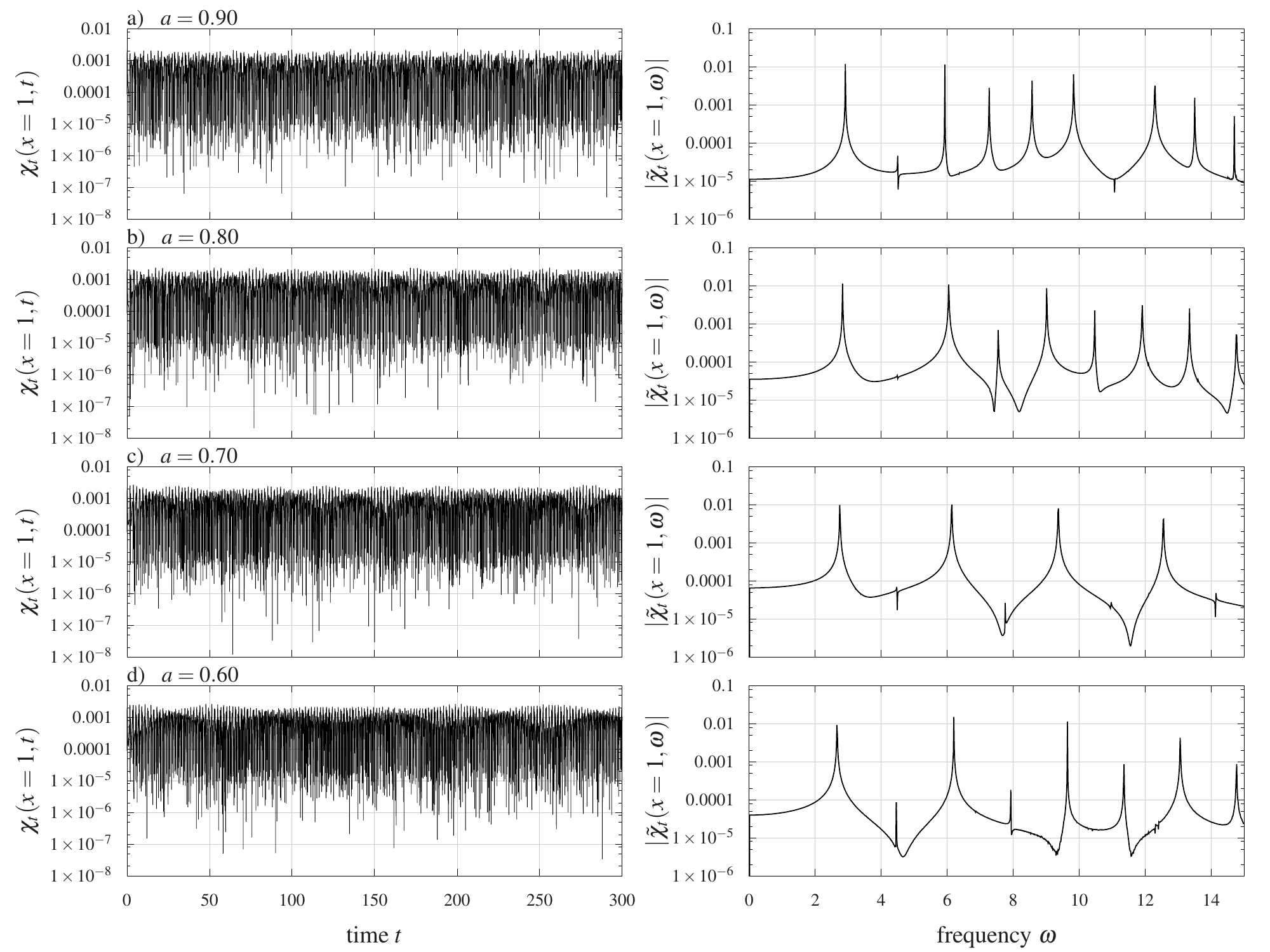}
\caption{Evolution of the field and power spectra for potentials $\mathcal{U}=\chi^{2a}$ for $a<1$ (compacton case). The amplitude of oscillations 
does not seem to change showing that hard wall approximation provides an appropriate description. The peaks correspond to bound states of the compacton.}\label{Figurescanasmalla}
\end{figure}

~~\textit{Compacton}~~We have also tested numerically our conjecture about compacton stiffness. We have found that compactons are not rigid bodies 
and, when perturbed, they pulse and radiate (Figure~\ref{Figurecompacton}). However, the nature of the radiation is non-linear and strongly depends on the 
amplitude of the perturbation. This is, in fact, not a surprise, since the linear approximation does not work quite well in the case of compactons. 
For each amplitude of small perturbation, there is a region near the surface of the compact skyrmion where the perturbation is larger than the 
background field around which the linearization is made.  
Near the surface, nonlinearities start to dominate the evolution of the system, no matter how small the perturbation was.
Formally, the linearization has a zero radius of convergence.
Despite all this, we have found that the linearization works extremely well for quite a large range of the perturbation amplitude and the parameter $a$, see Figure \ref{Figurescanasmalla}. 
Usually, there exists some critical amplitude of the perturbation, above which the compacton starts to radiate quite intensely. Below that amplitude, the 
radiation can be negligible. Moreover, compactons with larger $a<1$ have a softer surface, and it is easier for the perturbation to leave the skyrmion. For $a$ sufficiently close to 1, the radiation leaving the compacton behaves rather similar to the $a=1$ case and may be analysed by a perturbation in $\epsilon = 1-a$, see appendix C.
The smaller $a$, on the other hand, the better our approximation works. 

\begin{figure}
\includegraphics[height=12cm]{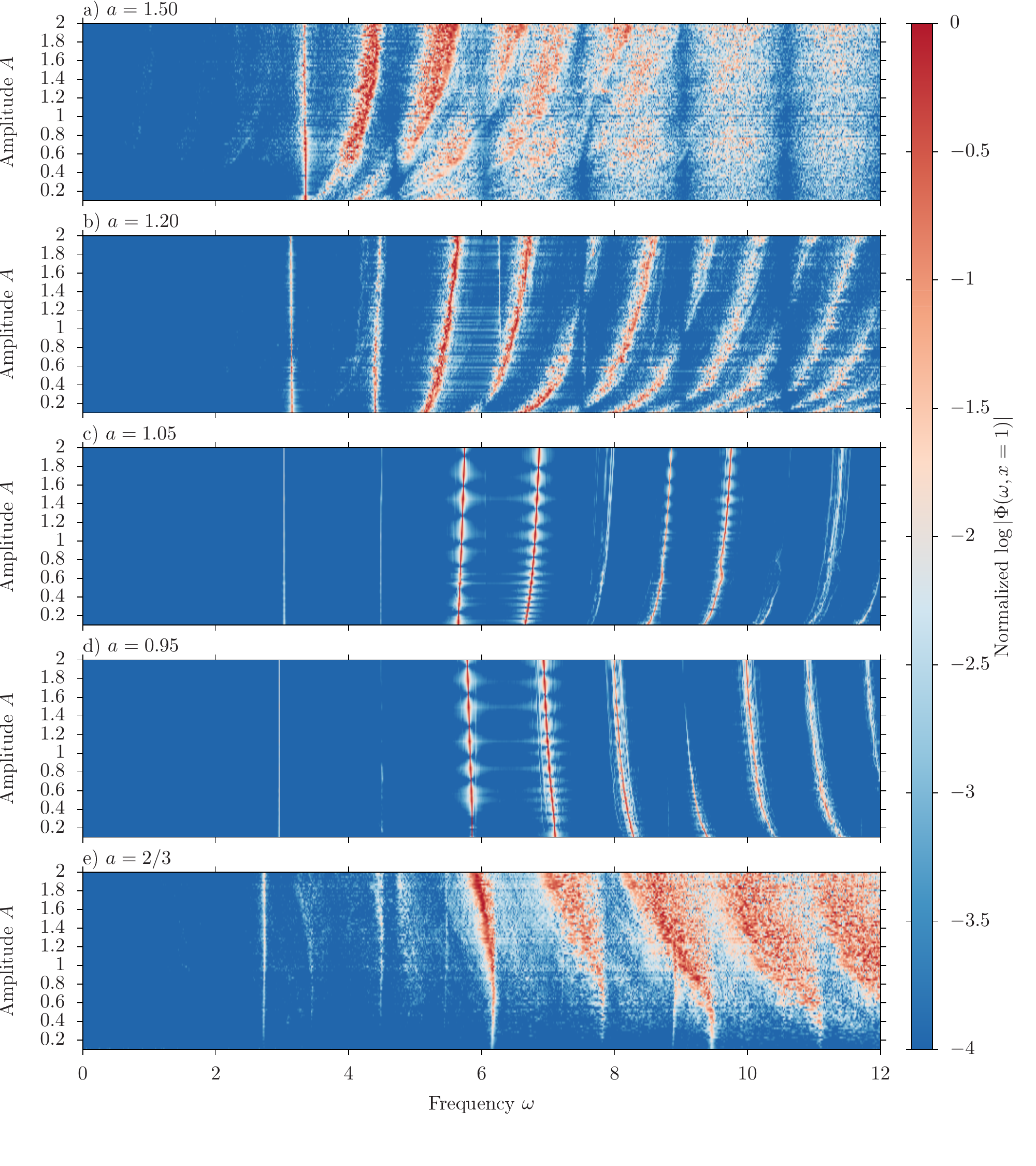}
\caption{Frequency shift with amplitude due to the nonlinearity.}\label{Figurenonlinearity}
\end{figure}

~~\textit{Nonlinearity}~~As we discussed above, the nonlinearity plays an extremely important role in case of compactons. We have also checked how it 
influences the frequencies of bound modes and resonances. We expect that the frequencies will change when we change the amplitude of the perturbation, which is a  well known effect. For potentials for which the solutions are known analytically (such as the harmonic oscillator or Coulomb potentials), it is quite easy 
to find how the energy depends on the perturbation parameter. Here, we only present the scan of how the dominating frequencies change with the
perturbation amplitude (Figure~\ref{Figurenonlinearity}). For $a>1$, the resonance frequency grows as we increase the amplitude of the perturbation for 
all resonances except for the first one. $a=1$ is a linear case, and the frequencies do not depend on the amplitude. For $a<1$, the frequencies decrease as the
amplitude grows.

\section{Conclusions}
The main result of the present paper is a detailed study of radial oscillations and resonances (quasi-normal modes) of skyrmions in the BPS Skyrme model, where we performed both a linearized and a full nonlinear analysis. Although we did most of our calculations for a particular class of Skyrme potentials $\mathcal{U} \sim |\chi|^{2a}$, our results are, in fact, much more general, because the resulting modes mainly depend on the vacuum approach of the potential, $\mathcal{U}\sim \xi^b$ for $\xi \to 0$, and not on the particular potential. Concretely (remember $b=6a$), 
\begin{itemize}
\item for $b \in [0,6)$ we have compact skyrmions. The effective potential tends to (plus or minus) infinity at the compacton boundary. This results in an infinite, discrete oscillating spectrum. 
\item For $b=6$ we have stronger-than-exponentially localised skyrmions ($\simeq \exp (-r^3)$). The effective potential tends to infinity at spatial infinity. Hence, the spectrum is still infinite and discrete. 
\item For $b\in (6,\infty)$ we have power-like localised skyrmions. It is a rather surprising feature of the BPS Skyrme model that the resulting effective potential always tends to 0 (for arbitrary $b>6$), independently of the particular asymptotics of the potential. Hence, there is no discrete spectrum, i.e., no oscillating modes.  On the other hand, resonance modes may appear. We found resonance modes for all the potentials we considered. Concretely, for $a>1$ but close to 1 we found a large number of narrow resonances, whereas for larger $a$ all resonances except for the first one disappear, and the first resonance gets much broader.
\end{itemize}
All these results of the linearized analysis are fully confirmed by the nonlinear calculations. We find that the linearized method works well even in the case of compactons, which is somewhat surprising because of the intrinsically nonlinear character of compactons. 

Concerning physical applications of our results, first of all, we have to take into account the fact that the BPS Skyrme model is only a submodel, whereas a detailed description of physical properties of hadronic or nuclear matter requires the presence of further terms (the general Skyrme model). Within the BPS submodel it is, therefore, wiser to limit oneself to estimates rather than detailed fits in a first step. With this in mind, we take advantage of the fact that the first excitation frequency $\omega_1$ is described rather well by the collective mode frequency $\omega_c$ for a rather wide range of values for $a$, both for true oscillations $a\le1$ and for resonances $a>1$ (more precisely $\omega_c \sim \Omega_1 = {\rm Re}(\omega_1)$ in this case). But the collective mode frequency is related to the energy RMS radius $R_E$ by the very simple relation $\omega_c = (3/R_E)$ in all cases, so to find a physical value for $\omega_c$ we just have to find (or assume) one for $R_E$. 
Assuming, e.g., the nuclear radius per baryon number, $R_E = 1.25 \, {\rm fm}$, we get for the excitation energy $E_1 = \hbar \omega_1 \simeq \hbar \omega_c$,
\be
E_1 \simeq \frac{3 \cdot 197.3}{1.25}\, {\rm MeV} = 473.5 \, {\rm MeV}.
\ee
This compares reasonably well with the excitation energy of the Roper resonance, 
$$\Delta E_{\rm rop} = M_{\rm rop} -M_n = (1440 - 939) \, {\rm MeV} = 501 \, {\rm MeV}.$$
The fact that for a wide class of potentials we find several resonances is interesting from a physical perspective, as well. Indeed, together with the Roper resonance, the existence of at least two higher excitations (higher Roper resonances) is experimentally established. On the other hand, there are strong numerical indications that in the standard Skyrme model (without the sextic term) only one resonance exists. It is, therefore, plausible to conjecture that the inclusion of the sextic term (i.e., considering the full generalized Skyrme model of Eq. (\ref{gen-lag})) might induce higher resonances and lead to a realistic description of Roper resonances. This will be further investigated in a forthcoming publication.
 
On the other hand, the axially symmetric multi-skyrmion solutions considered here do not provide a realistic description of the giant monopole resonances. The excitation energy of the giant monopole resonance for sufficiently large nuclei extracted from different experiments is approximately
\be
E_{{\rm gm},B} \simeq 80 B^{-\frac{1}{3}} \, {\rm MeV},
\ee
see, e.g., \cite{blaizot}, which extrapolates for $B=1$ to 80 MeV.  The excitation energies calculated in the present paper  have the same scaling behaviour with the baryon number $B$, i.e., $E_{1,B} = B^{-\frac{1}{3}} E_1$, but with a value for $E_1$ which is about six times the value for the giant monopole resonance $E_{{\rm gm},B=1} \simeq 80 \, {\rm MeV}$. We think that this result just demonstrates that the axially symmetric ansatz (\ref{ax-sym}) leading to BPS skyrmions with spherically symmetric energy densities for arbitrary $B$ is not adequate for the giant monopole resonance. This is, in fact, not so surprising. As said, a detailed and quantitatively reliable description of hadronic and nuclear matter requires the full generalized Skyrme model, and the field equations of the full model (or of the original Skyrme model without the sextic term) are not compatible with the axially symmetric ansatz (\ref{ax-sym}). In the original Skyrme model and variants thereof, substructures (e.g., $B=1$ substructures, i.e., individual nucleons, or $B=4$ substructures, i.e., alpha particles) are visible in higher $B$ Skyrmions, and these substructures will most likely still be present in the solutions of the full model, owing to the huge symmetries of the BPS submodel (which is essentially compatible with any shape). But skyrmions with substructures contain regions with small energy density also in their interior and should, therefore, be softer w.r.t. compressions with small amplitudes and lead to smaller excitation energies. We remark that a similar hint that higher $B$ skyrmions with particle-like substructures are more adequate even for the BPS submodel may be found already in its application to nuclear binding energies (as commented in more detail in \cite{review}). Indeed, for axially symmetric BPS skyrmions, where the individual nucleons are completely dissolved, the isospin moments of inertia $I$ grow like $B^\frac{5}{3}$, leading to too small isospin excitation energies for large $B$ (large nuclei). For skyrmions with substructures, on the other hand, the total isospin moment of inertia is the sum of the contributions of the substructures, leading 
 to $I\sim B$, and the resulting isospin excitation energies essentially match the ``asymmetry energy" contribution of the Weisz\"acker formula.
(Even without these corrections, however, the BPS submodel already leads to much more accurate binding energies than the original model because of its BPS property.)

Let us remark that, while the BPS submodel is particularly stiff with respect to the monopole (radially symmetric) excitations considered in this paper, it gives rise to much softer modes. The static energy density is invariant under SDiff transformations (volume-preserving diffeomorphisms) on physical space, so the corresponding modes may be excited with arbitrarily small frequencies. In the full generalized Skyrme model the sextic term, therefore, will soften vibrations for modes which are close to SDiffs, whereas it will stiffen vibrations of modes which significantly change the volume of the original soliton (like the monopole excitation). In other words, the sextic term has a rather interesting and nontrivial influence on the spectrum of vibrations of skyrmions, which might lead to applications and new results in the description of baryon resonances. A detailed discussion of these issues is, however, beyond the scope of the present paper.

\begin{figure}
\includegraphics[height=12cm]{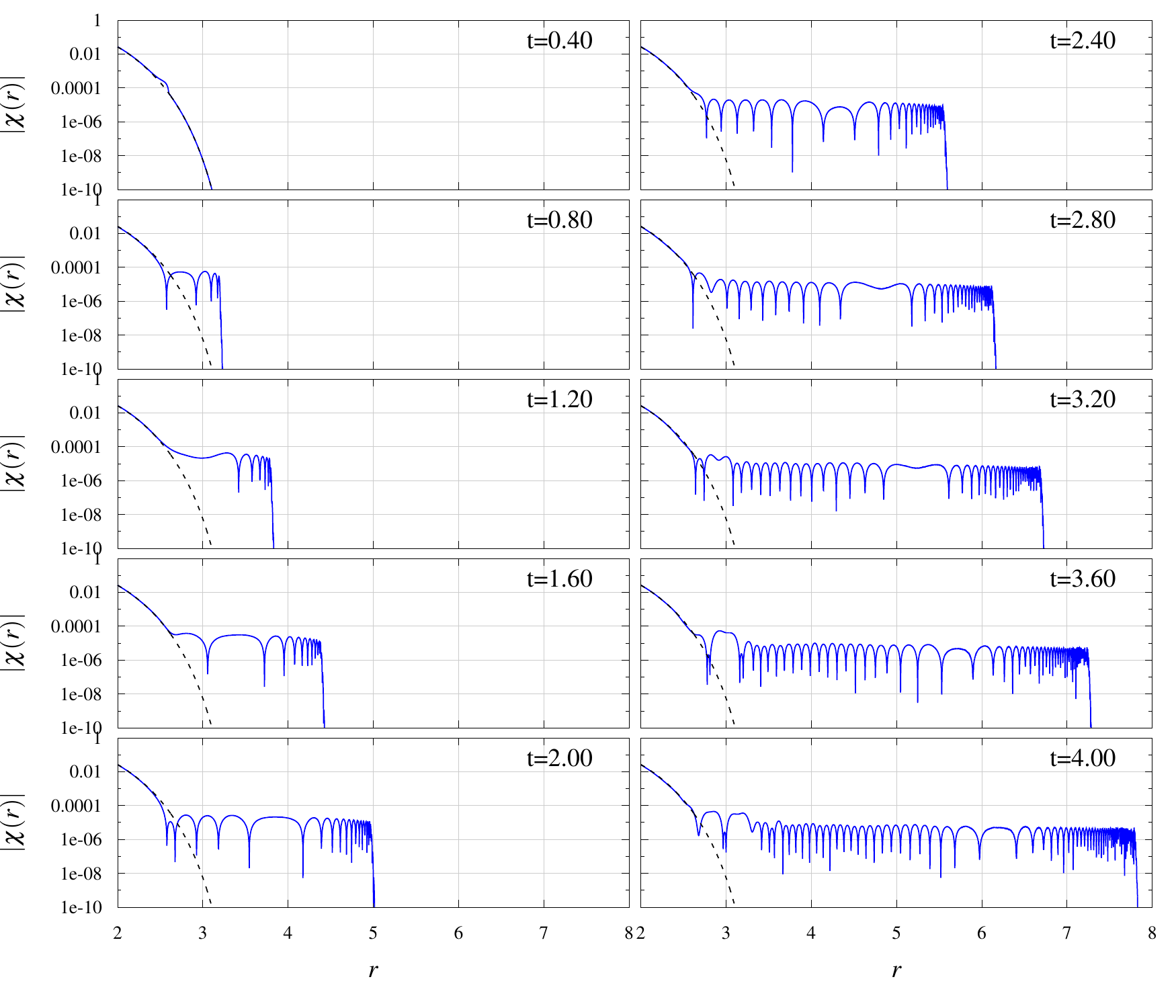}
\caption{Evolution of the perturbed compacton, with the perturbation induced by a time-dependent (energy non-conserving) boundary condition inside the compacton (at $r=2$), $\chi(2,t) = \chi_0(2)+A \sin (\omega t)$, for
 $A=0.001$, $\omega = 4$, and $a=0.95$. The dashed line is the unperturbed compacton. It can be seen that some radiation escapes the compacton although, at least for the chosen parameter values, the amplitude of the radiation is very small (observe the logarithmic scale; in the log plot, we actually plot $|\chi |$). Also, the frequency of the radiation outside the compacton seems to increase with distance. The radiation no longer ``freezes" at a finite distance, because of the energy non-conserving boundary condition which effectively pumps energy into the system.}\label{Figurecompacton3}
\end{figure}

\begin{figure}
\includegraphics[height=12cm]{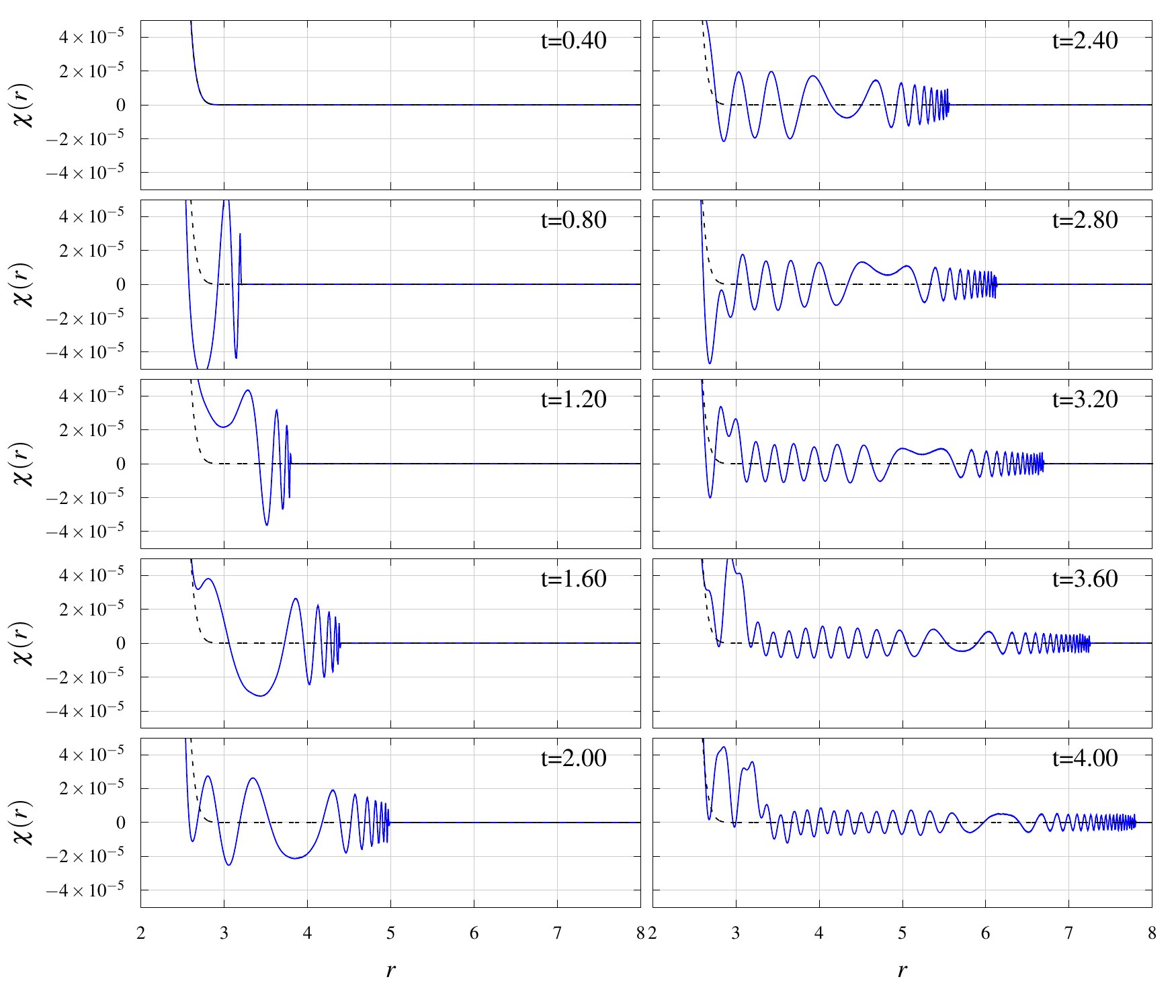}
\caption{Evolution of the perturbed compacton, as in Figure \ref{Figurecompacton3}, but for magnified radiation zone and for a non-log scale.}\label{Figurecompacton3b}
\end{figure}

Finally, we want to emphasize that the methods developed and applied in this article are of more general interest, because they may be applied to a much wider class of models than the ones considered here. They may be used for any nonlinear field theory where the field equations are compatible with spherical symmetry, in any dimensions. Some additional, rather general findings concern the issue of compactons. We found that, strictly speaking, the propagation of small perturbations is not confined to the interior of the compacton, so that, strictly speaking, the linear analysis is not correct. It results, however, that only a tiny amount of radiation escapes the compacton (in Figure \ref{Figurecompacton}, e.g., for a perturbation with an amplitude of $0.1$ the amplitude of the radiation outside the compacton is about $10^{-6}-10^{-8}$). Also, the radiation seems to ``freeze", while increasing its frequency, at some finite distance from the compacton radius, i.e., it does not escape to infinity. 
For comparison, we show further cases of the evolution of perturbed compacons in Figs. \ref{Figurecompacton3}, \ref{Figurecompacton3b}.
The behaviour is similar to the previous case, although the radius where the radiation freezes grows with time, because of the energy non-conserving boundary condition used here. A more detailed discussion of the radiation escaping compactons is given in appendix C.
We conclude that, effectively, the assumption that the compacton confines perturbations seems
to work. 
This can have some applications in the brane world mechanism for compact branes \cite{branes}, and it also
supports the 
stability of compactons \cite{stab-comp}.

\section*{Acknowledgements}
The authors acknowledge financial support from the Ministry of Education, Culture, and Sports, Spain (Grant No. FPA 2014-58-293-C2-1-P), the Xunta de Galicia (Grant No. INCITE09.296.035PR and Conselleria de Educacion), the Spanish Consolider-Ingenio 2010 Programme CPAN (CSD2007-00042), and FEDER. 
The work of M.H. is supported by MEXT-Supported Program for the Strategic Research
Foundation at Private Universities Topological Science (Grant No. S1511006). T.R. thanks P. Bizon for helpful discussions. 

\appendix
\section{The Sturm-Liouville problem}

It is convenient to transform the fluctuation equation into the form of a Sturm-Liouville (SL) problem
on a segment $[0,R]$, where R is finite (for compact solitons) or 
$R=\infty$ for usual infinitely extended solitons. 
The are two standard forms for the SL problem:
\begin{enumerate}
\item  The canonical form
\be
-\frac{d}{dr} \left( p(r) \frac{d\eta}{dr} \right) + q(r) \eta = \lambda s(r) \eta
\ee
For the fluctuation equation of the BPS Skyrme model we, thus, have
\be
p(r)=s(r)=\frac{1}{r^2}, \;\;\; q(r)=\frac{2r^2}{B^2 \boldsymbol{l}^6} \mathcal{U}_{\chi \chi}|
\ee
and $\lambda = \omega^2$.
\item The normal form
\be
-\frac{d^2u}{dx^2} +Q(x) u = \lambda u .
\ee
The transition from the canonical form to the normal form is given by 
\be
x=\int \sqrt{\frac{s(r)}{p(r)}} dr
\ee
\be
u(x)=f(x) \eta (r(x)), \;\;\; f(x)=\sqrt[4]{p(r(x)) s(r(x))}
\ee
and the potential 
\be
Q=\frac{q(r(x))}{s(r(x))} + \frac{f''}{f} .
\ee
Hence, for the BPS Skyrme model case
\be
r=x
\ee
\be
f(x)= \frac{1}{x}, \;\;\; u(x)=\frac{1}{x} \eta(x)
\ee
\be
Q=\frac{2}{x^2}+ \frac{2x^4}{B^2 \boldsymbol{l}^6} \mathcal{U}_{\chi \chi}|
\ee
This is the {\it effective potential} of the related SL problem. 
\end{enumerate}

\section{Resonance modes}
Our starting point is the complexified version of the linear fluctuation equation in normal form (see appendix A),
\begin{equation} \label{compl-lin}
 \psi_{tt}-\psi_{xx}+Q(x)\psi=0,
\end{equation} 
where $\psi$ is the complex linear fluctuation field, and $Q$ is the effective potential.
Resonance modes, also known as quasi-normal modes, play an important role in the long-time evolution of the system. Their origin is purely linear, so very often Green function techniques are applied in this context. Contrary to the bound or oscillational modes, the resonance modes decay in time even in the linearized problem. In principle, they can be characterized by a complex frequency $\omega=\Omega+i\Gamma$ where $\Omega$ and $\Gamma$ are real.   
Therefore, the time evolution has the form
\begin{equation}
 \psi(x,t)\approx e^{i\omega t}\phi(x) = e^{-\Gamma t} e^{i\Omega t}\phi(x) .
\end{equation} 
The characteristic time $T_{1/2}=\ln2/\Gamma$ is called half-life. The values $\Omega, \Gamma$ and the spatial profile $\phi(x)$ are obtained by solving equation (\ref{compl-lin}) with the purely outgoing wave condition $\phi(x)\sim e^{-ikx}$ as $x\to\infty$, where $k$ is a wave number where, for a massless field (i.e., in our case, for an effective potential $Q(x)$ which decays sufficiently fast for $x \to \infty$),  
$k=\omega$.
Some important physical applications of quasi-normal modes are radioactivity and gravitational waves after black-hole collisions.

We recall some basic properties of resonance modes after Kokkotas \cite{Kokkotas}. For the problem of solving equation (\ref{compl-lin})
with Cauchy initial data $\psi(x,0), \psi_t(x,0)$, it is best to apply the Laplace transform technique,
\begin{equation}
 \hat\psi(x,s)=\int_{0}^{\infty}e^{-st}\psi(x,t)\,dt ,
\end{equation} 
solve the appropriate inhomogeneous equation 
\begin{equation}
 s^2\hat\psi-\hat\psi_{xx}+V\hat\psi=s\psi(x,0)+\psi_t(x,0)
\end{equation} 
and apply the inverse Laplace transform:
\begin{equation}
 \psi(x,t)=\frac{1}{2\pi i}\int_{-\infty}^\infty e^{(a+is)t}\hat\psi(x, a+is) ds.
\end{equation} 
The real value $a$ is chosen in such a way that the real parts of all the poles of the integrand are less than $a$.
For solving the inhomogeneous equation, the Green function can be used,
\begin{equation}
 \hat\psi(x,s)=\int_{-\infty}^\infty G(x, x', s)j(s,x')dx'.
\end{equation} 
It can be written as
\begin{equation}
 G(x, x', s) = \frac{1}{W(s)}
 \begin{cases}
  \eta_1(x',s)\eta_2(x,s)&\qquad x'<x\\
  \eta_1(x,s)\eta_2(x',s)&\qquad x'>x
 \end{cases}
\end{equation} 
where $\eta_{1,2}$ are solutions of the homogeneous equation. We choose the Green function which gives bounded solutions for compact initial data.
The quasi-normal modes are those solutions for which the Green function has a pole for $|x|\to \infty$. 
This means that the Wronskian vanishes, and the two solutions $\eta_1$ and $\eta_2$ become linearly dependent. 
As a result, after applying the inverse Laplace transform for large $t$, we obtain a sum over all poles of the Green function,
\begin{equation}
 \psi(x,t)=\sum_ne^{i(\Omega_n+i\Gamma_n)t}\phi_n(x).
\end{equation} 
After sufficiently long time, only the resonance with the smallest $\Gamma$ will remain.
For
massive fields, just like for the double sine-Gordon case, one should use a dispersion relation
$\omega=\sqrt{k^2+m^2}$. This results in the appearance of poles at $\omega=\pm im$ and an additional
contribution in the form $e^{imt}/\sqrt{t}$ \cite{bizon2}.

Physically, one can use only real frequencies. The resonances manifest themselves as especially long-lived and much more localized states, compared to 
other scattering states. This feature can be used to localize the real parts of the resonant frequencies for narrow and, hence, the most important 
resonances. Another important feature of a resonance is that it appears as a peak in the scattering cross section.
For one channel scattering in 3d, the $\ell$-th partial cross section is defined as
\begin{equation}
 \sigma_{\ell}=\frac{4\pi}{k^2}(2\ell+1)\sin^2\delta_{\ell} ,
\end{equation} 
where the phase shift $\delta_{\ell}$ can be read out from the asymptotic form of the regular solution 
\begin{equation}
 \phi(x)\sim A_\ell\sin(kx-\ell\pi/2+\delta_\ell).
\end{equation} 
In our case $\ell=0$.
In the vicinity of a resonant frequency $\Omega$, the cross section can be written as
\begin{equation}
 \sigma_\ell(\omega)\approx\frac{4\pi}{k^2}(2\ell+1)\frac{\Gamma^2/4}{(\omega-\Omega)^2+\Gamma^2/4}.
\end{equation} 

Finding an exact resonance numerically can be a difficult task. The most straightforward method of finding a solution which satisfies purely outgoing 
wave condition can be used only for relatively small values of $\Gamma$. The profile of the resonance is a superposition of outgoing and ingoing wave 
$\phi(x)=Ae^{ikx}+Be^{-ikx}$ and, with complex values of $k$, one wave decreases exponentially and the other grows exponentially. Finding the exact 
values 
of $A$ and $B$ becomes an extremely difficult task. There exist some techniques which can improve the method by changing variables, 
hyperbolic foliations, or decomposing into phase and amplitude. 
Other techniques include the WKB approximation, matching to some known potentials, or simulating the linear evolution and fitting the values to obtained 
data (Prony's method). 
Most of the methods have their limitations. Generally, it is easier to find a single resonance with small (but not too small) $\Gamma$. For problems 
with a large number of wide resonances, most methods fail.

\textit{~~Direct solution.~~} 
We have found that the direct method of solving the linear equation to find the resonance mode can be applied only in the limiting case of very small $\Gamma$. We applied this 
method for resonances lying below $\omega_{\rm max}$. We have found that in order to obtain more or less $1\%$ precision, one can find the resonances for $\Gamma L<7$, where $L$ is the 
distance at which the potential $Q$ becomes insignificant (depending on the value of the parameter $a$, but usually one should use $L\simeq 10\ldots 1000$).

\textit{~~Linear evolution.~~} 
Sometimes it is easier to perform a simulation of the linear equation (\ref{compl-lin}). If there is a single mode or one most dominating one, it is easy 
to find its frequency and damping coefficient analysing the evolution. After some time, the resonance modes would be visible in the evolution 
$\psi(x,t)\sim e^{-\Gamma t}e^{i\Omega t}$. By fitting to the numerical data one can extract quite accurately the parameters describing the resonance. 
This method can be generalized to a larger number of resonances (Prony's method \cite{Prony}). The evolution of the linearized equation is not contaminated by the nonlinear coupling (increasing the 
process of decay of the resonance mode) nor by the interaction between modes or nonlinear tails.\\
In order to decrease the fitting errors we multiplied the results by $e^{\Gamma_0t }$ where $\Gamma_0$ is,  in principle, arbitrary but best when slightly less than the sought $\Gamma$.\\
However, there is still a problem with the mass threshold, which vanishes slower than the typical resonance. Usually, in the evolution we see some short 
perturbation dependent evolution, next, for intermediate time, the most narrow resonance and then the mass threshold $\xi(x,t)\sim e^{imt}/\sqrt t$. 
Sometimes the intermediate time is too short to fit the resonance. However, for the linearised equation we can add a constant term to the potential and 
hence change the mass threshold, in particular to set the mass to 0.
\begin{equation}\label{eq:linearPDE2}
 \ddot\psi-\psi''+\left(Q(x)-m^2\right)\psi=0.
\end{equation} 
With the above equation it is very easy to find a resonance if there is only a single mode or a very dominating one. The measured frequencies 
$\tilde\omega^2$ can be used to obtain the non-shifted frequency of the quasi mode: $\omega^2=\tilde\omega^2+m^2$.\\

\begin{figure}
\includegraphics[width=0.48\linewidth,angle=0]{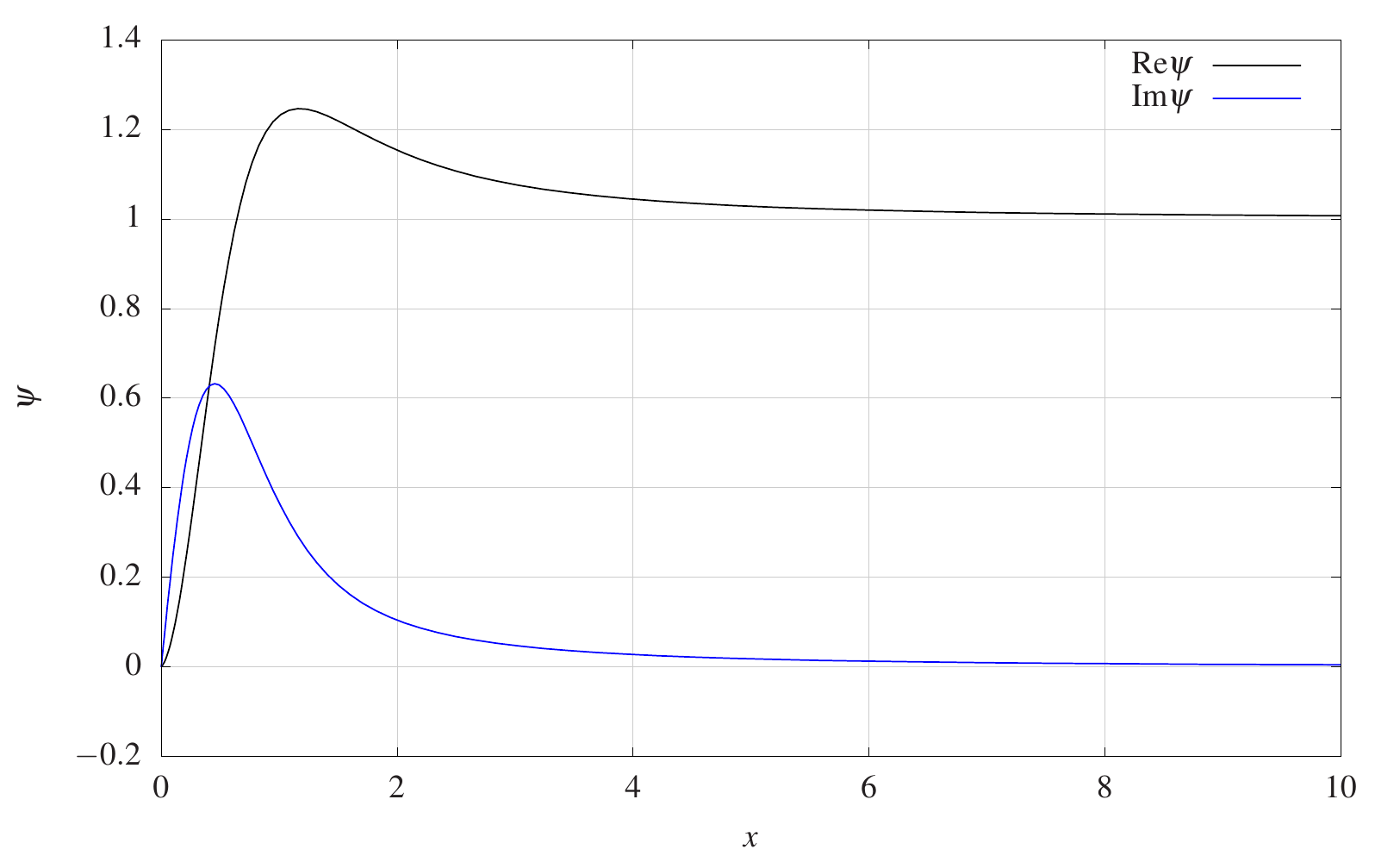}
\caption{\small A resonance mode found for massless double sine-Gordon $\epsilon=-1/4$ for frequency  $1.2665+0.4440i$ in WKB 
foliation using rational Chebyshev spectral method for semi infinite interval $x\in [0,\infty]$. The frequency is very close to the frequency found 
using linear simulation method. Note that the solution tends to 1 as $x$ grows to infinity without any oscillations.\\
}\label{Figuredsg1ResMode}
\end{figure}

\textit{~~Foliation.~~}
Instead of studying the solution along $t=\mbox{const}$ time slices, it is somewhat better to introduce a new time-like variable
\cite{bizon3}, \cite{Zenginoglu}
\begin{equation}
 \tau = t-h(x),\qquad\tau\to |x| \text{ as } |x|\to \infty .
\end{equation} 
The function $h(x)$ defines a new foliation of space-time, typically $h(x)=|x|$ (zero) or $h(x)=\sqrt{c^2+x^2}$ (hyperbolic foliation). 
The solution describing the resonance mode can now be written as
\begin{equation}
 \psi(x,\tau)\approx e^{i\omega\tau}\left(A_{in}e^{2i\omega x}+A_{out}\right).
\end{equation} 
This means that now we look for a solution which tends to a constant value as $x\to\infty$. The new coordinates trace the solution along with the 
characteristics of the outgoing wave. In a sense the outgoing wave is frozen. 
As long as $A_{in}$ is not negligible, the solution 
suffers the same errors as previously, but when we find a solution with $A_{in}=0$, we have a solution for which our method provides an accuracy 
$\epsilon_m$. This is equivalent to finding a soliton with vanishing derivative at spatial infinity and is very similar to finding 
bound states. 
Introducing the new function $\psi(x)=e^{i\omega h(x)}\phi(x)$ we obtain the following equation:
\begin{equation}
 -\phi''-2i\omega h'\phi'+\omega^2(h'^2-1)\phi-i\omega h''\phi+Q\phi=0.
\end{equation} 
Unfortunately, for long range potentials $\sim 1/x^2$ even zero or hyperbolic foliations are not enough. It is true that the solutions tend to a
constant plus an exponent which we want to get rid of, but the rate with which the solution tends to this asymptotics makes it sometimes difficult to treat it
numerically with high accuracy. Even far away from the potential center, the solution still oscillates. Such oscillations are very difficult to 
deal with using high accuracy spectral methods, which are excellent, but for smooth, nonoscillatory functions.
To avoid this behavior, we have decided to use different function $h(x)$, which will more suitably treat those oscillations. We assume that for almost 
flat, but non-zero potentials, the solution can be approximated as $\psi\approx Ae^{ikx}+Be^{-ikx}$, where $k^2=\omega^2-Q(x)$. On a short interval 
we  can assume $\omega h=ik$, so we can choose $h'(x)=k/\omega$, or $h(x)=kx/\omega$. We have chosen the first case so 
\begin{equation}
 h(x) = \int_0^x\sqrt{1-\frac{Q(x')}{\omega^2}}\;dx'.
\end{equation}
If $\omega>Q(x)$ the function $h(x)$ defines a new foliation, however when $\omega<Q(x)$, the function $h'$ becomes imaginary, so formally it is not a true foliation. It still provides
an excellent tool for studying resonances, and we would refer to this approach as a WKB foliation, pointing out its resemblance both to the WKB approximation and the foliations described 
above. 
The equation now takes the form:
\begin{equation}
 -\phi''-2ik\phi'+\frac{iQ'\phi}{2k}=0.
\end{equation}
We have obtained the solution solving directly the above equation for the double sine Gordon case
$\epsilon=-1/4$  with boundary
conditions $\phi(0)=0$ and $\phi(\infty)=1$. The solution is shown in Figure
\ref{Figuredsg1ResMode}.

\begin{figure}
\includegraphics[width=0.48\linewidth,angle=0]{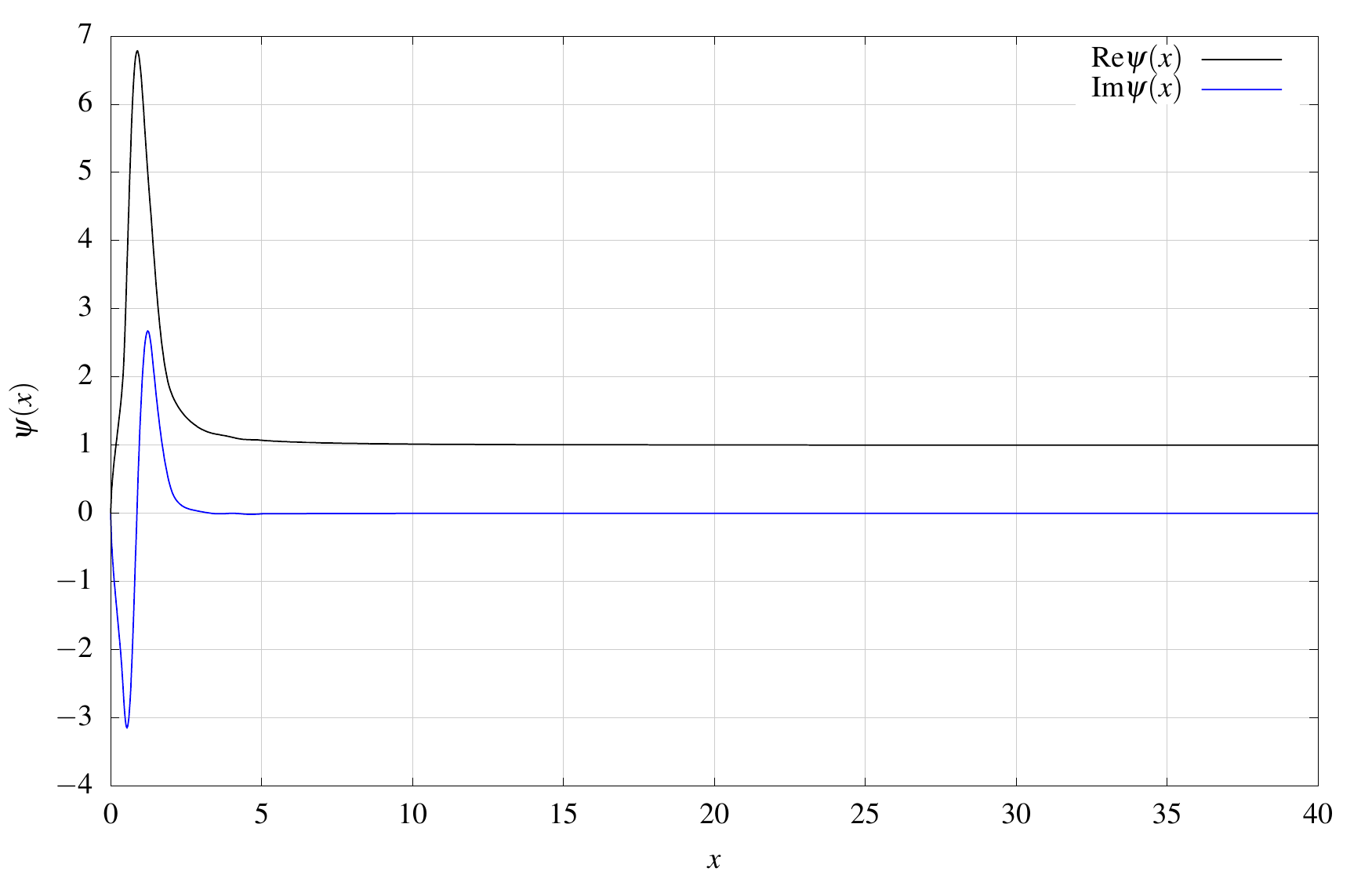}\includegraphics[width=0.48\linewidth,angle=0]{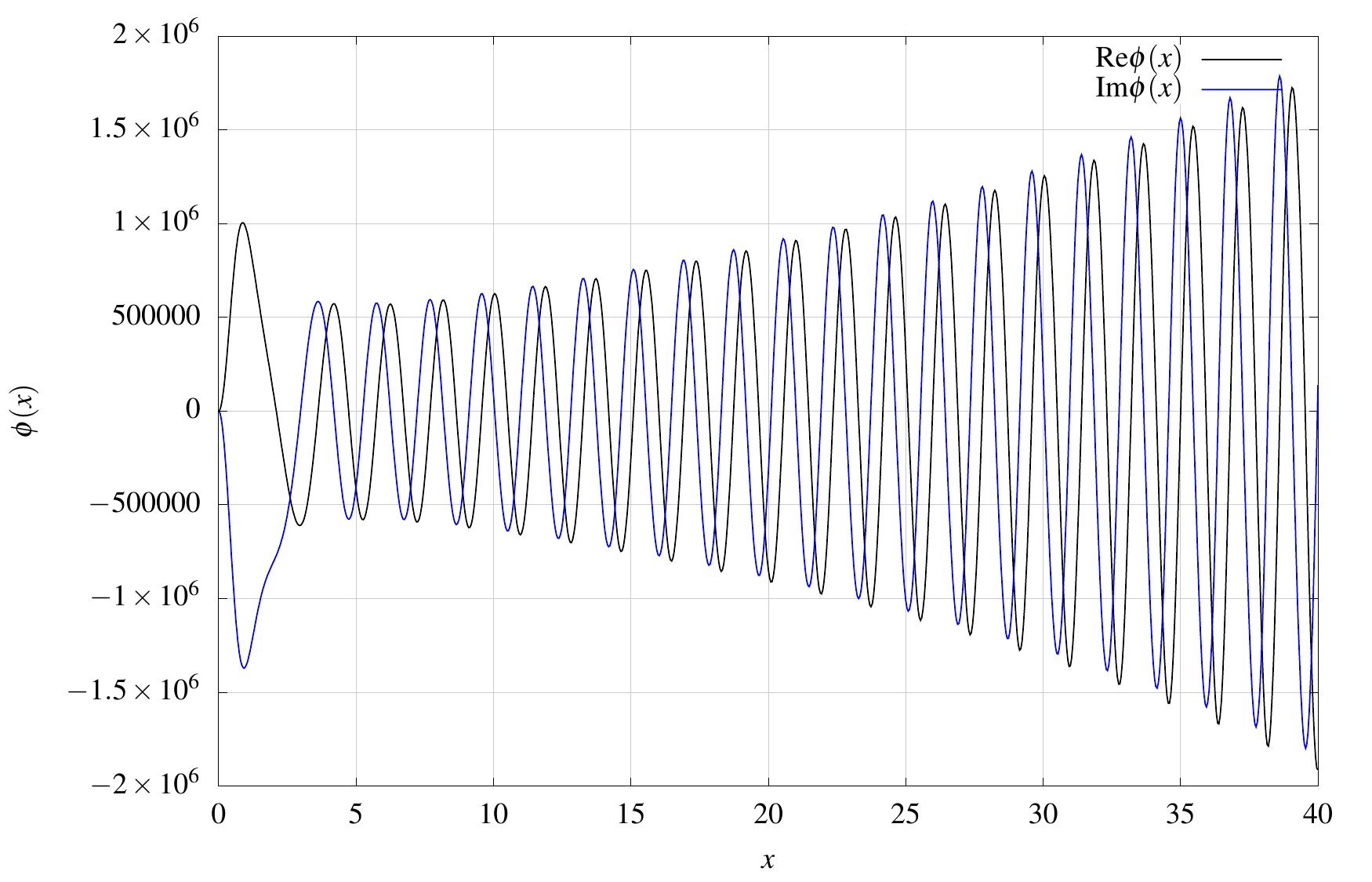}
\caption{\small Comparison between the solution in WKB foliation (left) and in standard $t=\mbox{const}$ time sheet 
(right). BPS model, $a=1.7$, $\omega =  3.4916 + 0.0364i$.}\label{FigureFoliationNonFoliation}
\end{figure}

\begin{figure}
\includegraphics[width=0.48\linewidth,angle=0]{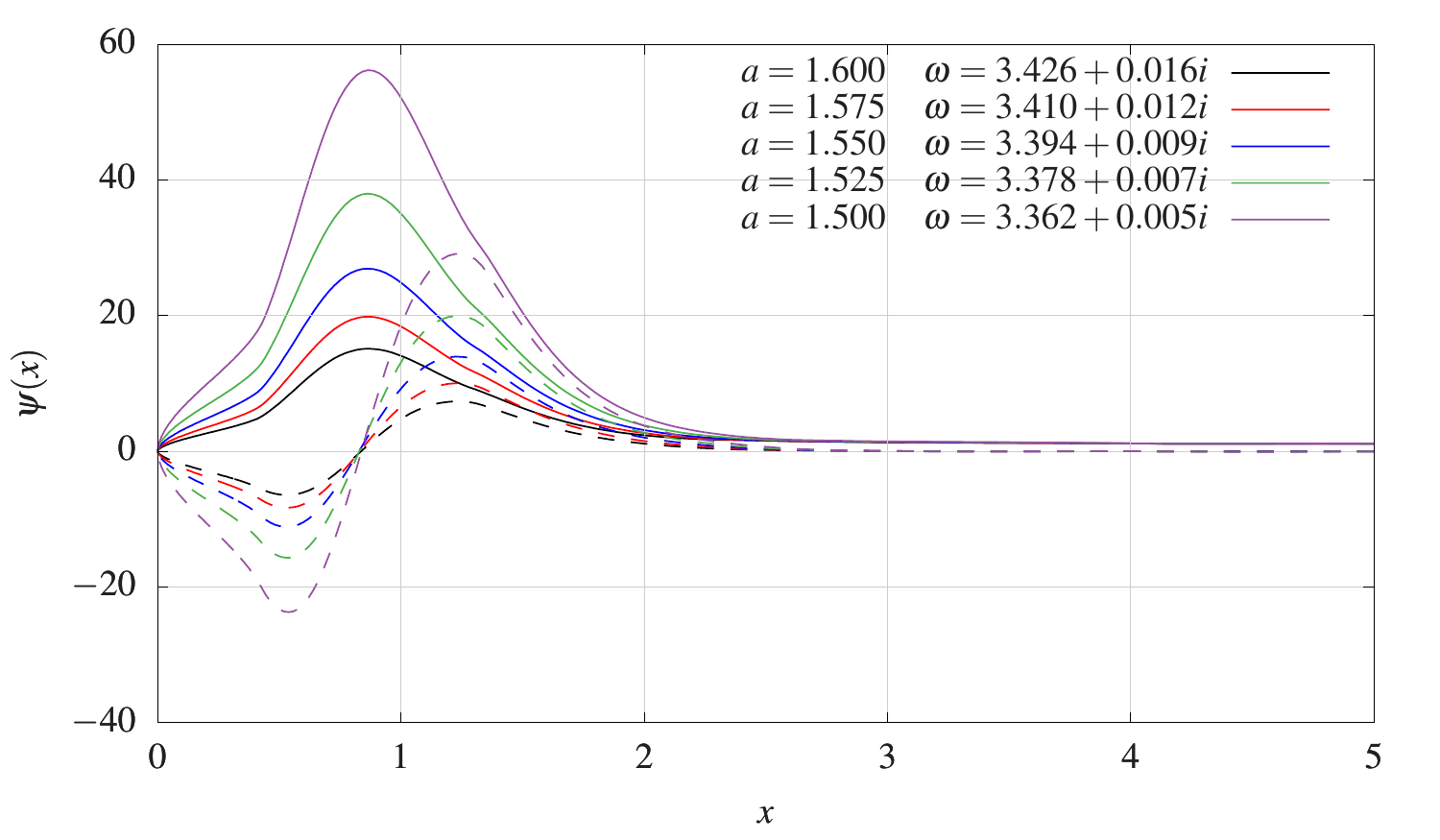}\includegraphics[width=0.48\linewidth,angle=0]{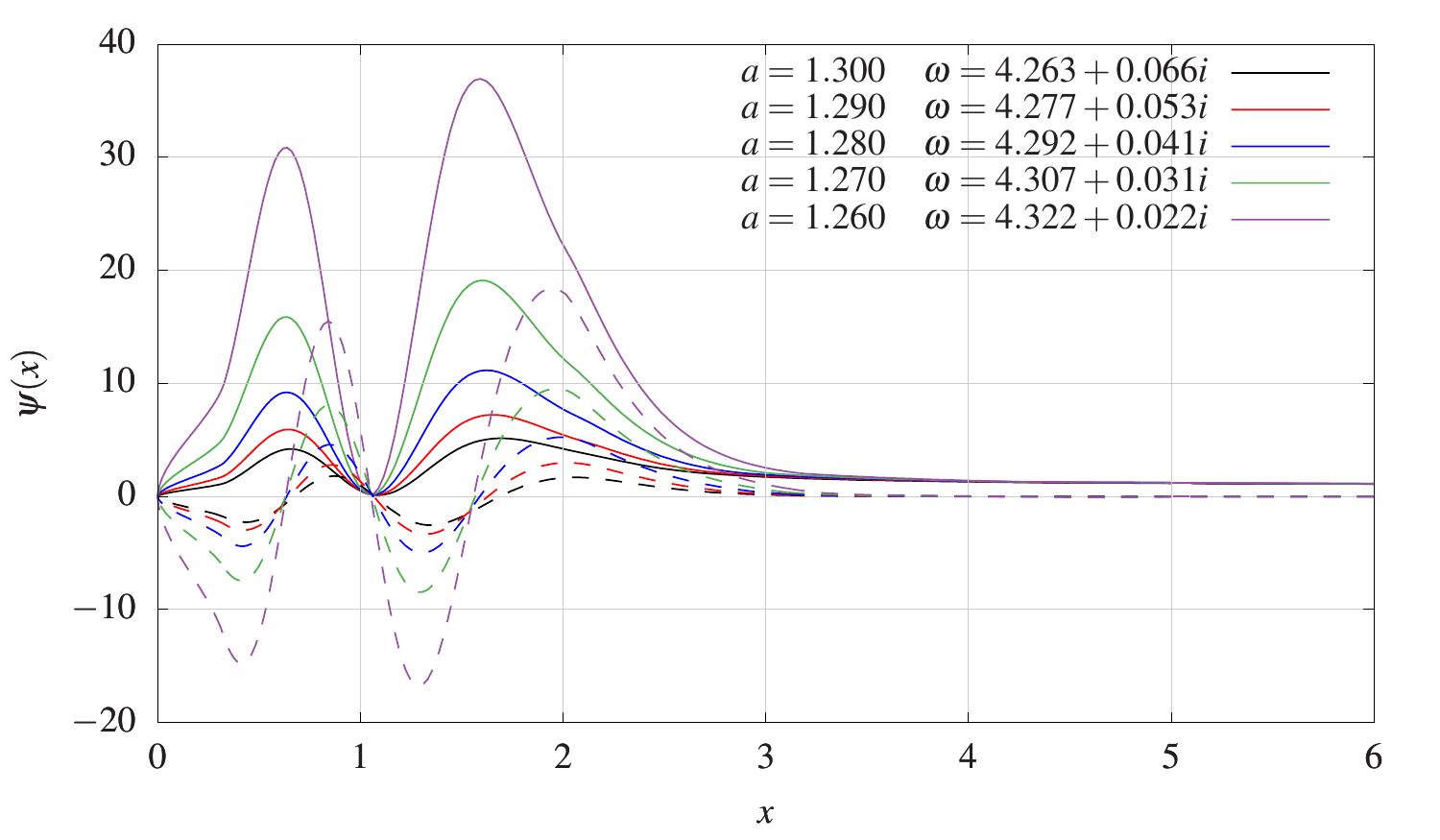}
\caption{\small Comparison between different stages of a single mode as $a$ changes and $\Gamma$ grows rapidly for the resonances formed from the 
first (left) and second (right) bound mode for $a=1$. The solid line is the real part of $\psi (x)$, and the dashed line is the imaginary part. The profiles with smaller $\Gamma$ are more 
localised. Profiles in WKB foliation normalized as $\psi(x\to\infty)\to 1$.
}\label{FigureBPSResMode2}
\end{figure}

The potential $Q(x)$ for the BPS case (and all potentials for higher than one dimensions) have a singularity at the origin. There are two independent 
solutions: regular $\psi(x)\approx x^2$ and singular $\psi(x)\approx x^{-1}$. For singular potentials, we have decided 
to find the solution piecewise. For $x\in[0,x_m]$ we used ordinary Chebyshev polynomials (with linear map $[-1,1]\to[0,x_m]$) and solved the 
standard linear problem for $\psi(x)$ with conditions $\psi(0)=0$ and $\psi(x_m)=1$. Next we solved the equation using the rational Chebyshev method 
with conditions $\phi(\infty)=\phi(x_m)=1$. The first solution was multiplied by $e^{i(h(x)-h(x_m))}$ to match with the second solution. The matching 
function $g(\omega)=\phi(x_m^{-})-\phi(x_m^+)$ vanishes when the functions can be smoothly matched. 
The result is shown in the plot
\ref{FigureFoliationNonFoliation}. For comparison, we have shown the resonance profile in the
original $t=\mbox{const}$ time slice. Note that the solution obtained from our method is almost constant
for large values of $x$ and hence much better treated numerically. Using this technique, we have
also compared the profiles for the first two resonances for different values of the parameter
$a$ Figure \ref{FigureBPSResMode2}.

\section{Skyrmion radiation}
\begin{figure}
\includegraphics[width=0.6\textwidth]{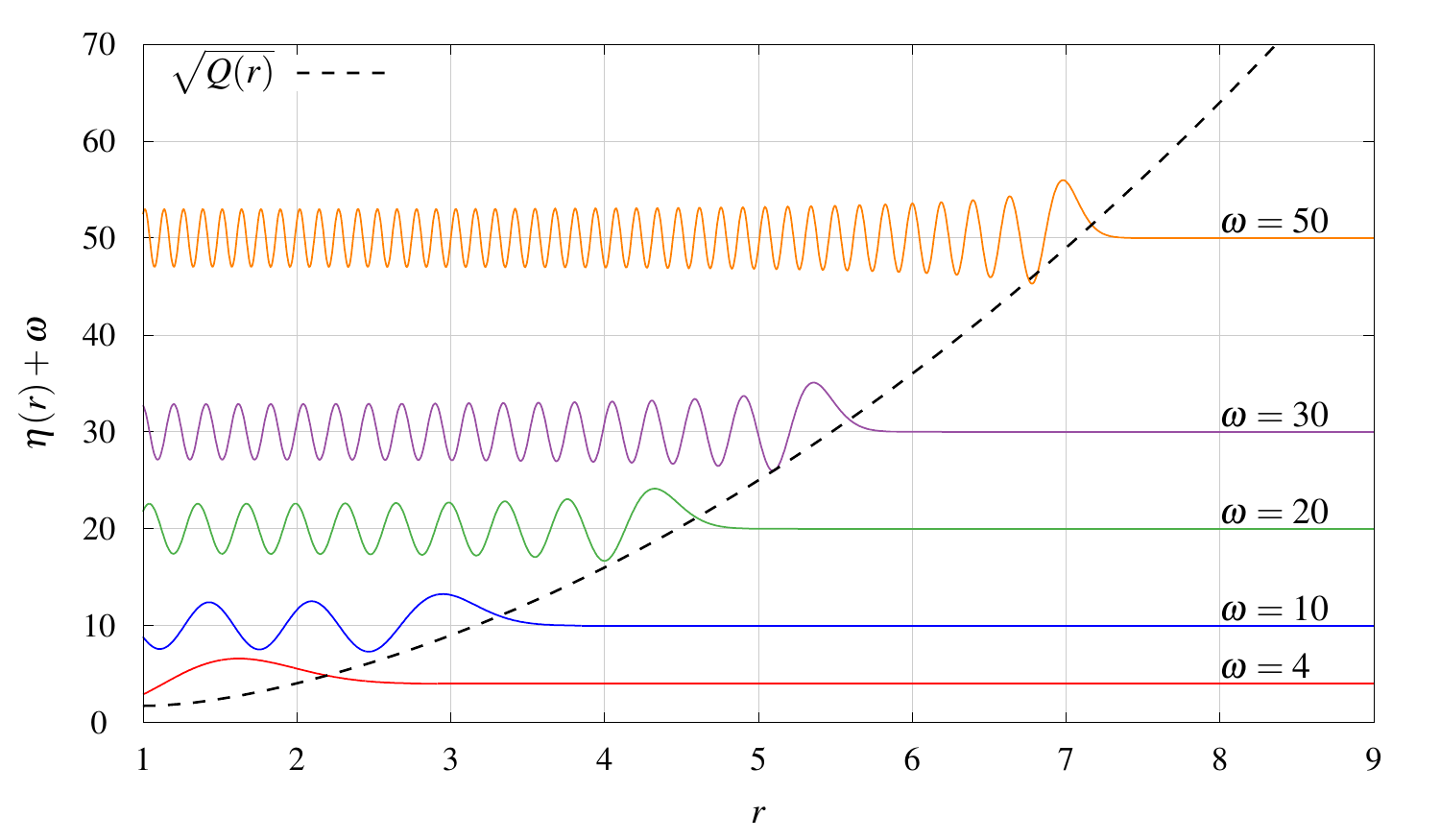}
\caption{Monochromatic wave ranges for scattering in the effective potential $Q(r)$ for the case $a=1$. }\label{fig:penetrators}
\end{figure}

\begin{figure}
\includegraphics[height=12cm]{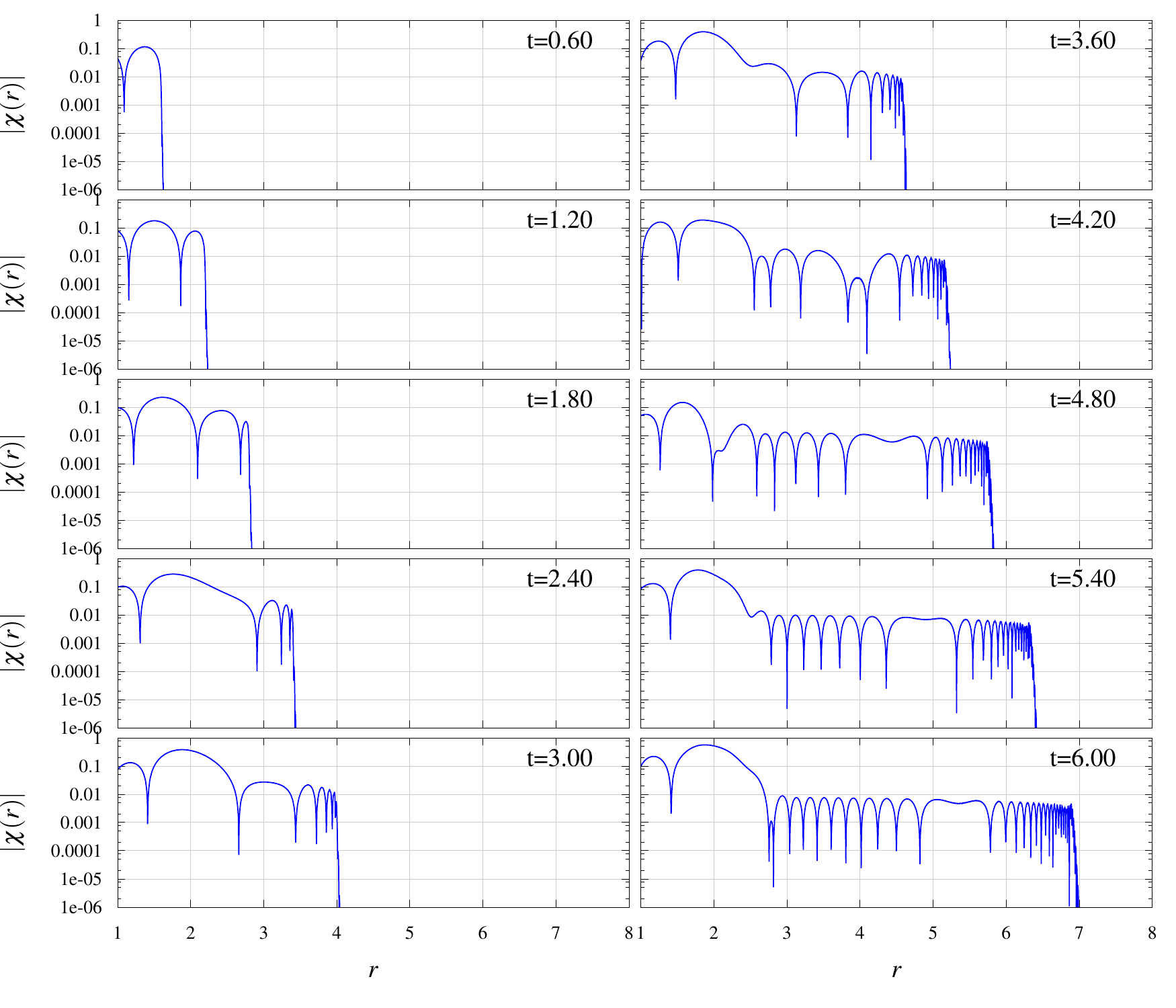}
\caption{Linear evolution of the perturbed skyrmion for $a=1$, with oscillating boundary conditions with the frequency $\omega=6$. For this frequency 
$R_{\rm crit}\simeq 2$.}\label{fig.Linear}
\end{figure}

Let us now consider the problem of radiation escaping from the surface of the skyrmion. 
As a starting point, we consider the case $a=1$. The equation of motion is linear, so there 
is no interaction among modes or between modes and the static solution. The skyrmion is a static solution satisfying the appropriate boundary conditions. 
This is a remarkable feature. Usually, the small perturbation scatters on the background static solution. In this case, the perturbation does not feel the field of 
the skyrmion. However, by choosing the appropriate symmetry for our ansatz, we have broken the symmetry. The effective potential of the Sturm-Liouville problem 
\begin{equation}
 Q(r) =\frac{2}{r^2} + \frac{2r^4}{B^2\boldsymbol{l}^6}
\end{equation} 
and, in particular, the presence of the baryon number $B$ in the second term, is a remnant of the symmetry breaking and the geometry of our ansatz. 
The first term, $2/r^2$, correspond to the dimensionality of physical space.
The effective potential $Q(r)$ grows to $\infty$ as $r\to \infty$ and confines the 
radiation. For each frequency there is a limit $R_\text{crit}(\omega)$ to which the radiation can propagate, defined by the relation $Q(R_\text{crit})=\omega^2$. Here,
 $R_\text{crit}$ is the second root of this equation, which for large amplitudes and $B^2\boldsymbol{l}^6=2$ can be approximated by
\begin{equation}
 R_\text{crit} \simeq \sqrt{\omega} .
\end{equation} 
Examples of ranges $R_{\rm crit}$ and wave profiles are shown in Fig.~\ref{fig:penetrators}.\\
For large frequencies one can think of the perturbation as a massive field propagating with the effective mass $m_{\rm eff}=2r^2/(B\boldsymbol{l}^3)$. Note that 
this mass grows as $r $ increases, so this approximation can be valid only for fast changing waves with large frequency (WKB approximation). 
There always exists some critical radius, say $R_{\rm crit}$ above which the wave cannot propagate any longer. Very similar conclusions can be drawn 
from the  Sturm-Liouville approach. The linearized equation is equivalent to a wave propagating in the unbounded potential $Q(r)$. Because the potential 
is unbounded, the wave can propagate only inside a certain sphere. Outside this sphere, only exponentially decaying tails can be found. The higher the frequency 
the further the wave can penetrate. For the wave packet it appears as if the radiation slows down at some distance and increases its wave 
number. 
An example for the evolution with oscillating boundary condition ($\omega=6$) is presented in Figure \ref{fig.Linear}. For this frequency, 
$R_{\rm crit}\simeq 2$. Clearly the wave propagates to this distance. Next the amplitude of the wave drops exponentially. Beyond that point only 
higher frequencies can propagate. Since the wave has a wave front, it consists of all frequencies, not only of the dominating one ($\omega=6$).  The higher 
the frequency the further the wave propagates.\\

\textit{~~Perturbation series.~~}
Let us now consider how the waves evolve around the static skyrmion solution for $a\neq 1$. We assume that the deviation $\epsilon=a-1$ from $a=1$ is small. We apply the 
perturbation series with $\epsilon$ as the expansion parameter. 
Let us assume that we know a solution describing radiation in the case $a=1$ which we denote $\tilde\chi(r,t)$. 

The solution for $\epsilon\neq 0$ can be written as
\begin{equation}
 \chi=\chi^{(0)}+\epsilon\chi^{(1)}+\ldots,
\end{equation} 
where $\chi^{(0)}(r,t)=\chi_0(r)+A\tilde\chi(r,t)$. Note that in the non-compacton case we can always find an $A$ such that $\chi^{(0)}(r,t)>0$.\\
We also expand the derivative of the potential:
\begin{equation}
 \mathcal{U}_{,\chi} = 2a\chi^{2a-1}=\mathcal{U}_{,\chi}^{(0)}+\epsilon\mathcal{U}_{,\chi}^{(1)}
\end{equation} 
where
\begin{equation}
 \mathcal{U}_{,\chi}^{(0)}=2\chi
\end{equation} 
is the  potential for the linear case $a=1$ and 
\begin{equation}
\mathcal{U}_{,\chi}^{(1)}=2\chi+4\chi\ln\chi.
\end{equation} 
In all orders of the perturbation series the equation takes the form
\begin{equation}
 \partial_{tt}{\chi^{(n)}}-\partial_{rr}{\chi^{(n)}}+\frac{2}{r}{\partial_{r}\chi^{(n)}}+\frac{2r^4}{B^2\boldsymbol{l}^6} \chi^{(n)} + 
g^{(n)}\left(\chi^{(n-1)}, \chi^{(n-2)},\ldots\chi^{(0)}\right)=0.
\end{equation} 
In the zeroth order $g^{(0)}=0$, so the solution is the solution to the linearized equation.\\
In the first order of the perturbation series, the inhomogeneous part is
\begin{equation}\label{eq:source}
 g^{(1)}\left(\chi^{(0)}\right) = 2\chi^{(0)}+4\chi^{(0)}\ln\left(\chi^{(0)}\right).
\end{equation} 
For the non-compacton case and $A=0$
\begin{equation}
\ln\left(\chi^{(0)}\right)=\frac{1}{\epsilon}\ln\left[\frac{\pi}{2}\left(1-\frac{r^3}{R^3}\right)\right],\;\;R^3 = \frac{3B \boldsymbol{l}^3 
}{2\epsilon}\left( 
\frac{\pi}{2}\right)^\epsilon .
\end{equation} 
Note that $\ln\left(\chi^{(0)}\right)=\ln(\pi/2)/\epsilon+\mathcal{O}(\epsilon^0)$. This term lowers the order of the equation, which means
that the perturbation series written in such a way is inconsistent. 
This is something we have actually expected since the skyrmion profile for $a>1$ has a power like behavior while the linear equation ($a=1$) gives 
either exponential approach or 
none (static solution, the linear equation has a superposition rule). 
The waves have significantly different properties in the $a=1$ and the $a>1$ cases. The $a=1$ wave can be decomposed as a linear 
combination of bound modes, each of them vanishing at spatial infinity. The $a>1$ waves behave similarly for small values of $r$, however, for large 
values of $r$, beyond the barrier they can move freely. For $a=1$ the spectrum is discrete and for $a>1$ it is continuous.\\
For $\epsilon<0$ and $a<1$, the static solution is a localized compacton with radius $R$. For $r<R$, we can use the same argument about the 
inconsistency of the perturbation series. Inside the compacton, the perturbation evolves obeying an appropriate linearized equation.  If the perturbation 
vanishes at the surface, there is no need to consider any radiation outside the compacton. The whole perturbation can be decomposed as a linear 
combination of the infinite but discrete spectrum of bound modes. However, if the perturbation does not vanish at the surface of the compacton, it can 
escape the skyrmion. For $r>R$, $\chi_{0}(r)=0$, which is consistent with the case of the linear static solution for $a=1$. This is the only case when our 
perturbation series can be valid. $\chi^{(0)}$ can now be chosen as the solution describing a wave which propagates up to a certain critical distance 
 $R_{\rm 
crit}$. Because $\chi^{(1)}$ obeys the same equation with the additional source term $g^{(1)}(\chi^{(0)})$, it has similar properties and cannot propagate 
to further distances. Thus,
$\chi^{(1)}$ is indeed a small correction to $\chi^{(0)}$. Note that it may happen that low frequencies cannot propagate outside the compacton if 
$R_{\rm crit}<R$. Please note the similarity between the linear evolution Figure \ref{fig.Linear} and the evolution outside the compacton (see, e.g., Figure \ref{Figurecompacton3}).\\

The above considerations can be summarised as follows.
\begin{itemize}
 \item For the case $a>1$, the waves are asymptotically free. When they cross the barrier they can move to infinity. The barrier's shape chooses 
certain frequencies which fit well inside the barrier forming quasi-normal modes.
\item For $a=1$, the waves can propagate only up to a certain distance. The higher the frequency the further the wave can reach. Beyond  the
critical value $r=R_{\rm crit}\simeq\sqrt{\omega}$, the wave can no longer propagate.
\item For compactons $a<1$, the wave propagating inside the skyrmion moves as the standard linearization method tells. However, if the radiation 
leaves the compacton, its evolution is less trivial. The propagation is described by a nonlinear equation. But for small deviations from $a=1$, the 
wave can be treated perturbatively. The most important feature which remains from the $a=1$ case is that the wave is trapped inside some critical radius $R_{\rm crit}$.
\end{itemize}

We remark that videos of the time evolution of perturbed compactons (for $a=0.55$ and $a=0.95$) and of the perturbed linear model ($a=1$) have been added as supplementary material to this eprint. The features described here can be clearly seen in the videos, and we urge the reader to view them.

\end{document}